\documentclass[a4paper, twocolumn, 11pt]{quantumarticle}
\pdfoutput=1
\usepackage[utf8]{inputenc}
\usepackage[english]{babel}
\usepackage{mathrsfs}
\usepackage[T1]{fontenc}
\usepackage[numbers]{natbib}
\usepackage{hyperref}
\usepackage{amsmath, amssymb, amsthm}
\usepackage{graphicx}
\usepackage{physics}
\usepackage{braket}
\usepackage{float}
\usepackage{tikz}
\usepackage{lipsum}
\usepackage{bbold}
\usepackage{bm}
\usepackage{multirow}
\usepackage{xcolor}
\usepackage{booktabs,multirow}
\usepackage{caption}
\usepackage{comment}
\usepackage{tikz}
\usetikzlibrary{positioning, fit, calc}
\newcommand{\pma}[1]{\textcolor{blue}{#1}}

\title{Magic for Hybrid Boson-Fermion Systems:\newline A Grassmann Phase-Space Approach}

\author{Matthieu Sarkis\href{https://orcid.org/0009-0002-5494-8406}{\includegraphics[scale=0.04]{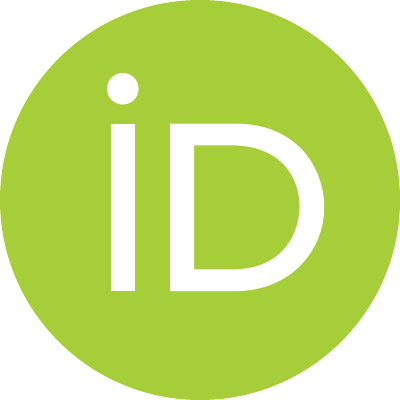}}}
\email[]{matthieu.sarkis@uni.lu}
\affiliation{Department of Physics and Materials Science, University of Luxembourg, L-1511 Luxembourg City, Luxembourg}

\author{Pablo Martinez-Azcona\href{https://orcid.org/0000-0002-9553-2610}{\includegraphics[scale=0.04]{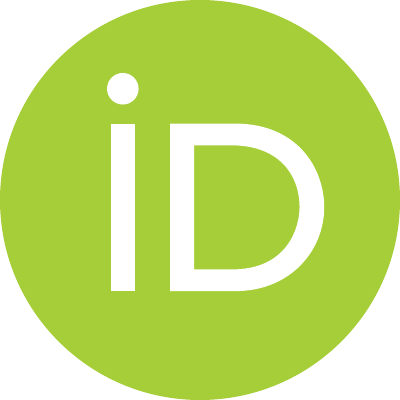}}}
\email[]{pablo.martinez-azcona@wmi.badw.de}
\affiliation{Department of Physics and Materials Science, University of Luxembourg, L-1511 Luxembourg City, Luxembourg}

\author{Alexandre Tkatchenko\href{https://orcid.org/0000-0002-1012-4854}{\includegraphics[scale=0.04]{./figures/orcidid.pdf}}}
\email[]{alexandre.tkatchenko@uni.lu}
\affiliation{Department of Physics and Materials Science, University of Luxembourg, L-1511 Luxembourg City, Luxembourg}

\begin{document}

\begin{abstract}
    Non-stabilizerness enables universality beyond Gaussian/Clifford dynamics, yet no resource theory exists for systems combining bosonic and fermionic degrees of freedom. Using the Grassmann approach of Cahill and Glauber, we develop a phase-space framework defining hybrid magic via the $\bm{L_p}$ norm of a hybrid Wigner function. We demonstrate it in the Holstein polaron, where phonon–electron coupling enhances magic growth, and in the fermionic Jaynes–Cummings model, examining dependence on atomic and cavity states. At the gate level, we define the non-stabilizer power of hybrid operations and derive a closed-form result for the conditional displacement gate. This establishes a unified quantification of non-stabilizerness in realistic hybrid systems.
\end{abstract}

\maketitle

\section{Introduction}

    Quantum computation derives its power from resources that transcend classical simulation, among which magic--or non-stabilizerness--plays a central role~\cite{bravyi2005universal,Veitch_2014,veitch2012negative,howard2014contextuality}. The resource theory of magic provides a quantitative framework for understanding the non-classicality of quantum states and operations, particularly in the context of fault-tolerant quantum computation, where Clifford operations alone are insufficient for universality~\cite{Veitch_2014,howell2022quantifying}.

    The foundational link between magic and the negativity of phase-space quasi-probability distributions, such as the Wigner function, has been established in both continuous-variable and later in finite dimensional spin systems~\cite{kenfack2004negativity,wootters1987wigner,gibbons2004discrete,mari2012positive}. In the context of continuous-variable systems, the negativity of the Wigner function is nicely captured by the Mana:
    \begin{equation}
        \textsc{Mana}(\rho) = \log\left[\int \left|\mathcal W_\rho(\bm\alpha)\right|\frac{\text{d}^{2M}\alpha}{\pi^M}\right]\,,
    \end{equation}
    where $W_\rho$ is the Wigner function of the state $\rho$. Negativity in these representations is not only a signature of non-classicality but also a necessary resource for quantum computational speedup~\cite{veitch2012negative,howard2014contextuality}. Recent works have further formalized the quantification of magic for multi-qubit operations~\cite{howell2022quantifying}, and extended resource-theoretic concepts to bosonic systems~\cite{kenfack2004negativity,mari2012positive} and fermionic Gaussian states~\cite{collura2412quantum,leone2022stabilizer}.

    A key challenge in developing a resource theory of magic for hybrid systems is the construction of appropriate monotones and operational tasks that reflect the hybrid structure. In the qubit and bosonic settings, resource monotones such as the robustness of magic and Wigner negativity have been shown to be closely related to classical simulability and contextuality~\cite{veitch2012negative,howard2014contextuality,Veitch_2014}. For qubit systems a key magic monotone is the stabilizer Rényi entropy (SRE)~\cite{leone2022stabilizer, leone2024stabilizer}, which can be easily computed in terms of expectation values of Pauli strings. For fermionic systems, recent work has explored the structure of the Majorana Clifford group~\cite{bettaque2024structure}, and the non-stabilizerness of the Sachdev-Ye-Kitaev model~\cite{bera2502non, jasser2025stabilizerentropyentanglementcomplexity, malvimat2026multipartite}.

    We now define the fermionic SRE in terms of Majorana strings as done in \cite{bera2502non}. Let $\gamma_1,\ldots,\gamma_{2N}$ be Majorana operators satisfying the Clifford algebra $\{\gamma_a,\gamma_b\}=2\delta_{ab}$ and let $\{\Gamma_\mu\}$ denote the orthonormal basis of all Majorana strings.  For a state $\rho$ (pure or mixed) define the coefficients
    \begin{equation}
      p_\mu = 2^{-N}\,\big|\mathrm{Tr}\big(\rho\,\Gamma_\mu\big)\big|^2,\qquad \sum_\mu p_\mu=1\,.
    \end{equation}
    The stabilizer $\alpha$-Rényi entropy is then defined as:
    \begin{equation}
    \label{eq:sre_def}
      \text{SRE}_\alpha(\rho)=\frac{1}{1-\alpha}\,\log\left(\sum_\mu p_\mu^\alpha\right) - N\log(2)\,.
    \end{equation}

    Beyond its definition, the SRE has been widely adopted to probe non-stabilizerness across diverse settings. It has been used to quantify magic growth and spreading in random quantum circuits and generic ergodic dynamics~\cite{turkeshi2025magic,tirrito2024anticoncentration}, to characterize non-stabilizerness in permutationally invariant models and in kinetically constrained Rydberg-atom arrays~\cite{passarelli2024nonstabilizerness,smith2406non}, and to study strongly interacting fermionic systems such as the Sachdev-Ye-Kitaev model and fermionic Gaussian states~\cite{bera2502non,collura2412quantum}. Related applications appear in quantum optics, for instance, tracking the dynamics of atomic magic in the Jaynes-Cummings model~\cite{shuangshuang2022dynamics}. In non-Hermitian settings, it has also been employed to design and diagnose protocols for producing magic steady states~\cite{martinezazcona2025magic}. More recently, the SRE has also been used in quantum chemistry to assess non-stabilizerness in molecular bonding~\cite{sarkis2025molecules}.
    Additional works include random-circuit and monitored-dynamics settings~\cite{bejan2024dynamical,niroula2024phase,zhang2024quantummagicdynamics,turkeshi2025magic,tirrito2024anticoncentration}, spin-model ground states and monotones~\cite{oliviero2022magic,haug2023stabilizer,rattacaso2023stabilizer,tarabunga2024critical,frau2024nonstabilizerness,robin2025stabilizeraccelerated,viscardi2025interplay}, non-integrable evolutions~\cite{odavic2024stabilizer}, chaos diagnostics~\cite{goto2021chaos}, connections between Pauli-Markov chains, criticality, and gauge theories~\cite{tarabunga2023many,falcao2025nonstabilizerness,santra2025quantumresources,esposito2025magic}, quantum chemistry and molecular simulation~\cite{gu2024zero,sarkis2025molecules}, and nuclear-physics applications~\cite{robin2024magic,brokemeier2025quantum}.

    A central contribution of this work is the \textit{derivation} of the expression for the fermionic SRE, as introduced by Leone et al. \cite{leone2022stabilizer}, from first principles using phase-space methods, cf. Sec. \ref{sec:fermionicMagic}. By starting from the structure of the fermionic Wigner function and its generalization to hybrid systems, we provide a principled motivation for the SRE as a natural resource monotone in the context of hybrid boson-fermion quantum systems. This approach not only clarifies the operational meaning of the SRE, but also demonstrates how phase-space techniques can unify and extend resource-theoretic concepts across different quantum platforms.

    Despite advances on both the finite-dimensional and continuous-variable frontlines, the extension of magic to hybrid systems—those comprising both bosonic and fermionic degrees of freedom—remains an open frontier. Such systems are not only of theoretical interest, as in supersymmetric quantum mechanics~\cite{abe2001theory}, but also arise in practical quantum simulation platforms and models of quantum matter~\cite{dalton2013grassmann,sanchez2025jaynes}. Hybrid boson-fermion settings thus provide a broad arena for quantum information science and quantum simulation, encompassing scenarios where both types of degrees of freedom coexist and interact. The development of phase-space methods for hybrid systems, including Grassmann-valued representations and supercoherent states, provides a promising foundation for this general approach~\cite{cahill1999density,dalton2013grassmann,abe2001theory}, enabling the study of non-classicality and resource theories across a wide range of experimental and theoretical platforms.

    In this work, we propose a unified phase-space approach to define and quantify magic in hybrid boson-fermion systems. Our framework leverages the structure of hybrid phase-space representations to generalize the notion of Wigner function negativity and resource monotones to the hybrid setting. We demonstrate how this approach naturally connects to operational tasks relevant in quantum information, such as classical simulability, contextuality, and the structure of quantum circuits~\cite{howard2014contextuality,Veitch_2014,bera2502non}. We also discuss potential applications to supersymmetric quantum mechanics as a notable example, as well as to quantum simulation and the study of non-classicality in generic hybrid quantum systems.

    In the present work, the term ``magic'' should be understood operationally as quantifying the departure from the classically simulable sector of a hybrid boson-fermion processor, or equivalently of a qubit-oscillator architecture when the fermionic degrees of freedom are encoded in qubits. This viewpoint is aligned with current hybrid oscillator-qubit processor architectures, where continuous-variable and finite-dimensional subsystems are combined within a single computational platform~\cite{liu2024hybrid}. In a forthcoming work~\cite{hybrid_resource}, we make this interpretation precise by providing a rigorous construction of hybrid non-stabilizerness for boson-fermion systems together with the corresponding resource-theoretic structure.

    We also stress that, for fermions, the resource theory of non-stabilizerness should be distinguished from a resource theory of non-Gaussianity. The latter is organized around fermionic Gaussian states and Gaussian transformations, whereas the former is organized around the Majorana-stabilizer/Clifford structure and the spread over Majorana strings. In the present paper, our focus is on fermionic non-stabilizerness, and on its hybrid extension to boson-fermion systems.

\begin{figure}[H]
\centering
\resizebox{\columnwidth}{!}{%
\begin{tikzpicture}[
  box/.style={
    draw, rounded corners=5pt, thick, align=center,
    inner sep=5pt, font=\scriptsize
  },
  halfs/.style={box, text width=3.15cm, minimum height=1.15cm},
]

\node[halfs, anchor=north west] (VA) at (0,0) {
  \textbf{\ref{sec:hybHO}}\\[1pt] Hybrid HO
};
\node[halfs, anchor=north west] (VB) at (3.55,0) {
  \textbf{\ref{sec:dressedCat}}\\[1pt] Dressed Cat
};
\node[halfs, anchor=north west] (VC) at (0,-1.4) {
  \textbf{\ref{sec:holstein}}
  \quad\footnotesize[CM]\\[1pt] Holstein model
};
\node[halfs, anchor=north west] (VD) at (3.55,-1.4) {
  \textbf{\ref{sec:JaynesCummings}}
  \quad\footnotesize[QO]\\[1pt] Jaynes-Cummings
};
\node[halfs, anchor=north west] (VE) at (0,-2.8) {
  \textbf{\ref{sec:TavisCummings}}
  \quad\footnotesize[QO]\\[1pt] Tavis-Cummings
};
\node[halfs, anchor=north west] (VF) at (3.55,-2.8) {
  \textbf{\ref{subsec:susy_magic}}
  \quad\footnotesize[HEP]\\[1pt] Supersymmetric QM
};

\node[
  draw, rounded corners=5pt, thick, dashed,
  fit=(VA)(VB)(VC)(VD)(VE)(VF),
  inner sep=8pt
] (secV) {};
\node[font=\scriptsize\bfseries, anchor=south]
  at ([yshift=6pt]secV.north)
  {Sec.~\ref{sec:applications}: Applications};

\path
  let
    \p1 = (secV.west),
    \p2 = (secV.east),
    \n1 = {\x2-\x1},
    \n2 = {0.5*\n1-5pt}
  in
  node[
    box,
    minimum width=\n1,
    text width=\dimexpr\n1-10pt\relax,
    minimum height=0.85cm,
    anchor=south
  ] (secIV) at ([yshift=28pt]secV.north) {
    \textbf{Sec.~\ref{sec:resource_theory}:}
    \quad\footnotesize[QI]\\[1pt]
    Resource-theoretic Properties
  }
  node[
    box,
    minimum width=\n2,
    text width=\dimexpr\n2-10pt\relax,
    minimum height=2.55cm,
    anchor=south west
  ] (secII) at ([yshift=10pt]secIV.north west) {
    \textbf{Sec.~\ref{sec:fermionicMagic}:}\\
    Fermionic SRE as\\
    Wigner negativity over\\
    Grassmann phase space\\
    Eq.~\eqref{eq:fermionicSRE_grassmann}
  }
  node[
    box,
    minimum width=\n2,
    text width=\dimexpr\n2-10pt\relax,
    minimum height=2.55cm,
    anchor=south east
  ] (secIII) at ([yshift=10pt]secIV.north east) {
    \textbf{Sec.~\ref{sec:generaliz_Hyb}:}\\
    Hybrid Boson-Fermion\\
    magic over hybrid\\
    Grassmann phase space\\
    Eq.~\eqref{eq:hybrid_magic}
  }
  node[
    box,
    minimum width=\n1,
    text width=\dimexpr\n1-10pt\relax,
    minimum height=0.85cm,
    anchor=north
  ] (secVI) at ([yshift=-10pt]secV.south) {
    \textbf{Sec.~\ref{sec:magic_power}:}
    \quad\footnotesize[QI]\\[1pt]
    Magic Power of Hybrid Gates
  };

\end{tikzpicture}%
}
\caption{Paper structure. The brackets denote the main field to which the section is relevant: Quantum Information [QI], Condensed Matter [CM], High Energy Physics [HEP] and Quantum Optics [QO].}
\end{figure}
    By bridging the gap between the resource theory of magic and the rich structure of hybrid quantum systems, our results open new avenues for foundational studies and practical applications in quantum information science. We anticipate that this framework will stimulate further research into the role of non-stabilizerness in a wide variety of hybrid systems, including supersymmetric models.

    In Sec. \ref{sec:applications}, we illustrate our framework through several representative examples. We begin with the free supersymmetric quantum harmonic oscillator, which provides a minimal model for hybrid boson-fermion systems. We then analyze the dressed cat state, a paradigmatic example of bosonic non-classicality coupled to a fermionic degree of freedom, and compare it to its purely bosonic counterpart. The Holstein model, a cornerstone of polaron physics in condensed matter, serves as a realistic setting where electron-phonon coupling gives rise to hybrid magic~\cite{hohenadler2007lang, rongsheng2002exact}. Lastly, in the context of quantum optics, we investigate the behavior of hybrid magic for the fermionic Jaynes-Cummings model~\cite{dalton2013grassmann,sanchez2025jaynes}, we study the dynamics and the maximum value of the hybrid magic for many different atomic and photonic initial states, highlighting the role of initial non-classicality. These examples demonstrate the versatility of our approach and its relevance to a broad range of physical platforms, studied in fields ranging from Quantum Information, Quantum Optics or Condensed Matter to High Energy Physics.

\section{Fermionic phase space and magic}\label{sec:fermionicMagic}

    In this section, we provide a first-principles derivation of the fermionic SRE. Suppose we are given $N$ fermionic modes, with annihilation and creation operators $\hat{c}_n$ and $\hat{c}_n^\dagger$ respectively, which satisfy the canonical anticommutation relations:
    \begin{equation}
        \left\{c_n, c_m^\dagger\right\} = \delta_{nm}\,.
    \end{equation}
    Following Cahill and Glauber \cite{cahill1969ordered}, we define the $s$-ordered displacement operator as:
    \begin{equation}
        D(\boldsymbol{\xi}; s) = \bigotimes_{n=1}^N \exp\left(c_n^\dagger\xi_n-\bar\xi_n c_n+\frac{s}{2}\bar\xi_n\xi_n\right)\,,
    \end{equation}
    where $\left\{\xi_n,\bar\xi_n\right\}_{n=1}^N$ are (complex) Grassmann variables. We refer the reader to the Supplemental Material for a very short overview of Grassmann variables. Given a quantum state $\rho$, we then define the generating function as:
    \begin{equation}
        \chi_\rho(\boldsymbol{\xi}; s) = \text{Tr}\left[\rho D(\boldsymbol{\xi} ; s)\right]\,.
    \end{equation}
    The Wigner function can then be defined naturally as the symplectic Fourier transform of the generating function:
    \begin{equation}
        W_\rho(\boldsymbol{\theta}; s) = \int\text{d}^{2N}\xi\, e^{\sum_n\left(\theta_n\bar\xi_n - \xi_n\bar\theta_n\right)} \chi_\rho(\xi; s)\,.
    \end{equation}
    One can alternatively understand the Wigner function as the expected value in the state $\rho$ of \textit{phase-point operators}:
    \begin{equation}
        W_\rho(\boldsymbol \theta; s) = \text{Tr}\left[\rho \Delta(\boldsymbol \theta; s)\right]\,,
    \end{equation}
    with
    \begin{equation}
    \label{eq:total_phase_point_operator}
        \Delta(\boldsymbol \theta; s) = \bigotimes_{n=1}^N \Delta_n(\theta_n; s)\,,
    \end{equation}
    where the local phase-point operators are given by:
    \begin{equation}
        \begin{aligned}
            &\Delta_n(\theta_n; s) = \int\text{d}^{2}\xi_n\\
            &\exp\left[\frac{s}{2}\,\bar\xi_n\xi_n+\left(c_n + \theta_n\right)\bar\xi_n - \xi_n(c_n^\dagger + \bar\theta_n)\right]\,.
        \end{aligned}
    \end{equation}
    A little bit of Grassmann gymnastics allows to rewrite the integrand as:
    \begin{equation}
        \begin{aligned}
            &\exp\left[\frac{s}{2}\,\bar\xi_n\xi_n+\left(c_n + \theta_n\right)\bar\xi_n - \xi_n(c_n^\dagger + \bar\theta_n)\right] = \\
            &= 1 + \left(c_n + \theta_n\right)\bar\xi_n - \xi_n(c_n^\dagger + \bar\theta_n) + \\
            & + \bar\xi_n\xi_n\left[(c_n^\dagger + \bar\theta_n)\left(c_n + \theta_n\right)-\frac{1-s}{2}\right]\,.
        \end{aligned}
    \end{equation}
    With the following convention (cf. Supplemental Material) for Berezin integration\footnote{Our convention differs in particular from~\cite{cahill1999density} by a global sign. This is a mere convention.}:
    \begin{equation}
        \int\text{d}^{2}\xi_n\, \bar\xi_n\xi_n = 1\,,
    \end{equation}
    we then obtain the following very simple expression for the local phase-point operators:
    \begin{equation}
    \label{eq:local_phase_point_operator}
        \begin{aligned}
            &\Delta_n(\theta_n; s) = (c_n^\dagger + \bar\theta_n)(c_n + \theta_n) + \frac{s-1}{2}\,.
        \end{aligned}
    \end{equation}

    The s-ordered Wigner function can be written as:
    \begin{equation}
        W_\rho(\boldsymbol \theta; s) = \text{Tr}\left[\rho \bigotimes_{n=1}^N \Delta_n(\theta_n; s)\right]\,.
    \end{equation}
    This expression is Grassmann-valued and can be expanded along the Grassmann directions. For instance, in the case of a single fermionic mode, we have:
    \begin{equation}
        \begin{aligned}
            W_\rho(\theta; s) &= \text{Tr}\left[\rho\left(c^\dagger c-\frac{1-s}{2}\right)\right]+\\
            &+\text{Tr}\left[\rho c^\dagger\right]\theta-\text{Tr}\left[\rho c\right]\bar\theta+\text{Tr}\left[\rho\right]\bar\theta\theta\,.
        \end{aligned}
    \end{equation}
    In order to establish connection with the fermionic Stabilizer Renyi Entropy, let us rotate to a real basis by defining the following real Grassmann variables and Majorana operators:
    \begin{equation}
        \begin{aligned}
            \gamma_1 &= c+c^\dagger\,,\ \ \ \gamma_2 = -i(c-c^\dagger)\,, \\
            \vartheta_1 &= \theta+\bar\theta\,,\ \ \ \ \vartheta_2 = -i(\theta-\bar\theta)\,.
        \end{aligned}
    \end{equation}
    One then obtains the following expression for the fermionic Wigner function:
    \begin{equation}
        \begin{aligned}
            &W_\rho(\theta; s) = \frac{1}{2}\Bigg\{\text{Tr}\left[\rho\left(i\gamma_1\gamma_2+s\right)\right]+\\
            &-\text{Tr}\left[\rho\gamma_2\right]i\vartheta_1+\text{Tr}\left[\rho\gamma_1\right]i\vartheta_2-
            \text{Tr}\left[\rho\mathbb 1\right]i\vartheta_1\vartheta_2\Bigg\}\,.
        \end{aligned}
    \end{equation}
    One recognizes the four Majorana strings pertaining to a single-mode fermionic system. In particular, one can foresee that selecting the symmetric or Weyl ordering ($s=0$) will allow to establish a connection with the fermionic Stabilizer Renyi Entropy, as defined in \cite{leone2022stabilizer, bera2502non}. The factors of $i$ sitting in front of the Grassmann monomials are simply here to ensure reality of the Wigner function.


    We now refer the reader to the Supplemental Material in which we provide the definition of the $L_p$ norm of a function of Grassmann variables. With this definition at hand, we are now equipped to define the $p$-fermionic magic as the $L_p$ norm of the fermionic Wigner function\footnote{The $p$-dependent prefactor in the definition will be justified soon.}:
    \begin{equation}
    \begin{aligned}
        \mathcal M_p(\rho; s) &= \frac{1}{1-\tfrac{p}{2}}\,\log\left(\frac{1}{2^N}\lVert W_\rho(\star; s)\rVert_p^p\right)\\
        &= \frac{1}{1-\tfrac{p}{2}}\,\log\left(\frac{1}{2^N}\sum_I \left|\frac{\mathrm{Tr}(\rho \Gamma_I)}{2^{N/2}}\right|^p\right).
    \end{aligned}
    \end{equation}
    The star simply indicates that we are taking the $L_p$ norm of $W_\rho$ viewed as a function of its first argument. For the previous example of a single-mode the fermionic magic with Weyl ordering $s=0$ thus reads
    \begin{equation}\label{eq:fermionicSRE_grassmann}
        \begin{aligned}
            \mathcal M_p = \frac{1}{1-\frac{p}{2}} \log \Biggl[
            &\frac{1  + |\braket{\gamma_1}|^p + |\braket{\gamma_2}|^p}{2^{\frac{p}{2}+1}}\\
            &\qquad + \frac{|\braket{i\gamma_1 \gamma_2}|^p}{2^{\frac{p}{2}+1}}
            \Biggr],
        \end{aligned}
    \end{equation}
    where $\braket{\bullet}=\Tr(\rho \bullet)$ denotes the expectation value over the state.

    Note that setting $p=1$ provides a reasonable definition of bosonic Wigner negativity and therefore of mana. Moreover, for $p=2k$, we have provided a first-principle derivation of the fermionic Stabilizer $k$-Renyi Entropy, as defined in \cite{leone2022stabilizer, bera2502non}, and recalled in Eq. (\ref{eq:sre_def}). The value $p=2k=1$ matches then the stabilizer norm of \cite{campbell2011catalysis}. Though not a magic monotone per se \cite{leone2024stabilizer}, the $p=1$ case will be relevant to us in the next section when we start including bosonic degrees of freedom.

\section{Generalization to hybrid boson-fermion systems}\label{sec:generaliz_Hyb}

    The generalization of the fermionic phase-space approach to hybrid boson-fermion systems is straightforward. We consider a system of $N$ fermionic modes, with annihilation and creation operators $c_n$ and $c_n^\dagger$ respectively, and $M$ bosonic modes, with annihilation and creation operators $a_n$ and $a_n^\dagger$ respectively. The total phase-point operator factorizes as:
    \begin{equation}
        \Delta(\bm\alpha, \bm\theta; \bm s) = \Upsilon(\bm\alpha; r)\otimes\Delta(\bm\theta; s)\,,
    \end{equation}
    where $\boldsymbol{s} = (r,s)$ is the set of ordering parameters for the bosonic and fermionic degrees of freedom, respectively, and $\Upsilon$ is the bosonic phase-point operator, as defined in the standard quantum optics literature:
    \begin{equation}
        \Upsilon(\bm\alpha; r) = \int \frac{d^{2M}\xi}{\pi^M}\,\exp\left(\bm\alpha\cdot\bm{\bar\xi}-\bm{\bar\alpha}\cdot\bm\xi\right)D(\bm \xi; r)\,,
    \end{equation}
    where both $\bm\alpha$ and $\bm\xi$ are standard c-numbers-valued vectors. Recall that the bosonic displacement operator is given by:
    \begin{equation}
        D(\bm\xi; r) =\exp\left(\bm\xi\cdot\bm{a^\dagger} - \bm{\bar\xi}\cdot \bm a+\frac{r}{2}\,\bm {\bar\xi}\cdot\bm\xi\right)\,,
    \end{equation}
    with of course the bosonic creation and annihilation operators satisfying the canonical commutation relations $[a_k, a_l^\dagger] = 1$. The fermionic phase-point operator is given in eqs. (\ref{eq:total_phase_point_operator}) and (\ref{eq:local_phase_point_operator}). The Wigner function is then a function of both the phase space variables $\alpha_n$ and the Grassmann variables $\theta_n$. Rotating to a real basis of Grassmann variables again, the coefficients of the real Grassmann expansion are now functions of the bosonic phase space variables\footnote{Again, the factor of $i$ simply ensures reality and we extract it from the definition of the coefficients $w_I$.}:
    \begin{equation}
        W_\rho(\bm\alpha, \bm\vartheta; \bm s) = \frac{1}{2^N}\sum_{I}i^{\omega(I)}w_{I}(\bm\alpha; \bm s)\,\vartheta_I\,,
    \end{equation}
    where the summation is a multi-index accounting for the whole collection of fermionic modes. We refer the reader to Sec. \ref{app:superLp} of the Supplemental Material for the generic definition of the $L_p$ norm of a function of both c-number (bosonic) and Grassmann variables.     One can then define the $p$-hybrid magic as the $L_p$ norm of the hybrid Wigner function:
    \begin{equation}
    \label{eq:hybrid_magic}
        \begin{aligned}
            &\mathcal M_p(\rho; \bm s)
            = \frac{1}{1-\tfrac{p}{2}}\,\log\left(\frac{1}{2^{N}}\lVert W_\rho(\star, \star; \bm s)\rVert_p^p\right)\\
            &= \frac{1}{1-\frac{p}{2}} \log \Biggl(
            \int_{\mathbb{C}^M} \sum_{I}
            |\mathrm{Tr}(\rho \Upsilon(\boldsymbol \alpha) \Gamma_I)|^p\frac{d^{2M}\boldsymbol \alpha}{2^N\pi^M}\Biggr)
        \end{aligned}
    \end{equation}

    Let us specialize to the case of a single fermionic mode and an arbitrary number of bosonic modes. The hybrid Wigner function is then given by:
    \begin{equation}
        \begin{aligned}
            &W_\rho(\bm\alpha, \bm\vartheta; \bm s) = \frac{1}{2}\Bigg\{\text{Tr}\left[\rho\Upsilon(\bm\alpha; r)\left(i\gamma_1\gamma_2+s\right)\right]+\\
            &-\text{Tr}\left[\rho\Upsilon(\bm\alpha; r)\gamma_2\right]i\vartheta_1+\text{Tr}\left[\rho\Upsilon(\bm\alpha; r)\gamma_1\right]i\vartheta_2+\\
            &-\text{Tr}\left[\rho\Upsilon(\bm\alpha; r)\mathbb 1\right]i\vartheta_1\vartheta_2\Bigg\}\,.
        \end{aligned}
    \end{equation}
    leading to the following expression for the $p$-hybrid magic:
    \begin{equation}\label{eq:hybrid_magic_single_mode}
        \begin{aligned}
            &\mathcal M_p(\rho; \bm s) = \frac{1}{1-\tfrac{p}{2}}\,\log\frac{1}{2}\int_{\mathbb C^M}\Bigg\{\left|\text{Tr}\left[\rho\Upsilon(\bm\alpha; r)\right]\right|^p+\\
            &+\left|\text{Tr}\left[\rho\Upsilon(\bm\alpha; r)\gamma_1\right]\right|^p+\left|\text{Tr}\left[\rho\Upsilon(\bm\alpha; r)\gamma_2\right]\right|^p+\\
            &+\left|\text{Tr}\left[\rho\Upsilon(\bm\alpha; r)\left(i\gamma_1\gamma_2+s\right)\right]\right|^p\Bigg\}\,\frac{\text{d}^{2M}\alpha}{\pi^M}\,.
        \end{aligned}
    \end{equation}

    We can also express the hybrid magic in the case of $N$ fermionic modes. For definiteness we fix the Weyl ordering $r=s=0$. For each subset $I=\{i_1<\cdots<i_{|I|}\}\subseteq\{1,\dots,2N\}$ write $\gamma_I=\gamma_{i_1}\cdots\gamma_{i_{|I|}}$ and define
    \begin{equation}
    \Gamma_I=i^{|I|(|I|-1)/2}\gamma_I.
    \end{equation}
    Then we have hermiticity $\Gamma_I^\dagger=\Gamma_I$ and each Majorana string squares to the identity $\Gamma_I^2=\mathbb{1}$.
    We then define the following \emph{dressed Wigner functions}:
    \begin{equation}\label{eq:wI}
    w_I(\bm\alpha)=\mathrm{Tr}\big(\rho\,\Upsilon(\bm\alpha;0)\otimes \Gamma_I\big)\,,
    \end{equation}
    In terms of which the hybrid magic reads:
    \begin{equation}
    \label{eq:hybrid_magic_explicit}
        \begin{aligned}
            &\mathcal M_p(\rho; \bm 0)=\\
            &\quad= \frac{1}{1-\frac{p}{2}}\log\left[
            \frac{1}{2^N}\sum_{I}\int_{\mathbb C^M}
            \left|w_I(\bm\alpha)\right|^p\,
            \frac{\text{d}^{2M}\alpha}{\pi^M}
            \right]
        \end{aligned}
    \end{equation}

    We see that the prescription measures the spread of the state in the Majorana basis (as measured by the $L_p$ norm). However, the Majorana expected values are evaluated with respect to a weighted measure, where the bosonic phase-point operators fix the weight. The presence of the bosonic degrees of freedom can, therefore, be interpreted from the point of view of the fermionic Wigner function as merely implementing a change of measure.

    Let us make a comment concerning the ordering prescription, which appears as a free parameter in the definition of the hybrid magic. From the bosonic point of view, the symmetric/Weyl ordering prescription appears as more natural, in the sense that it allows to recover the interpretation of magic as distribution of negativity in the Wigner function. From the fermionic point of view, symmetric/Weyl ordering prescription also appears as more natural, in the sense that it allows a direct connection with the definition of the Stabilizer Renyi Entropy, as defined in \cite{leone2022stabilizer, bera2502non}. Note however that the two ordering parameters are not prescribed by the construction, providing a two-parameter family of hybrid magic functions.

    Before proceeding to the abstract resource-theoretic considerations underlying our framework, let us make a last comment regarding the functional shape (\ref{eq:hybrid_magic_explicit}) of the hybrid magic that we have derived. Even though this quantity has its roots in the hybrid boson-fermion super-phase-space, we note that eq.~(\ref{eq:hybrid_magic_explicit}) remains perfectly meaningful upon replacing everywhere Majorana strings $\Gamma_I$ by Pauli strings $P_I$, therefore defining a hybrid qubit-oscillator magic. Note in particular that we naturally recover the matter-dressed Wigner functions studied for instance in \cite{Vlastakis2015, wang2016schrodinger}, providing further motivation for defining the hybrid magic.

\section{Resource-theoretic Properties of the Hybrid magic}
\label{sec:resource_theory}
We now collect the main structural properties of the hybrid magic introduced above. These properties support its interpretation as a natural quantifier of non-stabilizerness in boson-fermion systems and clarify the role of the free sector underlying the construction.

We begin by defining the set of free pure states. In analogy with the standard resource theory of magic, we take the free hybrid pure states to be products of a bosonic Gaussian state and a fermionic Majorana stabilizer state:
\begin{equation}\label{eq:Stab_set}
    \begin{aligned}
        &\textsc{hStab}
        =
        \Bigl\{
        \ket{\psi_{\mathrm g}}\otimes \ket{\phi}
        \;\Big|\;
        \\
        &\qquad
        \ket{\psi_{\mathrm g}}\in \textsc{Gauss},\;
        \ket{\phi}\in \textsc{mStab}
        \Bigr\}.
    \end{aligned}
\end{equation}
Here $\textsc{Gauss}$ denotes the set of pure bosonic Gaussian states, while $\textsc{mStab}$ denotes the set of pure Majorana stabilizer states.

This choice of free set is motivated by classical simulability. On the bosonic side, Gaussian states and Gaussian operations define the standard classically tractable sector, while on the fermionic side the analogous role is played by Majorana stabilizer states and Clifford transformations. It is therefore natural to identify the free pure hybrid states with products of these two efficiently simulable structures. From this perspective, nontrivial boson-fermion hybridization should be regarded as a genuine resource rather than as part of the free sector.

For mixed states, a natural extension of the free set is given by the convex hull of $\textsc{hStab}$,
\begin{equation}
    \mathrm{conv}(\textsc{hStab}),
\end{equation}
namely all probabilistic mixtures of free pure hybrid states. In the present work, however, we restrict the discussion of resource-theoretic properties to pure states.

The free operations on pure states are those that map $\textsc{hStab}$ into itself:
\begin{equation}\label{eq:stab}
    \mathfrak F
    =
    \left\{
    \hat F
    \;\middle|\;
    \hat F \ket{\psi}\in \textsc{hStab}
    \ \text{for all}\
    \ket{\psi}\in \textsc{hStab}
    \right\}.
\end{equation}
A distinguished subset is given by product operations of the form
\begin{equation}
    \hat G \otimes \hat C \in \mathfrak F,
\end{equation}
where $\hat G$ is a bosonic Gaussian unitary and $\hat C$ is a fermionic Clifford unitary. More generally, one can wonder whether free transformations compatible with the bosonic and fermionic phase-space structures factorize necessarily in this way. It is of course a non-trivial claim. A proof of this factorization statement, together with a more complete characterization of the free hybrid sector, will appear in a forthcoming paper~\cite{hybrid_resource}.

\subsubsection{Additivity}

A basic property of the hybrid magic is its additivity on product states. For a product state $\rho_{\mathrm b}\otimes \rho_{\mathrm f}$ one finds
\begin{equation}\label{eq:additivity_HybMag}
    \mathcal M_p(\rho_{\mathrm b}\otimes \rho_{\mathrm f}; \bm s)
    =
    \textsc{Mana}_p(\rho_{\mathrm b}; r)
    +
    \mathrm{SRE}_{p/2}(\rho_{\mathrm f}; s),
\end{equation}
where $\textsc{Mana}_p$ is the bosonic $L_p$-based magic measure,
\begin{equation}
    \begin{aligned}
        &\textsc{Mana}_p(\rho_{\mathrm b}; r)
        :=
        \\
        &=\frac{1}{1-\frac{p}{2}}
        \log
        \int\left|
        \mathcal W_{\rho_{\mathrm b}}(\bm\alpha; r)
        \right|^p
        \frac{d^{2M}\alpha}{\pi^M},
    \end{aligned}
\end{equation}
and $\mathrm{SRE}_{p/2}$ is the fermionic stabilizer Rényi entropy. Thus, on uncorrelated hybrid states, the hybrid magic decomposes exactly into a bosonic contribution and a fermionic contribution. In particular, if the bosonic or fermionic subsystem itself factorizes further, the corresponding term inherits the usual additivity properties of bosonic mana and fermionic stabilizer Rényi entropy.

\subsubsection{Faithfulness}

For the Weyl-ordered case and $p=1$, the hybrid magic is faithful on pure states. More precisely,
\begin{equation}
    \mathcal M_1(\ket{\psi})=0
    \qquad\Longleftrightarrow\qquad
    \ket{\psi}\in \textsc{hStab}.
\end{equation}
In other words, the hybrid magic vanishes if and only if the pure state is a product of a bosonic Gaussian state and a fermionic Majorana stabilizer state.

The implication
$
\ket{\psi}\in \textsc{hStab}\Rightarrow \mathcal M_1(\ket{\psi})=0
$
follows immediately from the additivity relation \eqref{eq:additivity_HybMag}, together with the faithfulness of bosonic mana on pure states and of the fermionic stabilizer Rényi entropy. The converse implication is more subtle in the hybrid setting, since one must exclude the possibility that boson-fermion correlations conspire to produce vanishing hybrid magic without the state belonging to the free set. A detailed proof of this pure-state faithfulness statement will be given in a forthcoming paper~\cite{hybrid_resource}. We provide here for the curious reader an outline of the idea of the prood in the Appendix \ref{app:faithfulness_outline}.

\subsubsection{Lower bounds}

The hybrid phase-space formalism also yields general lower bounds on the hybrid magic in the Weyl-ordered case. In particular, the hybrid Wigner function satisfies the $L_2$ identity
\begin{equation}
    \left\|W_\rho\right\|_{L_2(\mathbb R^{2M|2N})}^2
    =
    2^{M+N}\,\mathrm{Tr}(\rho^2),
\end{equation}
as well as the uniform bound
\begin{equation}
    \left\|W_\rho\right\|_{L_\infty(\mathbb R^{2M|2N})}
    \leq 2^M.
\end{equation}
Combining these two relations yields, for every $p\in[2,\infty]$,
\begin{equation}
    \left\|W_\rho\right\|_{L_p(\mathbb R^{2M|2N})}
    \leq
    2^{\,M(1-1/p)+N/p}\,
    \mathrm{Tr}(\rho^2)^{1/p},
\end{equation}
and therefore, for $p>2$,
\begin{equation}
    \mathcal M_p(\rho;\bm 0)
    \geq
    \frac{1}{1-\tfrac p2}
    \log\left(2^{M(p-1)}\,\mathrm{Tr}(\rho^2)\right).
\end{equation}
These bounds provide a useful baseline for the size of the hybrid magic and relate it directly to basic state characteristics such as purity. The derivation of these inequalities is given in the Appendix.

\subsubsection{Upper bounds}

Complementary upper bounds can also be obtained for the Weyl-ordered hybrid magic at the two values most relevant in this work, namely $p=1$ and $p=4$. The only cutoff required in the derivation concerns the bosonic phase-space variables: the fermionic part is already a finite sum over the $2^{2N}$ Majorana strings. If all dressed Wigner functions are supported, or effectively truncated, in a bosonic region $\Omega\subset\mathbb C^M$ of finite volume
\begin{equation}
    \mathcal V_{\mathrm B}(\Omega)
    =
    \int_\Omega
    \frac{\mathrm d^{2M}\alpha}{\pi^M},
\end{equation}
then
\begin{equation}
    \mathcal M_1(\rho;\bm 0)
    \leq
    \log\left(2^{M+N}\,\mathcal V_{\mathrm B}(\Omega)\,\mathrm{Tr}(\rho^2)\right),
\end{equation}
and
\begin{equation}
    \mathcal M_4(\rho;\bm 0)
    \leq
    \log\left(
    \frac{2^N\,\mathcal V_{\mathrm B}(\Omega)}
    {2^{2M}\,\mathrm{Tr}(\rho^2)^2}
    \right).
\end{equation}
For pure states, these inequalities reduce to
\begin{equation}
\begin{aligned}
    \mathcal M_1(\ket{\psi};\bm 0)
    &\leq
    \log\left(2^{M+N}\,\mathcal V_{\mathrm B}(\Omega)\right), \\
    \mathcal M_4(\ket{\psi};\bm 0)
    &\leq
    \log\left(2^{N-2M}\,\mathcal V_{\mathrm B}(\Omega)\right).
\end{aligned}
\end{equation}
As discussed in App.~\ref{app:upper_bounds_hybrid_magic}, the sign of the prefactor $1/(1-p/2)$ is important in the derivation: for $p=4$ it is negative, so an upper bound on $\mathcal M_4$ follows from a lower bound on the corresponding fourth moment. We refer to App.~\ref{app:upper_bounds_hybrid_magic} for the full proof. Without a finite bosonic phase-space volume, no state-independent upper bound of this type is obtained.

\subsubsection{Stability under free operations}
\label{subsubsec:free_ops}

For the subclass of product free operations $\hat G\otimes \hat C\in\mathfrak F$, the hybrid magic is invariant. This follows directly from the fact that bosonic Gaussian unitaries preserve the modulus structure of the bosonic Wigner function, while fermionic Clifford unitaries preserve the Majorana-string probability distribution entering the stabilizer Rényi entropy. Therefore, under these free product operations, the amount of hybrid magic is unchanged.

\subsubsection{Comments on non-monotonicity}

We do not claim that $\mathcal M_p$ is a genuine magic monotone for arbitrary $p$. The reason is intrinsic to the hybrid construction: the bosonic and fermionic sectors select different distinguished values of $p$. On the bosonic side, the natural choice is $p=1$, corresponding to Wigner negativity and mana, which is a bona fide monotone under Gaussian protocols, i.e.\ Gaussian channels/unitaries, Gaussian measurements with feed-forward, and tracing out of modes \cite{albarelli2018resource}. In particular, mana is faithful, additive on tensor products, and non-increasing under free operations. By contrast, for $p>1$ the bosonic $L_p$ norm is not normalized on the full set of Wigner-positive states, so it does not directly define a faithful bosonic magic monotone.

On the fermionic side, Majorana stabilizer states are eigenstates of maximally commuting sets of Majorana strings, and the free operations are generated by the Majorana group, i.e.\ the fermionic analog of the Clifford group, which preserves the stabilizer structure \cite{bravyi2002fermionic, collura2412quantum, mclauchlan2022fermion, mudassar2024encoding, bettaque2024structure}. The associated Rényi entropy of the Majorana-spectrum probability distribution generalizes the qubit stabilizer Rényi entropy \cite{leone2022stabilizer}. However, it is known to define a bona fide magic monotone only for $p\geq 4$ \cite{leone2024stabilizer}, although the cases $p=2$ and $p=1$ have also been used in the literature as useful magic proxies \cite{sarkis2025molecules, shuangshuang2022dynamics}.

Therefore, since the hybrid boson-fermion magic in Eq.~\eqref{eq:hybrid_magic} necessarily depends on a single choice of $p$, it cannot in general be claimed to be a monotone. It should instead be viewed as a physically well-motivated proxy for the magic of hybrid boson-fermion systems, as will be illustrated in the following section.

Finally note that one could generalize the notion of norm on superspace in order to achieve a larger flexibility, hence solving the tension related to the choice of $p$. We comment about that in the Appendix.

\subsubsection{Towards a resource theory of hybrid systems}

The reader well versed in quantum information theory may at this stage wonder what resource-theoretic foundations underlie our framework. A rigorous bottom-up construction of a resource theory of hybrid magic should start by specifying properly what the free states and operations are, and then provide measures of the corresponding resource. We have already pointed toward these core concepts by defining $\textsc{hStab}$ and discussing free operations in Sec.~\ref{subsubsec:free_ops}. One might also ask how this putative resource theory of hybrid non-stabilizerness interacts with quantum correlations. In a forthcoming paper~\cite{hybrid_resource}, we establish the mathematical foundations of this resource theory, clarify in depth its interaction with the resource theory of entanglement for hybrid systems, and analyze the implications of the resulting structure.

\subsubsection{Computational cost}

    It is a well-known fact that nonstabilizerness for qubit/fermionic systems is a hard quantity to compute. If one restricts to pure states the complexity of computing the SRE \cite{leone2022stabilizer} is exponential in system size $\mathcal O(8^{N})$ for a system of $N$ qubits or fermions, stemming from the exponential number $\mathcal O(4^{N})$ of expectation values of Pauli strings. When considering mixed states determining whether a state is a member of the stabilizer polytope is even more challening, with monotones such as the robustness of magic requiring optimization \cite{howard2017application}, in this case the complexity is superexponential $\mathcal O(\exp N^2)$ \cite{leone2026unbearablehardnessdecidingmagic}.

    On the bosonic side, if we consider a truncation level $d$  on the Fock space level and a number of grid points $n_{\rm grid}$ the complexity of computing the Wigner negativity for $M$ bosonic modes is $\mathcal O(d^{2 M} n_{\rm grid}^{2 M})$ for a generic mixed state.

    Overall, 
    the complexity of computing the hybrid magic is large as expected, of the order $\mathcal O(8^{N} d^{2 M} n_{\rm grid}^{2 M})$ and therefore it scales exponentially in the both the number of bosonic and fermionic degrees of freedom, as $\mathcal O(\exp(c_\textsc{f} N + c_\textsc{b} M))$, where $c_\textsc{f}, c_\textsc{b}$ are certain constants.

    There are several strategies to work around the exponential cost of computing the SRE using Monte Carlo sampling and tensor networks \cite{tarabunga2023many, tarabunga2024critical, frau2024nonstabilizerness, tarabunga_nonstab_24}, which enables the study of magic in many-body systems. Furthermore the hybrid case with Monte Carlo sampling has been considered in \cite{crew2025magic}.

\section{Applications}\label{sec:applications}

    \subsection{Free hybrid harmonic oscillator}\label{sec:hybHO}

        As a first simple example, let us consider the free supersymmetric quantum harmonic oscillator. Introduce bosonic operators $a,a^\dagger$ and fermionic $c,c^\dagger$, and the free Hamiltonian:
        \begin{equation}
            H = a^\dagger a + c^\dagger c\,,
        \end{equation}
        whose unique ground state $|\varnothing\rangle = |0\rangle_b\otimes|0\rangle_f$ has zero energy. We can directly compute the components of the hybrid Wigner function:
        \begin{equation}
            \begin{aligned}
                &\langle\varnothing|\Upsilon(\alpha; r)|\varnothing\rangle = \mathcal W(\alpha; r)\,,\\
                &\langle\varnothing|\gamma_1\Upsilon(\alpha; r)|\varnothing\rangle = 0\,,\\
                &\langle\varnothing|\gamma_2\Upsilon(\alpha; r)|\varnothing\rangle = 0\,,\\
                &\langle\varnothing|(i\gamma_1\gamma_2+s)\Upsilon(\alpha; r)|\varnothing\rangle =-(1-s)\mathcal W(\alpha; r)\,,\\
            \end{aligned}
        \end{equation}
        where $\mathcal W(\alpha; r)$ is the $r$-ordered bosonic Wigner function of the Fock vacuum:
        \begin{equation}
            \mathcal W(\alpha; r) = \frac{2}{1-r}\,\exp\left(-\frac{2}{1-r}\,|\alpha|^2\right)\,.
        \end{equation}
        The $p$-hybrid magic is then given by:
        \begin{equation}
            \begin{aligned}
                \mathcal M_p(|\varnothing\rangle; \bm s) &= \frac{1}{1-\tfrac{p}{2}}\,\log\left[\lVert\mathcal W(\star; r)\rVert_p^p\right]+\\
                &+\frac{1}{1-\tfrac{p}{2}}\,\log\left[\frac{1+\left|1-s\right|^p}{2}\right]\,.
            \end{aligned}
        \end{equation}
        In the $p=1$ case, that mimicks the traditional notion of mana in quantum optics, we then have\footnote{In the $r\to 1$ limit, the $r$-ordered Wigner function of the Fock vacuum converges to its Glauber–Sudarshan
        P representation, which is a Dirac delta function.}:
        \begin{equation}
            \mathcal M_1(|\varnothing\rangle; \bm s=(r, 0)) = 0\,,
        \end{equation}
        as could be expected for a (hybrid) Gaussian state.

    \subsection{Dressed cat state}\label{sec:dressedCat}

\begin{figure}[h]
            \centering
            \includegraphics[width=0.39\textwidth]{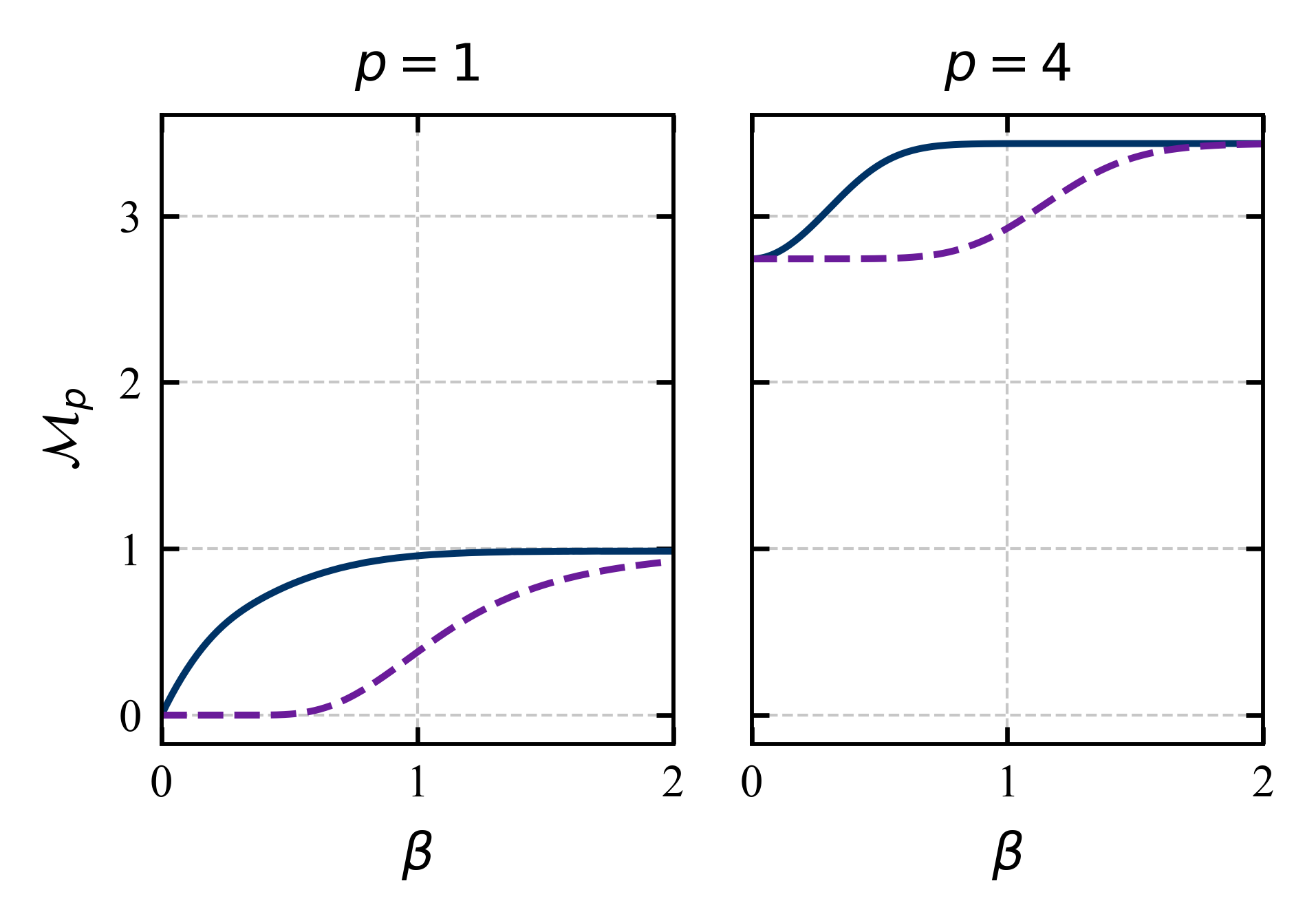}
            \caption{Comparison of the $p$-hybrid magic $\mathcal{M}_p$ of the dressed cat state and the pure bosonic cat state as a function of the parameter $\beta\in\mathbb R_{\geq 0}$ for different values of $p$. In each subplot, the solid blue curve corresponds to the dressed cat state, while the dashed orange curve corresponds to the pure bosonic cat state. We choose Weyl ordering $s=r=0$ for definiteness.}
            \label{fig:compare_dressed_pure}
        \end{figure}

        As before, we consider a Hilbert space of the form $\mathcal H = \mathcal H_\text{Fock}\otimes \mathbb C^2$. We define the following family of states:
        \begin{equation}
        \label{eq:dressed_cat_state}
            |\psi(\beta)\rangle = \frac{|\beta\rangle\otimes|0\rangle+|-\beta\rangle\otimes|1\rangle}{\sqrt{2}}\,,
        \end{equation}
        describing an even bosonic cat state dressed by a fermionic degree of freedom. $|0\rangle$ is defined by the fact that $c|0\rangle=0$, and we denote $|1\rangle = c^\dagger|0\rangle$. The reader will find in the Supplemental Material the details of the derivation of the hybrid Wigner function components. We depict in fig. \ref{fig:compare_dressed_pure} (solid lines) the $p$-hybrid magic as a function of the parameter $\beta$ for different values of $p$.

        Note however that the dressed cat state (\ref{eq:dressed_cat_state}) is unphysical in the sense that it is inhomogeneous from the Grassmann degree point of view. It is however instructive to compare the $p$-hybrid magic of the dressed cat state with the pure bosonic cat state \eqref{eq:bosonicCat}. We depict in fig. \ref{fig:compare_dressed_pure} the comparison for two different values of $p=1$, related to the mana, and $p=4$, related to the usual $\mathrm{SRE}_2$. We observe that the larger the displacement parameter $\beta$, the larger the measure of magic, as should be expected since for small $\beta$ the cat state does not show much Wigner negativity, and it saturates at large $\beta$ \cite{kenfack2004negativity}. We also observe that both types of cat states converge at large values of $\beta$ to the same asymptotic value of the magic. However, we note that very interestingly, the presence of the fermionic degree of freedom increases the speed of production of magic with respect to the pure bosonic cat state, this increase may be caused by the presence of entanglement between the photon and the qubit. We will come back to this point below, in the context of the Holstein model.

    \subsection{Polaron physics and the Holstein model}\label{sec:holstein}

        The spinless Holstein model describes fermionic particle coupled locally to an optical phonon mode on each lattice site.  It is defined by the Hamiltonian
        \begin{equation}
        \label{eq:holstein_hamiltonian}
            \begin{aligned}
                H &= -\tau \sum_{\langle i,j\rangle} \left(c_i^\dagger c_j + c_j^\dagger c_i\right)
                + \omega_0 \sum_i b_i^\dagger b_i
                + \\
                &+g \sum_i \left(b_i^\dagger + b_i\right)\,n_i\,,
            \end{aligned}
        \end{equation}
        where $c_i$ and $c_i^\dagger$ are creation and annihilation operators of for a spinless fermion on site $i$,
        $b_i$ and $b_i^\dagger$ are the local phonon creation and annihilation operators of frequency $\omega_0$,
        $n_i = c_i^\dagger c_i$ is the fermion number operator,
        $\tau$ is the nearest-neighbour hopping amplitude, and
        $g$ is the electron–phonon coupling strength. This model provides a minimal setting for studying polaron formation—the dressing of the fermion by a cloud of phonons—and for exploring the resulting crossover from weakly renormalized band motion to self-trapped small polarons.

        In the two-site case, one rewrites Eq.~\eqref{eq:holstein_hamiltonian} in terms of the center-of-mass and relative phononic normal modes. The only nontrivial coupling that drives polaron physics is to the relative coordinate, therefore leading to the following Hamiltonian:
        \begin{equation}
        \label{eq:holstein_hamiltonian_two_site}
            \begin{aligned}
                H &= -\tau\left(c_1^\dagger c_2 + c_2^\dagger c_1\right)
                + \omega_0\,b^\dagger b
                + \\
                &+g\left(b^\dagger + b\right)\,(n_2 - n_1)\,,
            \end{aligned}
        \end{equation}
        where $b$ denotes the bosonic annihilation operator associated to the relative normal mode. The interaction term modulates site energies between 1 and 2, enabling phonon-assisted hopping and genuine polaron formation. This two-site problem can be simplified by the Lang–Firsov canonical transformation:
        \begin{equation}
        \label{eq:lang_firsov_transformation}
            U = \exp\left[\frac{g}{\omega_0}\left(n_2 - n_1\right)\left(b^\dagger - b\right)\right]\,,
        \end{equation}
        which exactly cancels the linear coupling and renormalizes the hopping term. Indeed, conjugating the Holstein Hamiltonian with the Lang–Firsov transformation, one obtains:
        \begin{equation}
            \begin{aligned}
                H' &= U\,H\,U^\dagger\\
                &= -\tau\left(c_1^\dagger c_2\,X + X^\dagger\,c_2^\dagger c_1\right)
                \\
                &\quad + \omega_0\,b^\dagger b - \frac{g^2}{\omega_0}\,,
            \end{aligned}
        \end{equation}
        where
        \begin{equation}
        X = \exp\left[\frac{2g}{\omega_0}\left(b^\dagger - b\right)\right]\,.
        \end{equation}
        As an approximation, we project onto the phonon vacuum $|0\rangle_{\rm ph}$ and obtain the effective hopping amplitude:
        \begin{equation}
        \tau_{\rm eff}
        = \tau \,\langle 0|X|0\rangle
        = \tau \,\exp\left[-2\left(\frac{g}{\omega_0}\right)^2\right].
        \end{equation}
        The single-electron ground state in the transformed frame is:
        \begin{equation}
        |\psi'_0\rangle
        = \frac{1}{\sqrt{2}}\,|0\rangle_{\rm ph}\otimes\left(c_1^\dagger + c_2^\dagger\right)
        |0\rangle_{\rm el}\,.
        \end{equation}
        Transforming back to the original frame,
        \begin{equation}
        \label{eq:holstein_ground_state}
            \begin{aligned}
                |\psi_0\rangle
                &= U^\dagger\,|\psi'_0\rangle\\
                &= \frac{1}{\sqrt{2}}
                \Bigl[
                \left|\beta\right\rangle_{\rm ph}\otimes c_1^\dagger|0\rangle_{\rm el}
                \\
                &\qquad
                + \left|-\beta\right\rangle_{\rm ph}\otimes c_2^\dagger|0\rangle_{\rm el}
                \Bigr]\,.
            \end{aligned}
        \end{equation}
        where we denoted $\beta = -\tfrac{2g}{\omega_0}$ the bosonic displacement parameter. The corresponding ground-state energy is:
        \begin{equation}
        \label{eq:holstein_ground_state_energy}
            E_0 = -\frac{g^2}{\omega_0} - \tau_{\rm eff}\,.
        \end{equation}
        In the atomic limit $\tau=0$ in which the electron is fully localized on one site, or in the regime $\omega_0 \gg \tau$ in which phonons adjust almost instantaneously to electron motion, the solution (\ref{eq:holstein_ground_state}) becomes exact \cite{hohenadler2007lang}. Note that the exact two-site solution uses a displaced-Fock basis and leads to continued-fraction equations for $E_0$ that reduce to the Lang–Firsov result only when $\tau\to0$ or $\omega_0\gg \tau$ \cite{rongsheng2002exact}.

\begin{figure}[h]
            \centering
            \includegraphics[width=0.39\textwidth]{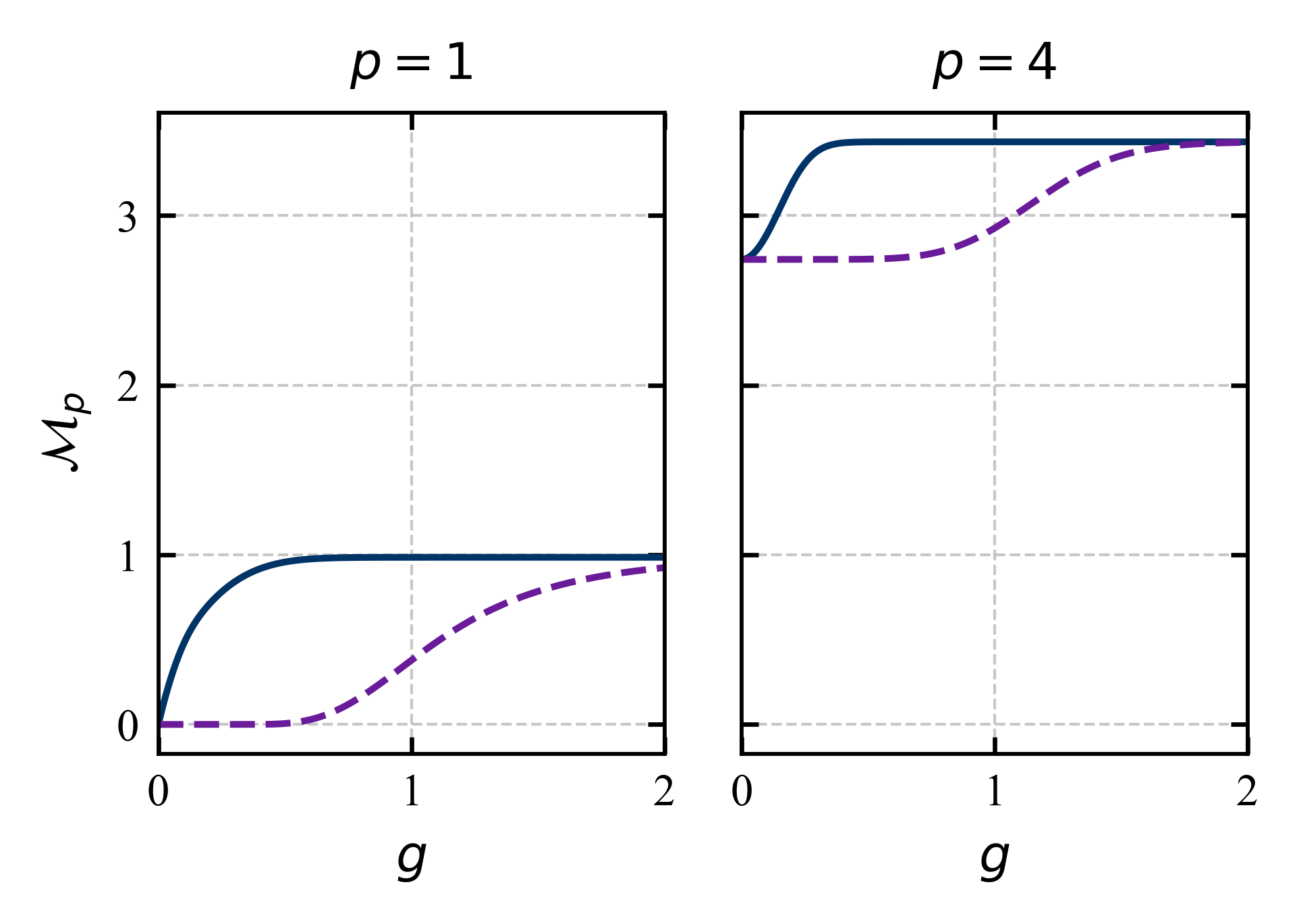}
            \caption{Comparison of the $p$-hybrid magic $\mathcal{M}_p$ of the Holstein model ground state and the pure bosonic cat state as a function of the coupling parameter $g$ (Holstein) and $\beta$ (bosonic cat) for different values of $p$. In each subplot, the solid blue curve corresponds to the Holstein model, while the dashed magenta curve corresponds to the pure bosonic cat state. We set $\omega_0=1$ and choose Weyl ordering for definiteness.}
            \label{fig:holstein_compare_pure}
        \end{figure}

        For illustrative purposes, we focus on the approximate ground state (\ref{eq:holstein_ground_state}). The reader will find in the Supplemental Material the details of the derivation of the hybrid Wigner function components for the Holstein model. We depict in fig. \ref{fig:holstein_compare_pure} the $p$-hybrid magic as a function of the parameter $\beta$ for different values of $p$.

        Note that the ground state (\ref{eq:holstein_ground_state}) is reminiscent of the dressed cat state (\ref{eq:dressed_cat_state}). However in this polaron context, the state under consideration is homogeneous from the Grassmann degree point of view, hence physical. In order to confirm the observation concerning the speed of production of magic discussed in the previous section, we depict in fig. \ref{fig:holstein_compare_pure} the comparison of the $p$-hybrid magic of the Holstein model ground state with the pure bosonic cat state. We observe that the presence of the electronic degree of freedom indeed increases the speed of production of magic with respect to a purely phononic cat state, indeed it even increases the speed of production with respect to the dressed cat state shown in fig. \ref{fig:compare_dressed_pure}.

       In the next section, for the sake of simplicity, we will study the hybrid magic with $p=1$, the generalization of the mana, the reason for this choice is that $\textsc{Mana}_p$ is non-zero for Gaussian states when $p > 1$.

    \subsection{Fermionic Jaynes-Cummings model}\label{sec:JaynesCummings}

        The Jaynes-Cummings model provides the simplest description of the interaction between a two-level atom and a single-mode cavity, and is defined by the Hamiltonian:
        \begin{equation}
            H = \omega_c a^\dagger a + \frac{\omega_a}{2}\,\sigma_z + g\left(a^\dagger\sigma_- + a\sigma_+\right)\,.
        \end{equation}
        Let us denote by $|g\rangle$ and $|e\rangle$ the eigenstates of $\sigma_z$:
        \begin{equation}
            \sigma_z = |e\rangle\langle e| - |g\rangle\langle g|\,.
        \end{equation}
        Then one has
        \begin{equation}
            \sigma_+ = |e\rangle\langle g| \quad \text{and} \quad \sigma_- = |g\rangle\langle e|\,.
        \end{equation}
        To each of the eigenstates of $\sigma_z$, we associate a fermionic mode, whose associated annihilation operator we denote $c_e$ and $c_g$. In terms of these operators, the bosonic raising and lowering operators can be naturally expressed as:
        \begin{equation}
        \label{eq:JC_fermionic_operators}
            \sigma_+ = c_e^\dagger c_g \quad \text{and} \quad \sigma_- = c_g^\dagger c_e\,.
        \end{equation}
        In terms of the fermionic number operators $n_e=c_e^\dagger c_e$ and $n_g=c_g^\dagger c_g$, the Pauli operators $\sigma_z$ reads:
        \begin{equation}
            \sigma_z = n_e - n_g = c_e^\dagger c_e - c_g^\dagger c_g\,.
        \end{equation}
        The Hamiltonian can then be written as:
        \begin{equation}
            H = \omega_c a^\dagger a + \frac{\omega_a}{2}\,\left(n_e - n_g\right) + g\left(a^\dagger c_g^\dagger c_e + a c_e^\dagger c_g\right)\,,
        \end{equation}
        where we see that in this fermionic language the interaction term is cubic and of Yukawa type. In order for this fermionic model to be equivalent to the bosonic Jaynes-Cummings model, we need to project on the subspace of the Hilbert space with $n_g + n_e = 1$. This is achieved by the following projection operator:
        \begin{equation}
            P = n_g + n_e - 2n_g n_e\,.
        \end{equation}
        The Hamiltonian can then be written in that subspace as\footnote{Up to an irrelevant additive fermionic vacuum energy $-\tfrac{\omega_a}{2}$.}:
        \begin{equation}
            H = \omega_c a^\dagger a + \omega_a c_e^\dagger c_e + g\left(a^\dagger c_g^\dagger c_e + a c_e^\dagger c_g\right)\,.
        \end{equation}
        Note that the Hamiltonian commutes with the total fermion number operator $N_\textsc{f} = n_g + n_e$. Therefore, if one initializes the system in a state with well-defined fermion number $N_\textsc{f} = 1$, then the constraint $N_\textsc{f} = 1$ will be satisfied at all times, ensuring equivalence with the bosonic (spin $1/2$) Jaynes-Cummings model. In this fermionic language, the ground state of the system reads:
        \begin{equation}
            |\varnothing\rangle = |0\rangle_c \otimes c_g^\dagger |00\rangle_a = |0\rangle_c \otimes |10\rangle_a\,,
        \end{equation}
        namely the cavity is in the Fock vacuum and the atom carries fermion number $1$ in the ground state. We will denote the kets with a subscript `$c$' the factor that pertains to the cavity (the bosonic sector), and by `$a$' the factors that pertains to the atom (the fermionic sector). We will also denotes interchangeably $|10\rangle_a\equiv|g\rangle$ and $|01\rangle_a\equiv|e\rangle$.

        Following the purely bosonic model, we note that the number operator $\mathcal N = a^\dagger a + c_e^\dagger c_e$ is a good quantum number on the $N_\textsc{f} = 1$ subspace, allowing to solve the system blockwise, as in the bosonic case. The fixed $\mathcal N=n$ subspace is spanned by:
        \begin{equation}
            \left\{|n\rangle_c\otimes|10\rangle_a, |n-1\rangle_c\otimes|01\rangle_a\right\}\,.
        \end{equation}
        The Hamiltonian can be written in this subspace as:
        \begin{equation}
            H_n = \begin{pmatrix}
                n\omega_c & g\sqrt{n} \\
                g\sqrt{n} & (n-1)\omega_c+\omega_a
            \end{pmatrix}\,.
        \end{equation}
        On a fixed $\mathcal N=n>0$ subspace, we define as usual the detuning and generalized Rabi frequency:
        \begin{equation}
            \Delta = \omega_c - \omega_a \quad \text{and} \quad \Omega_n = \sqrt{\Delta^2 + 4g^2 n}\,.
        \end{equation}

        Of relevance are the matrix elements of the $r$-ordered bosonic phase point operators in the Fock basis:
        \begin{equation}
            \mathcal O_{m,n}(\alpha; r) :=\langle m|\Upsilon(\alpha; r)|n\rangle\,.
        \end{equation}
        We provide an explicit expression of these overlap coefficients in the Supplemental Material. We obtain:
        \begin{widetext}
            \begin{equation}
            \label{eq:overlap_coefficients_JC}
                \mathcal O_{m,n}(\alpha; r) = \left(\frac{r+1}{r-1}\right)^m \sqrt{\frac{m!}{n!}} \left(\frac{2}{1-r}\right)^{n-m+1} \exp \left(-\frac{2 | \alpha | ^2}{1-r}\right)\bar\alpha^{n-m} L_m^{(n-m)}\left(\frac{4 | \alpha | ^2}{1-r^2}\right)\,,
            \end{equation}
        \end{widetext}\
        where $L_m^{(\alpha)}(x)$ denote the associated Laguerre polynomials.
        The coefficients of the hybrid Wigner function can then be expressed in terms of these overlap coefficients and their explicit expression can be found in the Supplemental Material.

        We consider the resonant case $\omega_c = \omega_a$, and we will study the behavior of the hybrid magic for different choices of initial states of the cavity and the atom.

        Given an initial product state of the form
        \begin{equation}
            |\psi(0)\rangle = \sum_{n=0}^\infty \gamma_n |n\rangle_c \otimes\Big(\mu_\text{g} |10\rangle_a + \mu_\text{e} |01\rangle_a\Big)\,,
        \end{equation}
        the state of the system at time $t$ reads:
        \begin{equation}
        \label{eq:JC_state}
            |\psi(t)\rangle = \sum_{n=0}^\infty |n\rangle_c \otimes\Big(\alpha_n(t) |10\rangle_a + \beta_n(t) |01\rangle_a\Big)\,,
        \end{equation}
        with
        \begin{equation}
            \begin{aligned}
                \alpha_n(t) &= e^{-in\omega_c t}\Big[\cos\left(\sqrt n gt\right)\alpha_n(0) \\
                &\quad\quad\quad\quad\quad- i\sin\left(\sqrt n gt\right)\beta_{n-1}(0)\Big]\,, \\
                \beta_n(t) &= e^{-in\omega_c t} \Big[-i\sin\left(\sqrt{n+1} gt\right)\alpha_{n+1}(0) \\
                &\quad\quad\quad\quad\quad+ \cos\left(\sqrt{n+1} gt\right)\beta_n(0)\Big]\,.
            \end{aligned}
        \end{equation}

\begin{figure*}
    \centering
    \includegraphics[width=.85\linewidth]{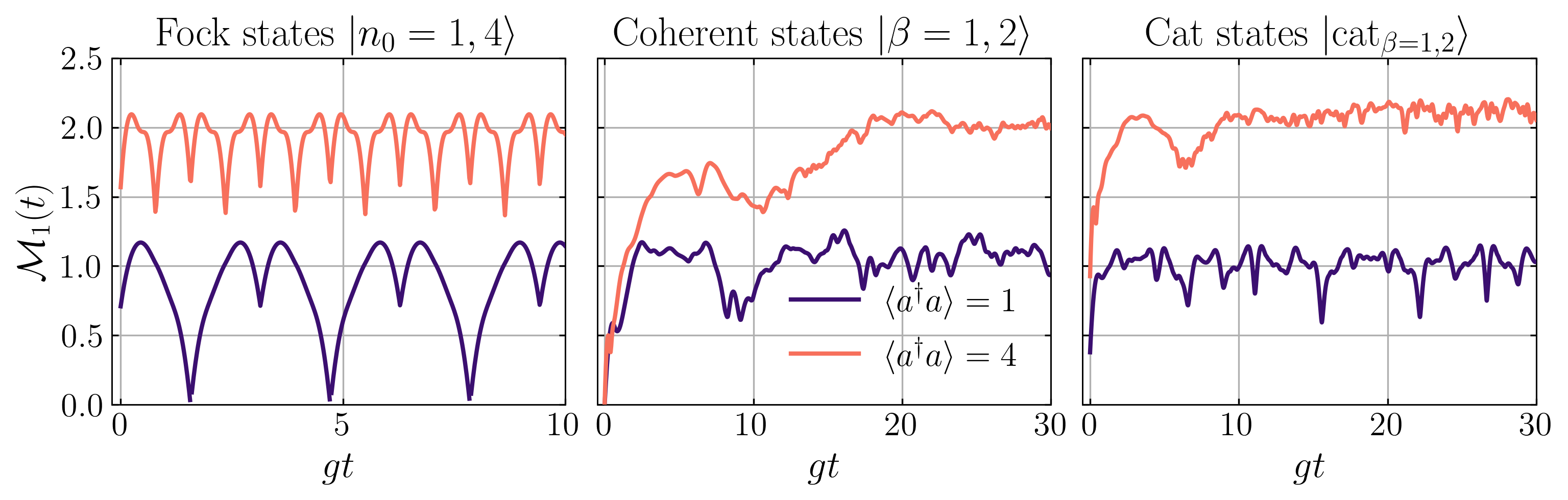}
    \caption{Hybrid magic $\mathcal M_1(t)$ as a function of time for different initial states of the cavity: Fock states $\ket{n_0}$ (left), coherent states $\ket{\beta}$ (center) and cat states $\ket{\mathrm{cat}_\beta}$ (right). The plots compare two different initial states with mean photon number $\braket{a^\dagger a} = 1$ (blue) and $\braket{a^\dagger a} = 4$ (light red). The coherent state and cat plot have a cutoff of the bosonic Hilbert space dimension of $d_{\rm max} = 15$. All the plots have $\omega_c = \omega_a = g =1$ and we set the Weyl ordering for the fermionic and bosonic degrees of freedom  $r = s = 0$. For all the plots, the initial atomic state is $\ket{g}_a$. }
    \label{fig:time_evol_HybMag_cavity}
\end{figure*}

        We will consider various initial states for the cavity and atom. We begin by studying the effect of the initial state of the cavity for the atom in its ground state $\ket{g} \equiv \ket{10}_a$, and consider that it has a definite number of photons $n_0$, thus starting in the Fock state $\ket{n_0}$. Since the JC model is on resonance the dynamics shows Rabi oscillations with frequency $\Omega_n = \sqrt{n_0} g$ between the states $\ket{g}\ket{n_0}$ and $\ket{e}\ket{n_0-1}$, if $n_0 \geq 1$. Fig. \ref{fig:time_evol_HybMag_cavity} (left) shows the time-evolution of the hybrid magic for two initial Fock states: $\ket{n_0=1}$ and $\ket{n_0=4}$. The initial magic starts from a given value, corresponding to the mana of the $\ket{n_0}$ state, then increases and decreases to the mana of the $\ket{n_0 -1}$ state, since after one full Rabi period the atomic state is a stabilizer state. During one of the Rabi periods, we see the hybrid magic increasing, which corresponds to magic in the entanglement between bosonic and fermionic degrees of freedom. Comparing the two initial Fock states we find that a higher number of photons gives a higher value of the magic, and that as $n_0$ grows the difference between $\textsc{Mana}(\ket{n_0})$ and $\textsc{Mana}(\ket{n_0-1})$ decreases, although the difference between the highest value and the mana does not necessarily decrease. We also observe that the period of oscillation of $\ket{n_0 = 4}$ is exactly half of that of $\ket{n_0 = 1}$, as expected, and that the behavior of magic in one of the periods slightly changes shape as we increase $n_0$. From the numerical simulation we can find the first maximum to lie in the interval $\sqrt{n_0} g t_{\rm max} \in [\frac{\pi}{7}, \frac{\pi}{6}]$, see the Supplemental Material. Interestingly, for high Fock numbers the hybrid magic shows two peaks, with a relative minimum at $\sqrt{n_0}g t=\frac{\pi}{4}$, with the second one being smaller than the first one for all the values of $n_0$ that we investigated.

        Fock states are known to be non-classical and, as such, they show non-zero mana for $n_0 \geq 1$. We now investigate the effect of considering an initial state which shows zero mana such as the coherent state
        \begin{equation}
            \ket{\beta} = \sum_{n=0}^\infty \frac{e^{-\frac{|\beta|^2}{2}}\,\beta^n}{\sqrt{n!}}\ket{n}.
        \end{equation}
        The mean photon number of the coherent state is given by $\braket{a^\dagger a}= |\beta|^2$.
        This state is highly classical, and as such, the initial hybrid magic starts at zero. Fig. \ref{fig:time_evol_HybMag_cavity} (center) shows the evolution for two different coherent states $\ket{\beta =1}$ and $\ket{\beta = 2}$. For both initial states the hybrid magic starts from zero, but it grows fast and reaches a value comparable to that obtained by a Fock state with the same mean photon number. The dynamics of the hybrid magic here does not show neat Rabi oscillations since many different Fock numbers contribute to the evolution, however this makes the hybrid magic reach a plateau, which ensures that the combined state is highly non-classical.

        Lastly, we study the behavior starting from a very non-classical cavity state, the Schr\"odinger's cat state
        \begin{equation}\label{eq:bosonicCat}
            \begin{aligned}
                \ket{{\rm cat}_\beta}
                &= \frac{\ket{\beta} + \ket{-\beta}}{\mathcal N}\\
                &= \frac{1}{\mathcal N}\sum_{n=0}^\infty
                \frac{\Bigl(1 + (-1)^n\Bigr)e^{-\frac{|\beta|^2}{2}}\beta^n}{\sqrt{n!}}
                \ket{n},
            \end{aligned}
        \end{equation}
        where $\mathcal{N} = \sqrt{2\left(1 + e^{-2|\beta|^2}\right)}$ is simply a normalization factor. Fig. \ref{fig:time_evol_HybMag_cavity} (right) shows the time-evolution of the hybrid magic, we find that it starts from a non-zero value, corresponding to the mana of the photonic cat state, but surprisingly, does not show a higher hybrid magic through the evolution than starting from a coherent state with the same $\beta$, or with a Fock state. This suggests that starting from a more non-classical state does not necessarily imply more production of hybrid magic than starting from a classical state.

                \begin{figure}[h]
            \centering
            \includegraphics[width=0.35\textwidth]{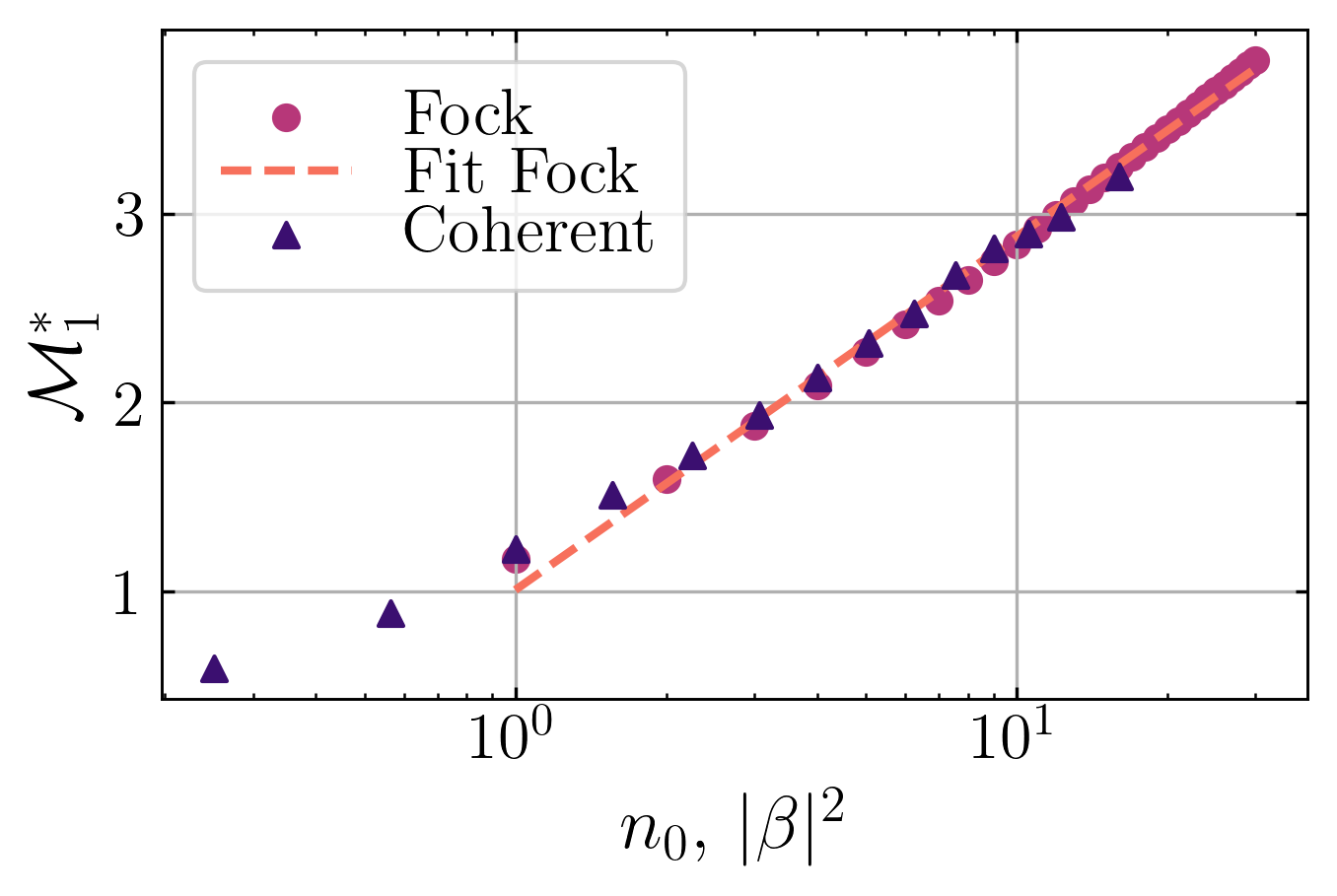}
            \caption{Maximum hybrid magic $\mathcal M^\star=\max_t \mathcal M_1(t)$ versus mean photon number $\bar n$ of the initial Fock (circle) or coherent (triangle) state. The dynamics is in resonance ($\omega_c=\omega_a,\ g=1$) with the atom initially in $|g\rangle$. The x axis is shown in logscale and thus the logarithmic fit $\mathcal M_1^* = \mathfrak a \log n_0 + \mathfrak b$ with $\mathfrak a\approx 0.81, \, \mathfrak b \approx 1.00$ looks linear. }
            \label{fig:fock_vs_coherent_comparison}
        \end{figure}

        Fig. \ref{fig:fock_vs_coherent_comparison} compares the maximum value of the magic $\mathcal M_1^* = \max_t \mathcal M_1(t)$ as a function of the average photon number of the initial Fock (circles) and coherent (triangles) states. Since the approximate location of the maximum is known for Fock states, see the Supplemental Material, we approximated the maximum hybrid magic as $\mathcal M_1^* \approx \mathcal M_1(\frac{\pi}{7 \sqrt{n_0} g})$. For Fock states we find that $\mathcal M_1^*$ grows logarithmically as $\mathcal M_1^* = \mathfrak a \log n_0 + \mathfrak b$ where $\mathfrak a\approx 0.81, \, \mathfrak b \approx 1.00$. The reason to expect a logarithmic growth can be motivated from the Wigner negativity for Fock states \cite{kenfack2004negativity}, which is close to the power law $\sqrt{n_0}/2$, the logarithm in the definition of the hybrid magic then ensures that this power law growth becomes a logarithmic function. Note that the maximum of the hybrid magic grows faster than the coefficient predicted by this power law since over a Rabi period the hybrid magic grows over $\textsc{Mana}(\ket{n_0})$ and $\textsc{Mana}(\ket{n_0-1})$. Surprisingly, we find that the maximum hybrid magic $\mathcal M^*_1$ starting from a very classical state such as the coherent state, scales similarly with the average photon number $|\beta|^2$. This implies that even starting from Gaussian and stabilizer states, the quantum time evolution can lead to highly non-classical states, although, as observed in Fig. \ref{fig:time_evol_HybMag_cavity}, these require a longer time to be reached than if the cavity starts from a non-classical state such as the Fock or cat states.

\begin{figure}[h]
            \centering
            \includegraphics[width=0.39\textwidth]{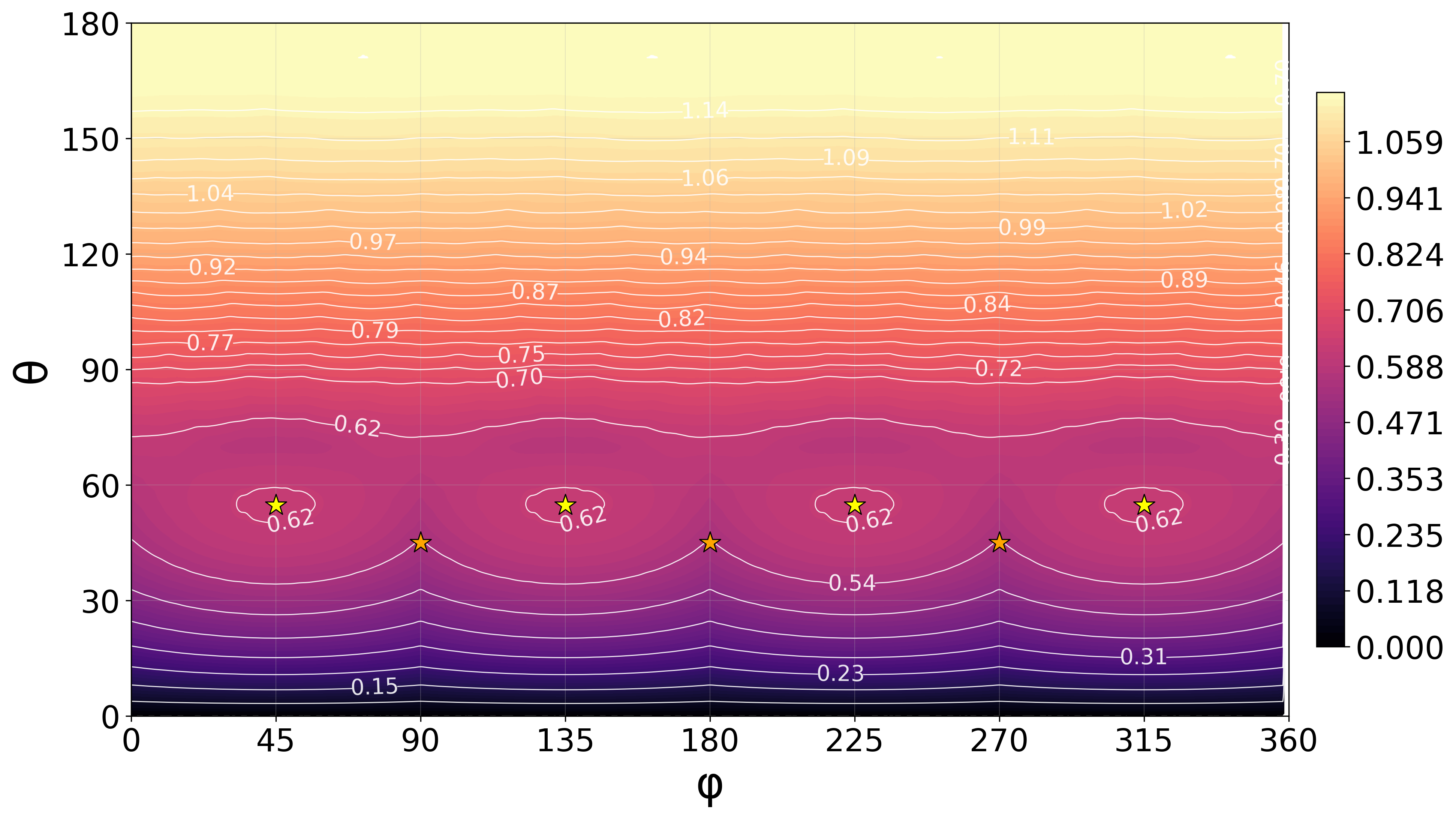}
            \caption{Two-dimensional colormap of $\mathcal M^\star$ over the Bloch sphere of initial atomic states for a vacuum cavity $(|0\rangle_c)$ under resonant JC dynamics ($\omega_c=\omega_a,\ g=1$). The distribution exhibits $\pi/2$-periodicity in the azimuthal angle $\phi$ and a non-monotonic dependence on the polar angle $\theta$, with pronounced local maxima at the location of the T-states (yellow stars). H-states are also depicted on the figure (orange stars).}
            \label{fig:bloch_sphere_2d_very_high_res}
        \end{figure}

\begin{figure}[h]
            \centering
            \includegraphics[width=0.39\textwidth]{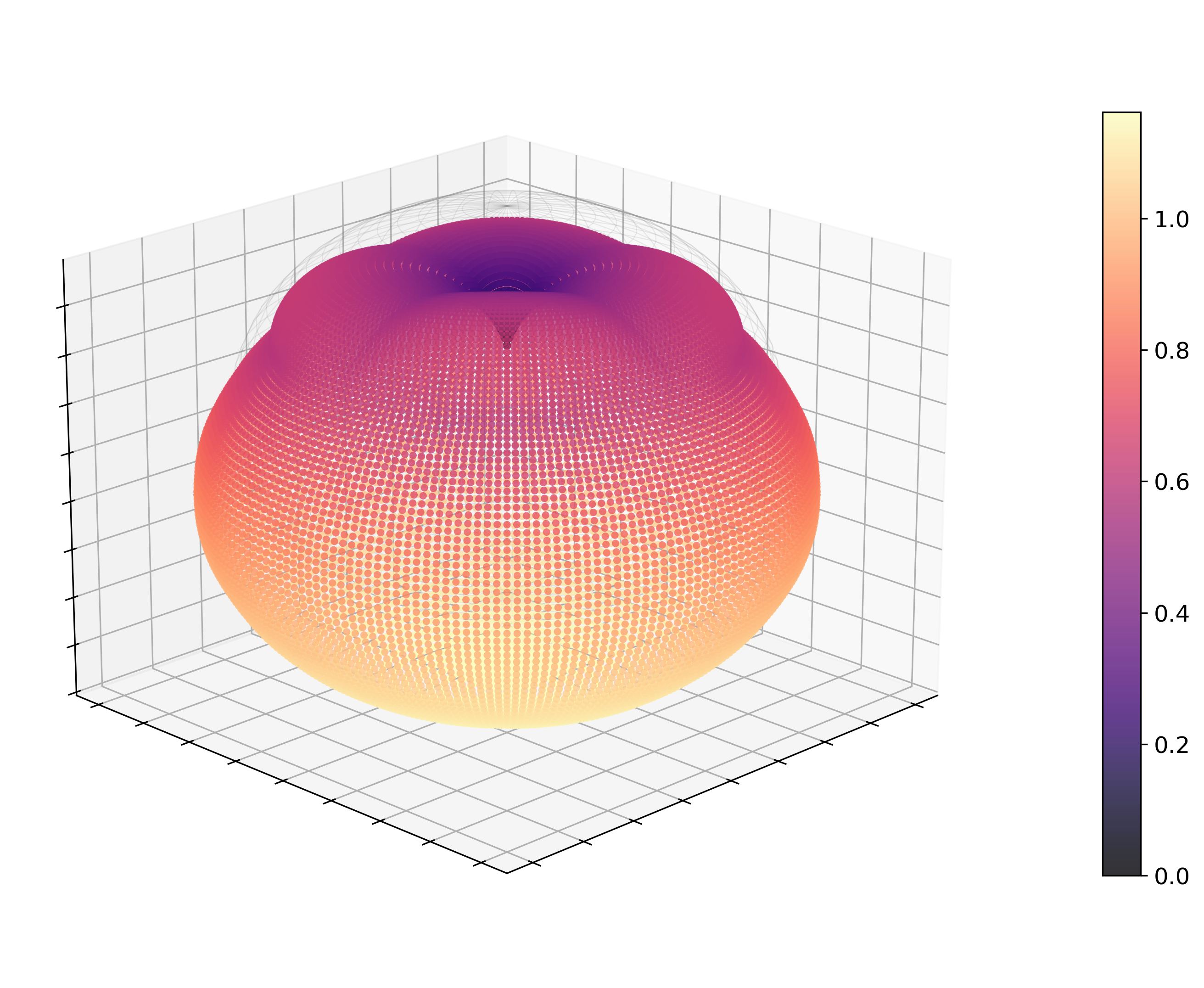}
            \caption{Three-dimensional Bloch-sphere rendering of Fig. \ref{fig:bloch_sphere_2d_very_high_res} with radial distance encoding $\mathcal M^\star$. The visualization emphasizes optimal regions away from the poles and illustrates the non-trivial interplay between atomic orientation and boson–fermion magic generation.}
            \label{fig:bloch_sphere_3d_very_high_res}
        \end{figure}

        We now study the dependence of the hybrid magic on the initial atomic state, thus setting the initial cavity state to always be the Fock vacuum $\ket{0}$. We find that starting in the atomic ground state $\ket{g}$ does not generate  any hybrid magic, since the dynamics stays in the sector with zero total excitations, however, as we start from a state closer to the $\ket{e}$ state the maximum magic grows. Indeed, the maximum magic starting from the $\ket{e}_a\ket{0}_c$ state is the same as starting from $\ket{g}_a\ket{1}_c$, cf. Fig. \ref{fig:time_evol_HybMag_cavity} (left). Figs. \ref{fig:bloch_sphere_2d_very_high_res} and \ref{fig:bloch_sphere_3d_very_high_res} show the maximum hybrid magic as a function of the starting initial state on the Bloch sphere, as a 2D color map and as a 3D plot scaling the radius of the Bloch sphere, respectively. We find that along the azimuthal angle, the maximum hybrid magic is $\pi/2$-periodic, however on the $\theta$ angle the behavior is more non-trivial, in the south pole of the Bloch sphere, when we are close to the $\ket{e}$ state, the maximum magic grows with $\theta$, however $\mathcal M_1^*$ shows relative maxima close to $\theta \approx \pi/3$. This means that in the northern hemisphere, the initial atomic state that produces the most hybrid magic is close to $\theta \approx \pi/3$ and $\varphi = \pi/4 + n \pi/2, \, n \in \mathbb Z$. These states happen to correspond to the T-state (and its Clifford orbit) defined by Bravyi and Kitaev \cite{bravyi2005universal}:
        \begin{equation}
             |T\rangle\langle T| = \frac{1}{2}\left[\mathbb 1 + \frac{\sigma_x + \sigma_y + \sigma_z}{\sqrt 3}\right]\,,
        \end{equation}
        which have $\varphi = \pi/4,\, \theta = \arccos \frac{1}{\sqrt{3}} \approx 0.955 \mathrm{rad}\approx  54.74^\circ$. Other highly magic states of a single qubit are the H-states:
        \begin{equation}
            |H\rangle\langle H| = \frac{1}{2}\left[\mathbb 1 + \frac{\sigma_x + \sigma_z}{\sqrt 2}\right]\,.
        \end{equation}
        This interesting profile of the hybrid magic illustrates the non-trivial interplay between atomic and bosonic magic, where a stabilizer state, such as $\ket{e}$, can lead to higher magic in the evolution than a highly magic qubit state. Note that both $|g\rangle$ and $|e\rangle$ are stabilizer states, one could therefore wonder why Figs. \ref{fig:bloch_sphere_2d_very_high_res} and \ref{fig:bloch_sphere_3d_very_high_res} show a lack of symmetry between these two states. This can be explained by the fact that the atom is coupled to the cavity whose spectrum is itself asymmetric in the sense that the tower of Fock states is bounded from below by $|0\rangle$.

        \subsubsection{Mutual Magic for the Jaynes-Cummings model}

        A hybrid system made up of bosons and fermions, especially the Jaynes-Cummings model considered here, has a natural bipartition between atomic and photonic degrees of freedom. Therefore it is natural to study what part of the hybrid magic can be attributed only to correlations between atom and cavity, and cannot be understood locally from atomic non-stabilizerness or photonic non-Gaussianity. We therefore define the \textit{mutual magic} as
        \begin{equation}
            \mathcal M_p^{\mathrm{(mut)}}(\rho) := \mathcal M_p(\rho) - \textsc{Mana}_p(\rho_c) - \widetilde{\mathrm{SRE}}_{\frac{p}{2}}(\rho_a),
        \end{equation}
        note that we consider the total cavity and atom state to be pure $\rho = \ket{\psi}\bra{\psi}$, but the reduced atomic $\rho_a = \Tr_c(\rho)$ and cavity $\rho_c = \Tr_a(\rho)$ density operators are in general mixed. The fact that $\rho_c$ is mixed is not problematic since the mana is still a ``\textit{genuine measure of non-stabilizerness}'' for mixed states \cite{tarabunga2024critical}, even allowing for a definition of \textit{mutual mana}. However, this is not the case for the atomic SRE, for this reason, we modify its definition as \cite{tarabunga2025efficientmutualmagicmagic}
\begin{align} \notag
    \widetilde{\mathrm{SRE}}_{\frac{p}{2}}(\rho) := &\frac{1}{1-\frac{p}{2}}\log\left(\sum_\Gamma \left|\frac{\Tr(\rho \Gamma)}{\sqrt{2^{N_\text{F}} \Tr(\rho^2)}}\right|^p \right)\\ &- \log(\Tr(\rho^2)) - N_\text{F} \log(2),
\end{align}
where $N_\text{F}=1$ is the number of fermionic degrees of freedom. Other possible definitions of mutual magic, such as subtracting the contribution from the mutual information \cite{hoshino2025stabilizer}, are equivalent to this one. Note that if the total state is a product state $\ket{\psi} = \ket{\psi}_a \otimes \ket{\phi}_c$ the mutual magic vanishes since the hybrid magic is additive \eqref{eq:additivity_HybMag} and the modified SRE reduces to the standard one $\widetilde{\mathrm{SRE}}(\ket{\psi}\bra{\psi}) = \mathrm{SRE}(\ket{\psi}\bra{\psi})$ for a pure state.

        As in the case of mutual information, this mutual-magic quantity should be interpreted as a correlation-sensitive excess-magic diagnostic, not as a fully isolated measure of irreducible shared magic. In particular, in mixed hybrid regimes it can contain contributions from correlation structure beyond purely local bosonic mana and fermionic stabilizer diagnostics. We therefore use it here as an informative witness of non-additive hybrid non-stabilizerness in the Jaynes--Cummings dynamics.

        \begin{figure}[h]
            \centering
            \includegraphics[width=.85\linewidth]{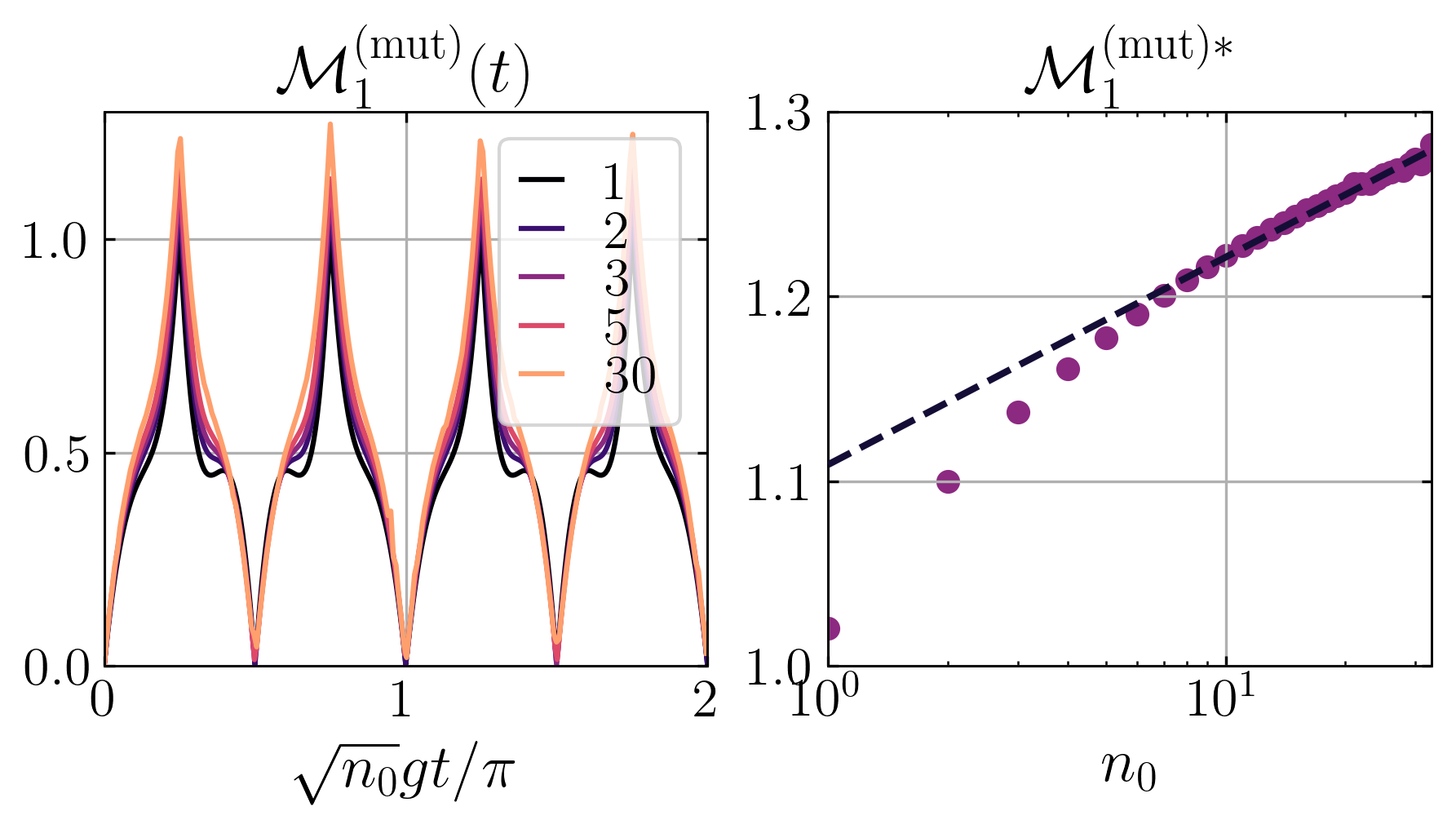}
            \caption{Mutual magic as a function of time for different initial Fock states $\ket{n_0}$.  }
            \label{fig:mutualMagic_time}
        \end{figure}

Fig. \ref{fig:mutualMagic_time} (left) shows the evolution of the mutual magic for different initial cavity Fock states $\ket{n_0}$ with $n_0 = 1, 2, 3, 5, 30$. We observe that the mutual magic inherits the periodicity with frequency $\Omega = \sqrt{n_0} g$ from the hybrid magic, but, interestingly, it shows maxima at $\sqrt{n_0}g t = \pi/4$, which does not correspond to the point of maximum hybrid magic for a Fock state. For $n_0 = 1$ we see that the mutual magic is asymmetric in each of the periods, e.g. in the first oscillation the decrease to zero is not monotonically but rather has a small dip. The small dip is not seen for $n_0\geq 2$, but the mutual magic is still asymmetric.

Fig. \ref{fig:mutualMagic_time} (right) investigates the scaling of the maximum value of the mutual magic $\mathcal M_1^{(\mathrm{mut})*}= \max_t \mathcal M_1^{(\mathrm{mut})}(t)$. We find that the maximum mutual magic also scales logarithmically with $n_0$ as $ \mathcal M_1^{(\mathrm{mut})*} \sim \mathfrak a_{\rm mut} \log(n_0) + \mathfrak b_{\rm mut}, $
where the coefficients are $\mathfrak a_{\rm mut} \approx 0.049, \, \mathfrak b_{\rm mut}\approx 1.109$, note that the logarithmic growth of the maximum mutual magic is much slower than that of the hybrid magic $ \mathfrak a_{\rm mut} \ll \mathfrak a$.

\subsection{Beyond single fermions: The Tavis-Cummings model}
\label{sec:TavisCummings}
\begin{figure*}
    \centering
    \includegraphics[width=.7\linewidth]{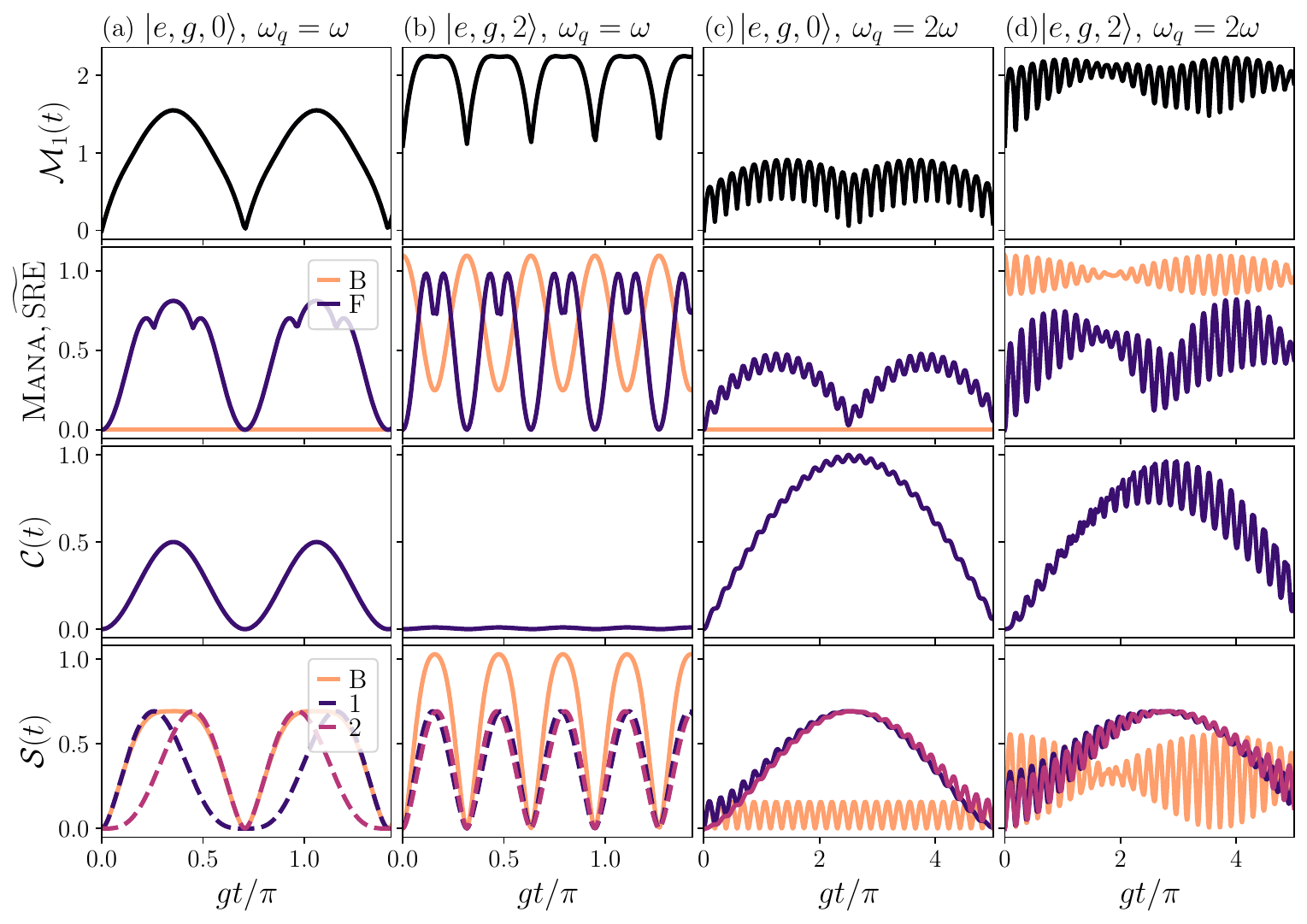}
    \caption{Dynamics for hybrid magic $\mathcal M_1$ along with other quantum resources: Fermionic $\widetilde{{\rm SRE}}_{1/2}$ and bosonic $\textsc{Mana}$, fermionic concurrence $\mathcal C$, and entanglement entropies $\mathcal S$ in the Tavis Cummings model. In all the plots we set $ g = 1, \omega = 10 g$. The different columns depict different initial states either with no initial photoncs $\ket{\psi_0} = \ket{e}_1 \otimes \ket{g}_2 \otimes \ket{0}_\textsc{b}$  (a,c) or with two initial photons $\ket{\psi_0} = \ket{e}_1 \otimes \ket{g}_2 \otimes \ket{2}_\textsc{b}$ (b,d) and the qubit frequency is either on resonance $\omega_q = \omega$ (a,b) or a multiple $\omega_q = 2 \omega$ (c,d). }
    \label{fig:TC_hybMagic}
\end{figure*}

The Tavis-Cummings model consists of several two level systems (2LS) interacting with a single mode photon. In this subsection for convenience we will use qubit language to represent the fermions, however this can be understood as a pair of fermions restricted to the single occupation sector, as explained in the previous section. Considering the case with two equal 2LS the Hamiltonian of the system is given by
\begin{equation}
    \hat H_\textsc{tc} = \omega_q\frac{\hat \sigma_z^1 + \hat \sigma^z_2}{2}  +\omega \hat a^\dagger \hat a + g \hat \sigma^+_1 \hat a + g \hat \sigma^+_2 \hat a +{\rm h.c. }
\end{equation}
where we consider that the coupling to both 2LS is the same. Figure \ref{fig:TC_hybMagic} shows the evolution of the hybrid magic of the time evolution generated by this model and compares it to several different measures of quantumness. The \textit{concurrence} for a two qubit system is an entanglement monotone defined as
\begin{equation}
    \mathcal C(\hat \rho) := \max(0, \lambda_1 - \lambda_2 - \lambda_3 - \lambda_4),
\end{equation}
where $\lambda_i$ are the decreasing eigenvalues of the matrix
\begin{equation}
    \hat R = \sqrt{\sqrt{\rho} (\hat \sigma_y\otimes \hat \sigma_y) \hat \rho^* (\hat \sigma_y\otimes \hat \sigma_y) \sqrt{\rho}},
\end{equation}
where $\bullet^*$ denotes complex conjugation in the $\hat \sigma_z$ eigenbasis. In this case, we will compute the concurrence of the reduced fermionic state $\hat \rho_F = \Tr_B(\ket{\Psi(t)}\bra{\Psi(t)})$. We will also compute the entanglement entropy
\begin{equation}
    \mathcal S(\hat \rho) = -\Tr(\hat \rho \log \hat \rho),
\end{equation}
and we will consider the reduced states of site 1 $\hat \rho_1 = \Tr_{2, B}(\ket{\Psi(t)}\bra{\Psi(t)})$, site 2 $\hat \rho_2 = \Tr_{1, B}(\ket{\Psi(t)}\bra{\Psi(t)})$ and the photon $\hat \rho_B = \Tr_{1, 2}(\ket{\Psi(t)}\bra{\Psi(t)})$.

In all the cases we consider the initial state to have an excitation in one of the atoms and no excitation in the second. Fig. \ref{fig:TC_hybMagic} (a) shows the time evolution in resonance starting from vacuum in the photon field. The system in resonance shows Rabi oscillations, and thus all the quantities oscillate periodically. Interestingly, the hybrid magic grows, reaching a maximum of $\mathcal M_1^* \approx 1.55$, and then decreases to a value close to zero, which contrasts with the behavior observed in Fig. \ref{fig:time_evol_HybMag_cavity} (a) where the oscillation was going to either 0 or a non-zero value which corresponded to either zero or one photon.  This is not the case here, as can be clearly seen in the case of the bosonic mana (orange line) which stays at zero. This means that the interaciton between the two modes is mediated by virtual photons which do not affect the non-classicality of the cavity state. This behavior contrasts with the evolution of the fermionic $\widetilde{{\rm SRE}}_{1/2}$ which oscillates with the same period. Interestingly it shows a smaller relative maximum, before showing the global maximum $\widetilde{{\rm SRE}}_{1/2} \approx 0.81$. The entanglement quantifiers also show interesting behavior, the concurrence of the reduced two fermion state $\Tr_B(\ket{\Psi(t)}\bra{\Psi(t)})$ oscillates reaaching a maximum value of $0.5$, while for the entanglement entropies we see that $\mathcal S_1$ grows first, at the same time as $\mathcal S_B$, then $\mathcal S_1$ decreases but $\mathcal S_2$ starts to grow, and $\mathcal S_B$ remains approximately constant between the times of maximum $\mathcal S_1$ and $\mathcal S_2$. The resonant case in which we start with two photons in the cavity is shown in Fig. \ref{fig:TC_hybMagic} (b). We see that the hybrid magic $\mathcal M_1$ now starts from a non-zero value, corresponding to the mana of a two photon state, it then grows and oscillates at a higher frequency than the no-photon case. We also see that the top of the oscillation flattens out, which means that the state has close to maximal hybrid magic for a longer period of time. The mana shows that the wigner function of the photon field starts with a high level of negativity, which subsequently decreases up to a smaller value. Therefore we see that now the interaction between the atoms mediated by the photon happens through real and not virtual photons which affect the degree of bosonic non-gaussianity. The fermionic SRE now shows periodic oscillations with two pronounced peaks per period. Interestingly, the reduced state of the two fermions has very low concurrence, much smaller than in the zero photon case, which means that the  fermions do not get entangled. Lastly the behavior of the entanglement entropies show that the fermionic $\mathcal S_1, \mathcal S_2$ oscillate at almost the same rate, with their peaks only slightly separated, while the photon entropy acquires higher values.

The case where the qubit frequency $\omega_q$ is double the frequency of the photon shows some more non-trivial dynamics. Figure \ref{fig:TC_hybMagic} (c) displays the case in which the initial state is the photon vacuum. The hybrid magic $\mathcal M_1$ shows some fast oscillations on top of a slower oscillation. Interestingly, it reaches smaller maximum values than in the resonant case with $\mathcal M_1^* \approx 0.91$. The bosonic mana stays at zero which implies that the interactions between the qubits happen through virtual photons, and the SRE shows a similar pattern to the hybrid magic, albeit with less amplitude in the fast oscillations. The plot of the concurrence shows a maximum close to unity at the point where the SRE vanishes, this implies that the dynamics is close to a stabilizer maximally entangled Bell state. Lastly, the entanglement entropies for the qubits show a similar behavior to the concurrence, with a maximum around the same time, while the photon entanglement entropy oscillates with a fixed amplitude. When the off resonant case starts from a state with two photons in the cavity, see Fig. \ref{fig:TC_hybMagic} (d) we see that the hybrid magic shows the fast and slow oscillations described previously, but the amplitude of the fast oscillation changes in times. This is also the case for the SRE. However, interestingly the bosonic mana stays at a high value with some fast oscillations and an envelope. The time at which the envelope vanishes seems to approximately coincide with the time at which the amplitud of the fast oscillations is minimal. The concurrence shows a similar behavior to the previous case, although it has more pronounced oscillations when it decreases. Lastly, the entanglement entropies of the qubit also behave similarly to the vacuum case, again with a higher amplitude in the oscillation. However, the bosonic entanglement entropy also shows the previously mentioned oscillations with an envelope.


\subsection{Supersymmetric quantum mechanics through the lens of hybrid magic}
\label{subsec:susy_magic}

Supersymmetric quantum mechanics (SUSY QM) is a natural testing ground for the hybrid magic introduced here, since it couples one bosonic degree of freedom to one fermionic one in the simplest nontrivial way. From a resource-theoretic viewpoint, SUSY QM is therefore a canonical genuinely hybrid system: neither purely bosonic nor purely fermionic magic alone captures its full phase-space structure. Hybrid magic is thus well suited to ask: \emph{does supersymmetry constrain, suppress, or reorganize non-stabilizerness in a characteristic way?}

We address this through two complementary diagnostics. First, we study a one-boson/one-fermion SUSY model and compare the exactly supersymmetric point with a family of softly broken theories, probing how hybrid-magic production changes as the supersymmetric structure is continuously lost. Second, we compare a standard non-supersymmetric quartic double well with its supersymmetric partner construction, isolating in a static setting the relation between supersymmetry, low-energy structure, and ground-state nonclassicality.

\paragraph*{Softly broken supersymmetric model.}
We begin with the standard one-dimensional SUSY-QM Hamiltonian
\begin{equation}
\hat H
=
\frac{p^2}{2}
+
\frac12 P(x)^2
+
\frac12 Q(x)\,\sigma_z ,
\label{eq:susy_hybrid_hamiltonian}
\end{equation}
where $\sigma_z$ acts on a single fermionic mode, equivalently on a two-dimensional Clifford/Majorana sector. At the supersymmetric point,
\begin{equation}
Q(x)=P'(x),
\end{equation}
and the Hamiltonian decomposes into the familiar partner pair
\begin{equation}
\hat H_\pm
=
\frac{p^2}{2}
+
V_\pm(x),
\qquad
V_\pm(x)
=
\frac12 P(x)^2 \pm \frac12 P'(x).
\label{eq:susy_partner_potentials_general}
\end{equation}
We choose
\begin{equation}
P(x)=\omega x + g x^3,
\label{eq:susy_superpotential_choice}
\end{equation}
for which
\begin{equation}
V_\pm(x)
=
\frac12(\omega x + g x^3)^2
\pm
\frac12(\omega + 3g x^2).
\label{eq:susy_partner_potentials_cubic}
\end{equation}
The supersymmetric zero mode in the $H_-$ sector is known analytically:
\begin{equation}
\begin{aligned}
\psi^{(-)}_0(x)
&\propto
\exp\left[-\int^x P(y)\,\mathrm dy\right]
\\
&=
\exp\left(-\frac{\omega x^2}{2}-\frac{g x^4}{4}\right),
\end{aligned}
\label{eq:susy_zero_mode}
\end{equation}
which is normalizable for $\omega>0$ and $g>0$. Thus \eqref{eq:susy_zero_mode} gives an exact zero-energy ground state in the unbroken theory, while the excited levels obey the usual pairing relation $E_{n+1}^{(-)}=E_n^{(+)}$.

We interpolate away from exact supersymmetry through the soft deformation
\begin{equation}
Q(x)=P'(x)+\epsilon x^2,
\label{eq:susy_soft_breaking}
\end{equation}
so that $\epsilon=0$ is supersymmetric, whereas $\epsilon\neq0$ breaks the partner relation while preserving the hybrid boson-fermion structure. The initial state is chosen to be free and hybrid,
\begin{equation}
|\Psi(0)\rangle
=
|\psi_G\rangle_{\rm b}\otimes |+\rangle_{\rm f},
\end{equation}
with $|\psi_G\rangle_{\rm b}$ Gaussian and $|+\rangle_{\rm f}$ a fermionic stabilizer state. The subsequent growth of $\mathcal M_1$ therefore directly measures the dynamical generation of hybrid non-stabilizerness.

\subsubsection{Mitigating the magic production rate}
For \eqref{eq:susy_hybrid_hamiltonian}, the dynamics generically entangles the bosonic and fermionic sectors. Bosonic mana would probe only the reduced bosonic sector, while fermionic stabilizer diagnostics would miss the bosonic phase-space distortion. The hybrid magic defined here is sensitive to both ingredients at once. This is precisely the regime where a hybrid phase-space treatment is meaningful: supersymmetry constrains the joint boson-fermion structure, and our proxy detects how this constraint appears in the hybrid quasiprobability representation.

\begin{figure}[ht]
    \centering
    \includegraphics[width=\linewidth]{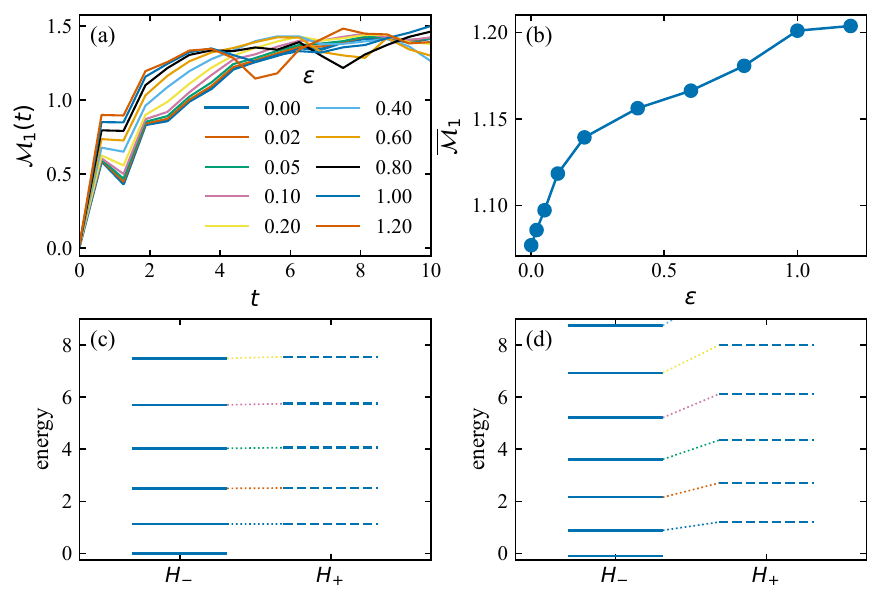}
    \caption{Supersymmetric hybrid model and soft breaking.
    (a) Time evolution of the hybrid magic $\mathcal M_1(t)$ for the Hamiltonian \eqref{eq:susy_hybrid_hamiltonian} with $P(x)=\omega x+g x^3$ and several values of the soft-breaking parameter $\epsilon$ in \eqref{eq:susy_soft_breaking}.
    (b) Time-averaged hybrid magic $\overline{\mathcal M}_1$ as a function of $\epsilon$.
    (c) Low-energy spectrum of the exactly supersymmetric theory, showing the zero mode in the $H_-$ sector and the characteristic pairing of positive-energy levels.
    (d) Low-energy spectrum at finite soft breaking, where the supersymmetric pairing is visibly degraded.
    Together, these panels show that exact supersymmetry suppresses the sustained production of hybrid magic and that this suppression is tied to the supersymmetric organization of the spectrum.}
    \label{fig:susy_soft_combined}
\end{figure}

The results are summarized in Fig.~\ref{fig:susy_soft_combined}. Panel~(a) shows the full time dependence of $\mathcal M_1(t)$ for a range of soft-breaking parameters $\epsilon$. The main qualitative point is that exact supersymmetry does not prevent hybrid-magic generation, but systematically suppresses its production rate. Panel~(b) makes this clearer: the time-averaged hybrid magic $\overline{\mathcal M}_1$ increases as one moves away from $\epsilon=0$. Thus exact supersymmetry does not trivialize the dynamics, but keeps the early-time evolution in a lower-magic regime.

Panels~(c) and~(d) give the spectral interpretation. At $\epsilon=0$, the spectrum shows the expected supersymmetric structure: a zero mode in the $H_-$ sector and approximate pairing of positive-energy levels between $H_-$ and $H_+$. For $\epsilon\neq0$, this organization is visibly altered. The broken theory in panel~(d) no longer displays the same protected low-energy structure, and the loss of spectral pairing is accompanied by larger time-averaged hybrid magic. Thus Fig.~\ref{fig:susy_soft_combined} shows that supersymmetry constrains how hybrid non-stabilizerness is dynamically generated: the more rigid the supersymmetric spectral organization, the smaller the magic-production rate.

\subsubsection{Quartic double well: how SUSY mitigates tunneling and ground state magic}
The previous analysis compared exact SUSY with a soft deformation of the same hybrid theory. It is also instructive to contrast a standard non-supersymmetric tunneling problem with its supersymmetric partner construction. We first consider the ordinary quartic double well
\begin{equation}
H_{\rm DW}
=
\frac{p^2}{2}
+
\frac12\bigl(x^2-a^2\bigr)^2.
\label{eq:quartic_dw}
\end{equation}
For large $a$, this Hamiltonian has the familiar tunneling structure: a parity-even ground state delocalized over the two wells, a parity-odd first excited state, and an exponentially small splitting $E_1-E_0$.

The supersymmetric counterpart follows from the superpotential derivative
\begin{equation}
W'(x)=x^2-a^2.
\label{eq:quartic_superpotential}
\end{equation}
The associated partner potentials are
\begin{equation}
V_\pm(x)
=
\frac12\bigl(x^2-a^2\bigr)^2 \pm x.
\label{eq:quartic_partner_potentials}
\end{equation}
Unlike the symmetric double well \eqref{eq:quartic_dw}, the partner potentials \eqref{eq:quartic_partner_potentials} are oppositely tilted. The low-energy states are therefore not organized as a symmetric tunneling doublet in one scalar potential, but as ground states of two distinct partner sectors. In this quartic example there is no normalizable supersymmetric zero mode, since
\begin{equation}
\psi_0^{(-)}(x)\propto
\exp\left[-W(x)\right]
=
\exp\left(-\frac{x^3}{3}+a^2x\right)
\end{equation}
is not square-integrable on the full line, and similarly for the formal $H_+$ zero mode. Supersymmetry is therefore broken in the strict spectral sense. Nevertheless, $H_\pm$ still provide the relevant SUSY partner structure, which we now contrast with the tunneling physics of the non-supersymmetric double well.

Hybrid magic gives a direct comparison. For branch states $|\psi\rangle\otimes|0\rangle$ and $|\psi\rangle\otimes|1\rangle$, the fermionic factor is stabilizer, so the hybrid magic reduces to the bosonic $p=1$ mana. Evaluating hybrid magic on the $H_-$ or $H_+$ ground states therefore compares the same phase-space negativity diagnostic across the non-supersymmetric and supersymmetric settings.

\begin{figure}[ht]
    \centering
    \includegraphics[width=\linewidth]{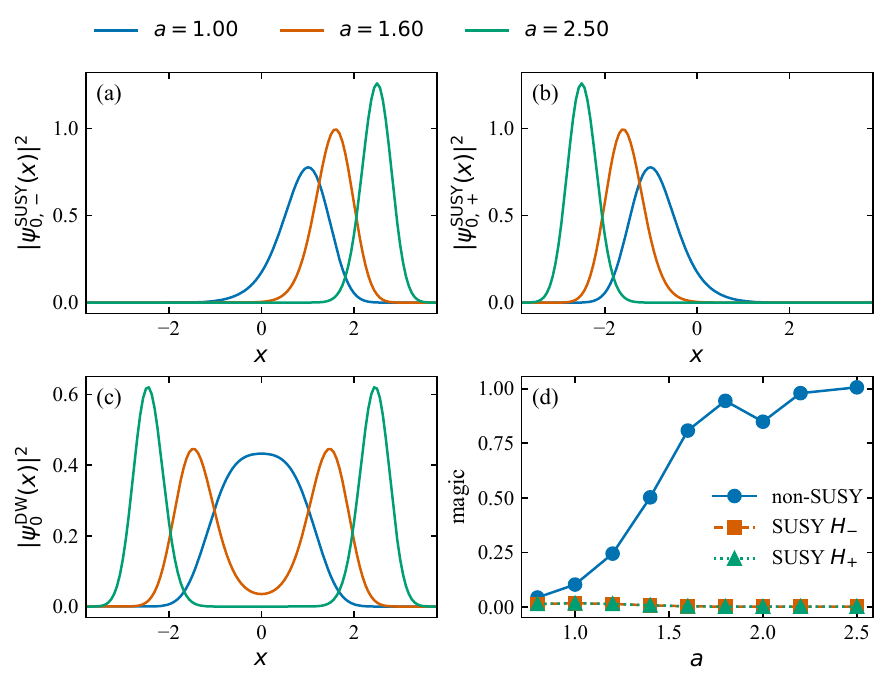}
    \caption{Quartic double well versus supersymmetric partner structure.
    (a) Ground-state density of the ordinary quartic double well \eqref{eq:quartic_dw} for several values of the well-separation parameter $a$. The state remains parity-symmetric and spreads over both wells, reflecting the usual tunneling picture.
    (b) Ground-state density in the $H_-$ sector of the supersymmetric partner construction \eqref{eq:quartic_partner_potentials}.
    (c) Ground-state density in the $H_+$ sector. The two branch densities are biased in opposite directions because the partner potentials are tilted.
    (d) Comparison of the ground-state bosonic mana in the non-supersymmetric double well with the hybrid magic of the $H_-$ and $H_+$ branch ground states.}
    \label{fig:susy_quartic_combined}
\end{figure}

The results are shown in Fig.~\ref{fig:susy_quartic_combined}. Panel~(c) displays the ground-state density of the ordinary double well \eqref{eq:quartic_dw}: the state remains parity-symmetric and spread over both wells, as expected from tunneling. Panels~(a) and~(b) show instead the ground states of the partner sectors \eqref{eq:quartic_partner_potentials}. The densities are biased in opposite directions because the partner potentials are tilted rather than symmetric. In this sense, supersymmetry mitigates tunneling by removing the usual symmetric tunneling-doublet mechanism and replacing it with a partner-sector organization of low-energy states.

Panel~(d) quantifies this statement in terms of magic. It plots the ground-state mana of the ordinary double well and the hybrid magic of the two supersymmetric branch ground states as functions of the well separation $a$. The three distinct curves show that the supersymmetric reorganization of the low-energy sector is not merely a reshuffling of wavefunctions, but is also reflected in the amount of phase-space magic.

Figs.~\ref{fig:susy_soft_combined} and \ref{fig:susy_quartic_combined} show that supersymmetry does not forbid magic generation, but constrains and reorganizes it. Dynamically, this appears as suppressed hybrid non-stabilizerness production relative to softly broken theories. Statically, SUSY replaces the standard tunneling-doublet structure by a partner-sector organization with a different pattern of phase-space negativity. SUSY QM is therefore a natural setting for studying magic resources.

\section{Magic power of hybrid bosonic-fermionic gates}
\label{sec:magic_power}
\begin{figure}[h!]
        \centering
        \includegraphics[width=0.4\textwidth]{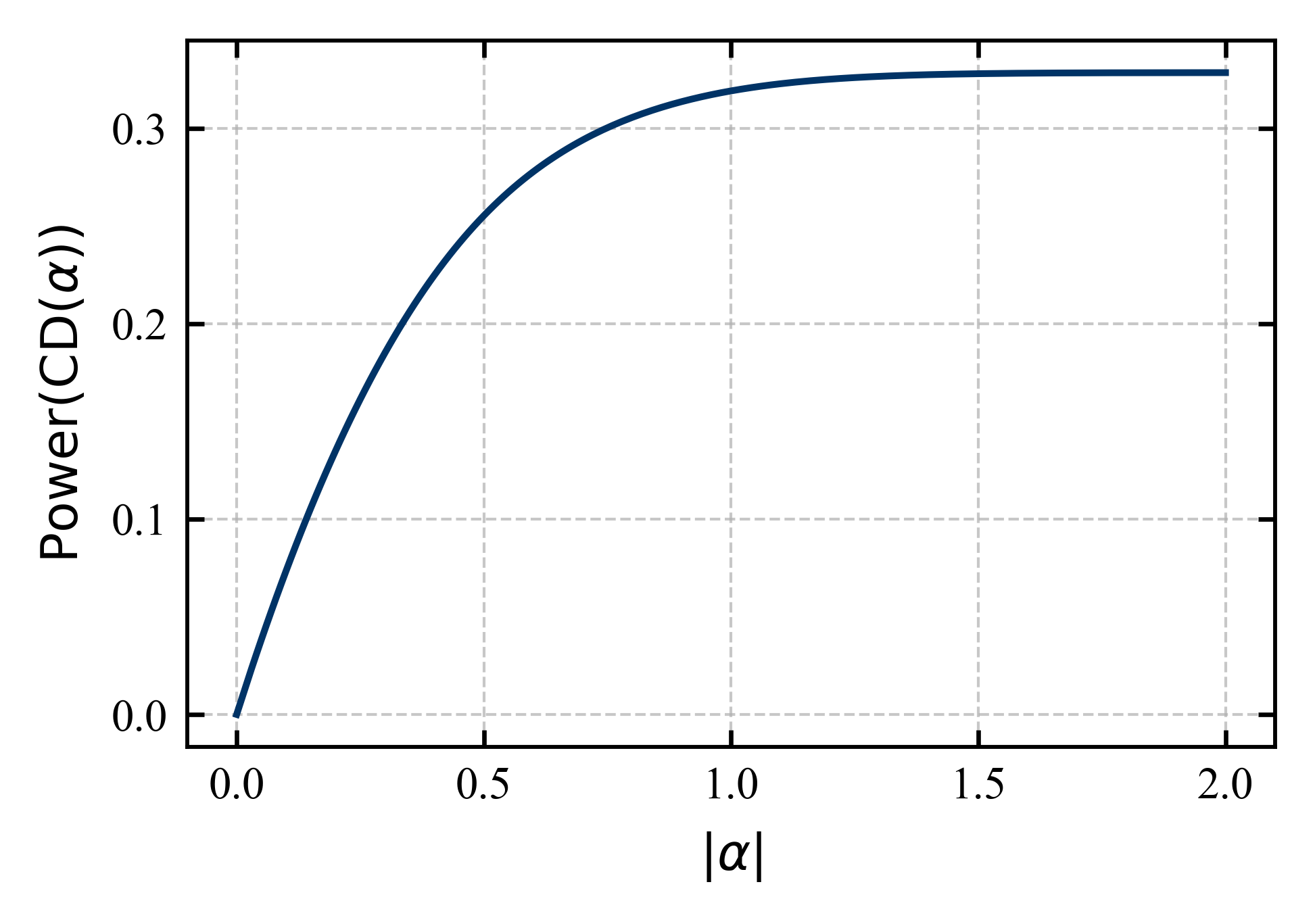}
        \caption{Non-stabilizer power of the conditional displacement gate $\mathrm{CD}(\alpha)$ as a function of $|\alpha|$.}
        \label{fig:power_cd_vs_alpha}
    \end{figure}

    A broad (if not the largest) class of interesting physical systems are composed ot both bosonic and fermionic degrees of freedom. Recently, a hybrid qubit-oscillator quantum computational paradigm was suggested \cite{liu2024hybrid}. In this paradigm, the qubit is represented by a fermionic mode, and the oscillator encodes a bosonic mode. Possible physical platforms that could encode these hybrid systems \cite{kbv4-jj51} as well as software stacks \cite{chen2025genesis, decker2025symbolic, singh2025towards} were studied short after. All the evident applications to gauge theories with matter and condensed matter systems also immediately followed \cite{crane2024hybrid, varona2024towards, kumar2025digital, saner2025realtimeobservationaharonovbohminterference}. A list a hybrid gates were proposed in \cite{liu2024hybrid}. One could instead of qubits consider fermionic modes and a corresponding set of hybrid boson-fermion gates after fermionizing like in eq. (\ref{eq:JC_fermionic_operators}). We refer the reader to the Supplemental Material where we provide such a list of hybrid gates directly translated from the gates of \cite{liu2024hybrid} to fermions. In that very same paper, they define universal sets of gates. Let us consider one of these sets, dubbed the \textit{Phase-Space} instruction set. It is composed of a fermionic rotation gate:
    \begin{equation}
        R_\varphi(\theta) = \exp\left[-i \frac{\theta}{2} \left( c_e^\dagger c_g\, e^{-i\varphi} + c_g^\dagger c_e\, e^{i\varphi} \right)\right]\,,
    \end{equation}
    a bosonic beam-splitter:
    \begin{equation}
        \mathrm{BS}(\theta, \varphi) = \exp\left[ -i \frac{\theta}{2} \left( e^{i\varphi} a^\dagger b + e^{-i\varphi} a b^\dagger \right) \right]\,,
    \end{equation}
    as well as a hybrid boson-fermion entangling gate, a conditional displacement gate:
    \begin{equation}
        \mathrm{CD}(\alpha) = \exp\left[(n_e - n_g) (\alpha a^\dagger - \alpha^* a)\right]
    \end{equation}
    Of interest to us is, of course, the hybrid gate, which, as we will see, not only produces entanglement between the bosonic and fermionic sectors of the circuit, but also injects magic into the system. We will, of course, use the hybric magic as a proxy for quantifying the magic present in the system.         Let us focus on the zero ordering parameters and $p=1$ case for concreteness. Following the definition of \cite{leone2022stabilizer}, we define the non-stabilizer power of a unitary transformation $U$ as the average hybrid magic over the set of transformed stabilizer states defined in eq. \eqref{eq:stab}.
    We consider, as a specific example, the case of the conditional displacement gate $U(\alpha) = \mathrm{CD}(\alpha) = \exp\left[(n_e - n_g) (\alpha a^\dagger - \alpha^* a)\right]$, which acts on a single bosonic mode and two fermionic modes. The set of two-mode Majorana stabilizer states can be easily characterized. Let us recollect the parity-preserving Majorana strings. In addition to the identity operator $\mathbb 1$ and the total parity $P=(i\gamma_1\gamma_2)(i\gamma_3\gamma_4)$, we have the following bilinears:
    \begin{equation}
        \begin{aligned}
        B_1 &= i\gamma_1\gamma_2\,,\quad B_2 = i\gamma_3\gamma_4\,,\quad B_3 = i\gamma_1\gamma_3\,,\\
        B_4 &= i\gamma_2\gamma_4\,,\quad B_5 = i\gamma_1\gamma_4\,,\quad B_6 = i\gamma_2\gamma_3\,.
        \end{aligned}
    \end{equation}
    For compactness, we will denote them collectively as $\Gamma = (\mathbb 1, P, B_1, B_2, B_3, B_4, B_5, B_6)$. The Majorana stabilizer states are then defined as the common eigenstates of the pairs of mutually commuting strings $(B_1, B_2)$, $(B_3, B_4)$ and $(B_5, B_6)$. For each given pair, there are 4 Majorana stabilizer states, and therefore 12 in total. They correspond to the four basis product states $\{|00\rangle, |01\rangle, |10\rangle, |11\rangle\}$, the four parity even Bell states $(|00\rangle + t|11\rangle)/\sqrt{2}$ (with $t\in\{\pm 1, \pm i\}$) and the four parity odd Bell states $(|01\rangle + s|10\rangle)/\sqrt{2}$ (with $s\in\{\pm 1, \pm i\}$). The unitary gate can be nicely decomposed in terms of the control operator $n_e - n_g$ eigenspaces as:
    \begin{equation}
        U(\alpha) = \sum_{m\in\{-1, 0, 1\}} D(m\alpha)\otimes \Pi_m\,,
    \end{equation}
    with $\Pi_m$ the projector on the $m$-th eigenspace of $n_e - n_g$:
    \begin{equation}
        \begin{aligned}
            \Pi_+ &= |01\rangle\langle 01|\,,\quad \Pi_- = |10\rangle\langle 10|\,,\\
            \Pi_0 &= |00\rangle\langle 00| + |11\rangle\langle 11|\,.
        \end{aligned}
    \end{equation}
    We are interested in computing the expectated values $\langle\Upsilon(\beta)\otimes\gamma\rangle_{U(\alpha)|\psi\rangle}$ for all $\beta\in\mathbb C$, $\gamma\in\Gamma$ and $|\psi\rangle\in\textsc{Stab}$. We refer the reader to the Supplemental Material for all the details concerning the derivations relevant to the current discussion. One obtains:
    \begin{equation}
        \begin{aligned}
            &\langle\Upsilon(\beta)\otimes\gamma\rangle\\
            &\quad=
            \sum_{m, n}
            e^{-2i(m-n)\text{Im}(\alpha\bar\beta)}
            \,G\left(x_{mn}; \psi_\text{g}\right)
            F_{m,n}(\gamma; \phi)
        \end{aligned}
    \end{equation}
    with the fermionic and bosonic kernels:
    \begin{equation}
        \begin{aligned}
            F_{m,n}(\gamma; |\phi\rangle) &= \langle\phi|\Pi_m\gamma\Pi_n|\phi\rangle \\
            G(\tau; |\psi_\text{g}\rangle) &= \langle\psi_\text{g}|\Upsilon(\tau)|\psi_\text{g}\rangle\,, \\
            x_{mn} &= \beta-\frac{m+n}{2}\,\alpha
        \end{aligned}
    \end{equation}
    Again, the reader will find in the Supplemental Material the explicit expressions for the fermionic and bosonic kernels. Note that the generic single-mode bosonic Gaussian state can be written as a displaced squeezed state:
    \begin{equation}
        |\psi_\text{g}\rangle = D(\delta)S(\zeta)|0\rangle\,.
    \end{equation}
    We will denote $\mu = \cosh(r)$ and $\nu = e^{i\phi}\sinh(r)$ with $\zeta = re^{i\phi}$.
    Equipped with the bosonic and fermionic kernels, we can now explicitely compute $\sum_\gamma \lVert\langle\Upsilon(\beta)\otimes\gamma\rangle_{U(\alpha)|\psi\rangle}\rVert_1$ for all Majorana stabilizer states. One can then extract the hybrid magic and average over the Majorana stabilizer states. Let us introduce a finite measure $\mathfrak{m}$ on the space of pure Gaussian states (that we parametrized by the displacement $\delta$ and the squeezing $\zeta$). We finally obtain the non-stabilizer power of the conditional displacement gate:
    \begin{equation}
        \begin{aligned}
            &\text{Power}_\mathfrak{m}(\text{CD}(\alpha))\\
            &\quad=
            \frac{2}{3}\,\mathbb E_\mathfrak{m}\log\left\{
            \frac{1+\text{erf}\left(y_{\mu \nu}\right)
            + \mathbb E_{\Theta}\left[z_\Theta\right]}{2}
            \right\},
        \end{aligned}
    \end{equation}
    where $y_{\mu \nu} = \sqrt 2|\mu\alpha+\nu\bar\alpha|$ and $z_\Theta = \left|\sin\Theta\right|+\left|\cos\Theta\right|$
    The space of pure Gaussian states being, of course, non-compact, one needs to adjoin a physical cutoff in order to be able to define a finite measure. One can, for instance, use a cutoff on the squeezing parameter, corresponding to an energy constraint. The cutoff can be a hard cutoff or a quickly decaying smooth cutoff. One can allow for displacement or not. Note that the law of $\Theta$ itself depends on the bosonic Gaussian state instance. For simplicity, let us take a Dirac measure picked on the Fock vacuum. After dust settles down, we finally obtain very explicitly:
    \begin{equation}
    \label{eq:power_cd_vs_alpha}
        \begin{aligned}
            \text{Power}_{\text{CD}(\alpha)}
            &=
            \frac{2}{3}\log\Biggl\{
            \frac{1+\text{erf}\left(\sqrt 2|\alpha|\right)}{2}
            +\frac{2}{\pi}
            \\
            &\qquad
            -\frac{4}{\pi}\sum_{n=1}^{\infty}
            \frac{e^{-8n^2|\alpha|^2}}{16 n^{2}-1}
            \Biggr\}
        \end{aligned}
    \end{equation}
    We can see that for $\alpha=0$, for which one can resum the series into $(4-\pi)/8$, we obtain
    $\text{Power}(\text{CD}(0)) = 0\,,$
    as it should. We can define two $\alpha$-independent scalar quantities associated to the non-stabilizer power, namely the scaling property for an infinitesimal displacement and the asymptotic value of the power. Indeed, a slightly more careful analysis allows to extract the behavior of the power for small $\alpha$, we obtain:
    \begin{equation}
        \text{Power}(\text{CD}(\alpha)) = \frac{2}{3}\sqrt{\frac{2}{\pi}}\,|\alpha| + \mathcal O(|\alpha|^2)\,.
    \end{equation}
    The reader will find in Fig. \ref{fig:power_cd_vs_alpha} the power (\ref{eq:power_cd_vs_alpha}) of the conditional displacement gate as a function of $|\alpha|$.
    We can see that the power converges fast to its asymptotic value:
    \begin{equation}
        \text{Power}(\text{CD}(\alpha)) \xrightarrow[|\alpha| \to \infty]{} \frac{2}{3}\,\log\left\{1+\frac{2}{\pi}
        \right\}\simeq 0.3284\,.
    \end{equation}

    The non-stabilizer power introduced above quantifies, at the \emph{gate} level, how efficiently a hybrid operation injects boson–fermion magic when averaged over a natural family of free (Gaussian $\times$ Majorana-stabilizer) inputs. For the conditional displacement $\text{CD}(\alpha)$ we obtained a fully explicit expression together with two informative limits: a linear small-amplitude scaling, $\mathrm{Power}(\text{CD}(\alpha)) \propto |\alpha|$, and a finite asymptote at large $|\alpha|$. Operationally, the linear regime captures the \emph{hybrid magic yield per unit drive} around a Gaussian state, whereas the saturation shows that a single CD gate has a bounded magic-injection capacity—beyond which further increase of $|\alpha|$ does not boost hybrid non-stabilizerness. This makes clear that large hybrid magic requires either sequences of non-Gaussian steps (depth) or compositions with Gaussian resources (e.g., squeezing) that reshape the input distribution seen by the gate.

    Our calculation also highlights a subtle but important modeling choice: the power is defined by averaging over a measure on the non-compact manifold of pure Gaussian states. Different, physically motivated measures (e.g., energy-constrained ensembles with or without displacement) weight phase-space directions and squeezing axes differently, leading to distinct quantitative values while preserving the same qualitative picture (linear onset and bounded asymptote). In particular, with squeezing the power becomes direction-dependent in phase space: it depends on the combination $\mu\,\alpha + \nu\,\bar{\alpha}$. Put simply, if the displacement $\alpha$ is aligned with the squeezed quadrature, $\text{CD}(\alpha)$ produces more hybrid magic for the same amplitude $ |\alpha|$.

    Conceptually, gate power complements state-level metrics (mana/SRE and their hybrid $p$-norm proxies) by isolating the intrinsic, input-averaged magic-generation capability of a transformation. This makes it a convenient primitive for \emph{magic budgeting} in hybrid circuits in the sense that it suggests natural design heuristics: (i) pre/post Gaussian shaping (displacements/squeezers) to align with the gate’s most magic-productive quadrature; (ii) interleaving $\text{CD}(\alpha)$ with fermionic Clifford rotations to steer which Majorana sectors are populated; and (iii) distributing moderate $|\alpha|$ across multiple layers (benefitting from the linear regime) rather than pushing a single layer deep into saturation.

\section{Discussion and Outlook}

    The phase-space framework for magic in hybrid boson-fermion systems developed in this work opens several promising directions for future research and applications.

    Supersymmetric quantum mechanics provides a natural playground for hybrid systems, as it unifies bosonic and fermionic degrees of freedom within a rich geometric framework \cite{witten1982supersymmetry}. The interplay between supersymmetry, phase-space structure, and non-stabilizerness may yield new insights into the geometry of quantum resources and their operational meaning.

    Beyond supersymmetry, hybrid models relevant to quantum optics, such as various generalizations of the Jaynes-Cummings model, offer a rich testbed for exploring the role of magic in light-matter interactions. The extension of phase-space methods to these settings could clarify the connection between non-classicality, contextuality, and quantum advantage in experimentally accessible systems.

    A particularly exciting avenue is the application of hybrid magic to realistic condensed matter and molecular systems, where both bosonic (e.g., phonons) and fermionic (e.g., electrons) excitations play a central role. Recent work has begun to explore the non-stabilizerness of molecular bonding~\cite{sarkis2025molecules}, suggesting that resource-theoretic concepts may provide new perspectives on quantum chemistry and materials science. A further promising direction relevant for physical chemistry is the application of the hybrid magic to fermionic systems coupled to the background quantum elctrodynamics gauge field. A concrete realization of these systems would be molecular cavity QED.

    The phase-space approach can also be generalized to hybrid boson-spin systems, such as those described by the Jaynes-Cummings Hamiltonian~\cite{sanchez2025jaynes}, further broadening the scope of resource theories for magic. This could enable the study of magic in a variety of hybrid quantum platforms, including trapped ions, superconducting circuits, and cavity QED systems.

    A major open question is the systematic development of a resource theory for magic in hybrid systems, including the identification of free operations, maximally magic states, and protocols for hybrid magic state distillation. The operational significance of hybrid magic, its relation to classical simulability, and its role in quantum error correction and computation remain to be fully elucidated~\cite{Veitch_2014,howard2014contextuality,bravyi2012magic,heinrich2019robustness,hamaguchi2024handbook}.

    Considering time evolution, the dynamics of hybrid magic—its evolution, transport, and possible localization in interacting boson-fermion systems—will represent an important direction. Understanding how magic propagates and transforms under hybrid dynamics could shed light on fundamental questions in quantum information and many-body physics~\cite{turkeshi2025magic, tirrito2024anticoncentration, smith2406non, passarelli2024nonstabilizerness}.

    A very interesting application would also be to lattice gauge theories, such as quantum electrodynamics (QED) on the lattice. In these models, both bosonic gauge fields and fermionic matter are naturally present, and the interplay of non-stabilizerness with gauge constraints and local symmetries could yield new insights into the computational complexity and simulation of high-energy physics models~\cite{homeier2023realistic,tagliacozzo2013simulation,mathis2020toward}. Recent advances in quantum simulation platforms have enabled the study of lattice gauge theories with dynamical matter, opening the door to experimental investigations of resource-theoretic properties in these systems~\cite{homeier2023realistic,tagliacozzo2013simulation}.
    Further recent work on digital quantum simulation of non-Abelian lattice gauge theories, fermion-qudit and qudit-based lattice-gauge processors, hadron dynamics, and systematic links between lattice discretizations and quantum simulation algorithms provides additional context for connecting hybrid resource questions to high-energy simulation efforts~\cite{calajo2024digital,meth2023simulating,farrell2023preparations,farrell2024quantum}.

    We anticipate that the unified hybrid definition proposed in this work will stimulate further research into the structure, dynamics, and applications of magic in a wide variety of hybrid quantum systems.
    Furthermore, since the dressed Wigner functions have already been measured experimentally \cite{Vlastakis2015, wang2016schrodinger}, the hybrid magic could be a measure that bridges between the resource theory of nonstabilizerness based on quantum gates, and the hardness of implementation on real quantum  hardware.

    \emph{Note added:} While this work was being completed, we came across \cite{crew2025magic} that follows a similar phase space approach, for spins rather than fermions, which generalizes the quantity from the SRE rather than from the Wigner function.

\section*{Acknowledgements}

    We thank Aur\'elia Chenu for her feedback after carefully reading the manuscript. The authors acknowledge funding via the FNR-CORE Grant ``BroadApp'' (FNR-CORE C20/MS/14769845), FNR Attract (FNR, Attract grant QOMPET 15382998), and ERC-AdG Grant ``FITMOL'' (Grant No. 1110643).

\bibliographystyle{quantum}
\bibliography{references}

@preamble{"\DoNotMakeWarningsErrors"}

@article{howell2022quantifying,
  title={Quantifying magic for multi-qubit operations},
  author = {Seddon, James R. and Campbell, Earl T.},
  journal = {Proc. R. Soc. A},
  volume = {475},
  pages = {20190251},
  year = {2019},
  number = {2227},
  doi = {10.1098/rspa.2019.0251},
}

@article{kenfack2004negativity,
  title={Negativity of the Wigner function as an indicator of non-classicality},
  author={Kenfack, Anatole and {\.Z}yczkowski, Karol},
  journal={J. Opt. B: Quantum Semiclass. Opt.},
  volume={6},
  number={10},
  pages={396--404},
  year={2004},
  doi = {10.1088/1464-4266/6/10/003},
}

@article{dalton2013grassmann,
  title={Grassmann phase space theory and the Jaynes--Cummings model},
  author={Dalton, BJ and Garraway, Barry M and Jeffers, John and Barnett, SM},
  journal={Annals of Physics},
  volume={334},
  pages={100--141},
  year={2013},
  publisher={Elsevier},
  doi = {10.1016/j.aop.2013.03.010},
}

@article{cahill1999density,
  title={Density operators for fermions},
  author={Cahill, Kevin E and Glauber, Roy J},
  journal={Physical Review A},
  volume={59},
  number={2},
  pages={1538},
  year={1999},
  publisher={APS},
  doi = {10.1103/PhysRevA.59.1538},
}

@article{abe2001theory,
  title={Theory of super phase-space representations and supercoherent states},
  author={Abe, Sumiyoshi},
  journal={International Journal of Theoretical Physics},
  volume={40},
  pages={1647--1655},
  year={2001},
  publisher={Springer},
  doi = {10.1023/A:1011698206460},
}

@article{collura2412quantum,
  title={The non-stabilizerness of fermionic Gaussian states},
  author={Collura, Mario and De Nardis, Jacopo and Alba, Vincenzo and Lami, Guglielmo},
  journal = {Quantum},
  year = {2026},
  volume = {10},
  pages = {2036},
  doi = {10.22331/q-2026-03-23-2036},
  eprint = {2412.05367},
  archivePrefix = {arXiv},
}

@article{howard2014contextuality,
  title={Contextuality supplies the ‘magic’ for quantum computation},
  author={Howard, Mark and Wallman, Joel and Veitch, Victor and Emerson, Joseph},
  journal={Nature},
  volume={510},
  number={7505},
  pages={351--355},
  year={2014},
  publisher={Nature Publishing Group},
  doi = {10.1038/nature13460},
}

@article{Veitch_2014,
  title={The resource theory of stabilizer quantum computation},
  volume={16},
  ISSN={1367-2630},
  url={http://dx.doi.org/10.1088/1367-2630/16/1/013009},
  DOI={10.1088/1367-2630/16/1/013009},
  number={1},
  journal={New J. Phys.},
  publisher={IOP Publishing},
  author={Veitch, Victor and Hamed Mousavian, S A and Gottesman, Daniel and Emerson, Joseph},
  year={2014},
  month=jan, pages={013009}
}

@article{bera2502non,
  title={Non-stabilizerness of Sachdev-Ye-Kitaev model},
  author={Bera, Surajit and Schir{\`o}, Marco},
  journal = {SciPost Phys.},
  year = {2025},
  volume = {19},
  number = {6},
  pages = {159},
  doi = {10.21468/SciPostPhys.19.6.159},
  eprint = {2502.01582},
  archivePrefix = {arXiv},
}

@article{bettaque2024structure,
  title={The structure of the Majorana Clifford group},
  author={Bettaque, Val{\'e}rie and Swingle, Brian},
  year={2024},
  eprint = {2407.11319},
  archivePrefix = {arXiv},
}

@article{veitch2012negative,
  title={Negative quasi-probability as a resource for quantum computation},
  author={Veitch, Victor and Ferrie, Christopher and Gross, David and Emerson, Joseph},
  journal={New J. Phys.},
  volume={14},
  number={11},
  pages={113011},
  year={2012},
  publisher={IOP Publishing},
  doi = {10.1088/1367-2630/14/11/113011},
}

@article{mari2012positive,
  title={Positive Wigner functions render classical simulation of quantum computation efficient},
  author={Mari, Andrea and Eisert, Jens},
  journal={Phys. Rev. Lett.},
  volume={109},
  number={23},
  pages={230503},
  year={2012},
  publisher={APS},
  doi = {10.1103/physrevlett.109.230503},
}

@article{wootters1987wigner,
  title={A Wigner-function formulation of finite-state quantum mechanics},
  author={Wootters, William K.},
  journal={Ann. Phys.},
  volume={176},
  number={1},
  pages={1--21},
  year={1987},
  publisher={Elsevier},
  doi={10.1016/0003-4916(87)90176-X},
}

@article{gibbons2004discrete,
  title={Discrete phase space based on finite fields},
  author={Gibbons, Kathleen S and Hoffman, Matthew J and Wootters, William K},
  journal={Phys. Rev. A},
  volume={70},
  number={6},
  pages={062101},
  year={2004},
  publisher={APS},
  doi = {10.1103/physreva.70.062101},
}

@article{leone2022stabilizer,
  title={Stabilizer r{\'e}nyi entropy},
  author={Leone, Lorenzo and Oliviero, Salvatore FE and Hamma, Alioscia},
  journal={Phys. Rev. Lett.},
  volume={128},
  number={5},
  pages={050402},
  year={2022},
  publisher={APS},
  doi = {10.1103/PhysRevLett.128.050402},
}

@article{bravyi2005universal,
  title={Universal quantum computation with ideal Clifford gates and noisy ancillas},
  author={Bravyi, Sergey and Kitaev, Alexei},
  journal={Phys. Rev. A},
  volume={71},
  number={2},
  pages={022316},
  year={2005},
  publisher={APS},
  doi = {10.1103/physreva.71.022316},
}

@article{howard2017application,
  title={Application of a resource theory for magic states to fault-tolerant quantum computing},
  author={Howard, Mark and Campbell, Earl},
  journal={Phys. Rev. Lett.},
  volume={118},
  number={9},
  pages={090501},
  year={2017},
  publisher={APS},
  doi = {10.1103/physrevlett.118.090501},
}

@article{heinrich2019robustness,
  title={Robustness of magic and symmetries of the stabiliser polytope},
  author={Heinrich, Markus and Gross, David},
  journal={Quantum},
  volume={3},
  pages={132},
  year={2019},
  publisher={Quantum Publishing},
  doi = {10.22331/q-2019-04-08-132},
}

@article{hamaguchi2024handbook,
  title={Handbook for quantifying robustness of magic},
  author={Hamaguchi, Hiroki and Hamada, Kou and Yoshioka, Nobuyuki},
  journal={Quantum},
  volume={8},
  pages={1461},
  year={2024},
  publisher={Quantum Publishing},
  doi = {10.22331/q-2024-09-05-1461},
}

@article{haug2023stabilizer,
  title={Stabilizer entropies and nonstabilizerness monotones},
  author={Haug, Tobias and Piroli, Lorenzo},
  journal={Quantum},
  volume={7},
  pages={1092},
  year={2023},
  publisher={Quantum Publishing},
  doi = {10.22331/q-2023-08-28-1092},
}

@article{bravyi2012magic,
  title={Magic-state distillation with low overhead},
  author={Bravyi, Sergey and Haah, Jeongwan},
  journal={Phys. Rev. A},
  volume={86},
  number={5},
  pages={052329},
  year={2012},
  publisher={APS},
  doi = {10.1103/physreva.86.052329},
}

@article{jasser2025stabilizerentropyentanglementcomplexity,
  title={Stabilizer entropy and entanglement complexity in the Sachdev-Ye-Kitaev model},
  author={Jasser, Barbara and Odavi{\'c}, Jovan and Hamma, Alioscia},
  journal={Phys. Rev. B},
  volume={112},
  number={17},
  pages={174204},
  year={2025},
  doi={10.1103/rz86-47h3},
  eprint={2502.03093},
  archivePrefix={arXiv},
  primaryClass={quant-ph},
}

@article{oliviero2022magic,
  title={Magic-state resource theory for the ground state of the transverse-field Ising model},
  author={Oliviero, Salvatore FE and Leone, Lorenzo and Hamma, Alioscia},
  journal={Phys. Rev. A},
  volume={106},
  number={4},
  pages={042426},
  year={2022},
  publisher={APS},
  doi = {10.1103/physreva.106.042426},
}

@article{smith2406non,
  title={Non-stabilizerness in kinetically-constrained Rydberg atom arrays},
  author={Smith, R and Papic, Z and Hallam, A},
  journal = {Phys. Rev. B},
  year = {2025},
  volume = {111},
  pages = {245148},
  doi = {10.1103/jz4d-vdhj},
  eprint = {2406.14348},
  archivePrefix = {arXiv},
}

@article{odavic2024stabilizer,
  title={Stabilizer entropy in non-integrable quantum evolutions},
  author={Odavi{\'c}, Jovan and Viscardi, Michele and Hamma, Alioscia},
  journal = {Phys. Rev. B},
  year = {2025},
  volume = {112},
  pages = {104301},
  doi = {10.1103/y9r6-dx7p},
  eprint = {2412.10228},
  archivePrefix = {arXiv},
}

@article{passarelli2024nonstabilizerness,
  title={Nonstabilizerness of permutationally invariant systems},
  author={Passarelli, Gianluca and Fazio, Rosario and Lucignano, Procolo},
  journal={Phys. Rev. A},
  volume={110},
  number={2},
  pages={022436},
  year={2024},
  publisher={APS},
  doi = {10.1103/physreva.110.022436},
}

@article{tarabunga2024critical,
  title={Critical behaviors of non-stabilizerness in quantum spin chains},
  author={Tarabunga, Poetri Sonya},
  journal={Quantum},
  volume={8},
  pages={1413},
  year={2024},
  publisher={Quantum Publishing},
  doi = {10.22331/q-2024-07-17-1413},
}

@article{goto2021chaos,
  title={Chaos by magic},
  author={Goto, Kanato and Nosaka, Tomoki and Nozaki, Masahiro},
  year={2021},
  eprint = {2112.14593},
  archivePrefix = {arXiv},
}

@article{mclauchlan2022fermion,
  title={Fermion-parity-based computation and its Majorana-zero-mode implementation},
  author={McLauchlan, Campbell K and B{\'e}ri, Benjamin},
  journal={Phys. Rev. Lett.},
  volume={128},
  number={18},
  pages={180504},
  year={2022},
  publisher={APS},
  doi = {10.1103/physrevlett.128.180504},
}

@article{mudassar2024encoding,
  title={Encoding Majorana codes},
  author={Mudassar, Maryam and Chien, Riley W and Gottesman, Daniel},
  journal={Phys. Rev. A},
  volume={110},
  number={3},
  pages={032430},
  year={2024},
  publisher={APS},
  doi = {10.1103/physreva.110.032430},
}

@article{tirrito2024anticoncentration,
  title={Anticoncentration and nonstabilizerness spreading under ergodic quantum dynamics},
  author={Tirrito, Emanuele and Turkeshi, Xhek and Sierant, Piotr},
  year={2024},
  eprint = {2412.10229},
  archivePrefix = {arXiv},
}

@article{turkeshi2025magic,
  title={Magic spreading in random quantum circuits},
  author={Turkeshi, Xhek and Tirrito, Emanuele and Sierant, Piotr},
  journal={Nat. Commun.},
  volume={16},
  number={1},
  pages={2575},
  year={2025},
  publisher={Nature Publishing Group UK London},
  doi = {10.1038/s41467-025-57704-x},
}

@article{gu2024zero,
  title={Zero and finite temperature quantum simulations powered by quantum magic},
  author={Gu, Andi and Hu, Hong-Ye and Luo, Di and Patti, Taylor L and Rubin, Nicholas C and Yelin, Susanne F},
  journal={Quantum},
  volume={8},
  pages={1422},
  year={2024},
  publisher={Verein zur F{\"o}rderung des Open Access Publizierens in den Quantenwissenschaften},
  doi = {10.22331/q-2024-07-23-1422},
}

@incollection{hohenadler2007lang,
  title={Lang-Firsov approaches to polaron physics: From variational methods to unbiased quantum Monte Carlo simulations},
  author={Hohenadler, Martin and von der Linden, Wolfgang},
  booktitle={Polarons in Advanced Materials},
  pages={463--502},
  year={2007},
  publisher={Springer},
  doi = {10.1007/978-1-4020-6348-0_11},
}

@article{rongsheng2002exact,
  title={Exact solutions for the two-site Holstein model},
  author={Han, Rongsheng and Lin, Zijing and Wang, Kelin},
  journal={Physical Review B},
  volume={65},
  number={17},
  pages={174303},
  year={2002},
  publisher={APS},
  doi = {10.1103/physrevb.65.174303},
}

@article{sarkis2025molecules,
  title={Nonstabilizerness in molecular bonding},
  author={Sarkis, Matthieu and Tkatchenko, Alexandre},
  journal={Phys. Rev. Research},
  volume={8},
  number={2},
  pages={L022048},
  year={2026},
  doi={10.1103/f62c-47jb},
  eprint={2504.06673},
  archivePrefix={arXiv},
}

@article{sanchez2025jaynes,
  title={The Jaynes-Cummings model in Phase Space Quantum Mechanics},
  author={Sanchez-Cordova, Mar and Berra-Montiel, Jasel and Molgado, Alberto},
  year={2025},
  eprint = {2506.23386},
  archivePrefix = {arXiv},
}

@article{homeier2023realistic,
  title={Realistic scheme for quantum simulation of Z 2 lattice gauge theories with dynamical matter in (2+ 1) D},
  author={Homeier, Lukas and Bohrdt, Annabelle and Linsel, Simon and Demler, Eugene and Halimeh, Jad C and Grusdt, Fabian},
  journal={Communications Physics},
  volume={6},
  number={1},
  pages={127},
  year={2023},
  publisher={Nature Publishing Group UK London},
  doi = {10.1038/s42005-023-01237-6},
}

@article{tagliacozzo2013simulation,
  title={Simulation of non-Abelian gauge theories with optical lattices},
  author={Tagliacozzo, L and Celi, A and Orland, P and Mitchell, MW and Lewenstein, M},
  journal={Nature communications},
  volume={4},
  number={1},
  pages={2615},
  year={2013},
  publisher={Nature Publishing Group UK London},
  doi = {10.1038/ncomms3615},
}

@article{mathis2020toward,
  title={Toward scalable simulations of lattice gauge theories on quantum computers},
  author={Mathis, Simon V and Mazzola, Guglielmo and Tavernelli, Ivano},
  journal={Physical Review D},
  volume={102},
  number={9},
  pages={094501},
  year={2020},
  publisher={APS},
  doi = {10.1103/physrevd.102.094501},
}

@article{albarelli2018resource,
  title={Resource theory of quantum non-Gaussianity and Wigner negativity},
  author={Albarelli, Francesco and Genoni, Marco G and Paris, Matteo GA and Ferraro, Alessandro},
  journal={Physical Review A},
  volume={98},
  number={5},
  pages={052350},
  year={2018},
  publisher={APS},
  doi = {10.1103/physreva.98.052350},
}

@article{bravyi2002fermionic,
  title={Fermionic quantum computation},
  author={Bravyi, Sergey B and Kitaev, Alexei Yu},
  journal={Annals of Physics},
  volume={298},
  number={1},
  pages={210--226},
  year={2002},
  publisher={Elsevier},
  doi = {10.1006/aphy.2002.6254},
}

@article{leone2024stabilizer,
  title={Stabilizer entropies are monotones for magic-state resource theory},
  author={Leone, Lorenzo and Bittel, Lennart},
  journal={Phys. Rev. A},
  volume={110},
  number={4},
  pages={L040403},
  year={2024},
  publisher={APS},
  doi = {10.1103/physreva.110.l040403},
}

@article{shuangshuang2022dynamics,
  title={Dynamics of atomic magic in the Jaynes--Cummings model},
  author={Fu, Shuangshuang and Li, Xiaohui and Luo, Shunlong},
  journal={Quantum Inf. Process.},
  volume={22},
  number={1},
  pages={7},
  year={2023},
  publisher={Springer},
  doi = {10.1007/s11128-022-03756-7},
}

@article{campbell2011catalysis,
  title={Catalysis and activation of magic states in fault-tolerant architectures},
  author={Campbell, Earl T},
  journal={Phys. Rev. A},
  volume={83},
  number={3},
  pages={032317},
  year={2011},
  publisher={APS},
  doi = {10.1103/physreva.83.032317},
}

@article{liu2024hybrid,
  title={Hybrid oscillator-qubit quantum processors: Instruction set architectures, abstract machine models, and applications},
  author={Liu, Yuan and Singh, Shraddha and Smith, Kevin C and Crane, Eleanor and Martyn, John M and Eickbusch, Alec and Schuckert, Alexander and Li, Richard D and Sinanan-Singh, Jasmine and Soley, Micheline B and others},
  journal = {PRX Quantum},
  year = {2026},
  volume = {7},
  pages = {010201},
  doi = {10.1103/4rf7-9tfx},
  eprint = {2407.10381},
  archivePrefix = {arXiv},
}

@article{varona2024towards,
  title={Towards quantum computing Feynman diagrams in hybrid qubit-oscillator devices},
  author={Varona, S and Saner, S and B{\u{a}}z{\u{a}}van, O and Araneda, G and Aarts, G and Bermudez, A},
  year={2024},
  eprint = {2411.05092},
  archivePrefix = {arXiv},
}

@article{kbv4-jj51,
  title = {Hybrid quantum simulations with qubits and qumodes on trapped-ion platforms},
  author = {Araz, Jack Y. and Grau, Matt and Montgomery, Jake and Ringer, Felix},
  journal = {Phys. Rev. A},
  volume = {112},
  issue = {1},
  pages = {012620},
  numpages = {19},
  year = {2025},
  month = {Jul},
  publisher = {American Physical Society},
  doi = {10.1103/kbv4-jj51},
  url = {https://link.aps.org/doi/10.1103/kbv4-jj51}
}

@article{crane2024hybrid,
  title={Hybrid oscillator-qubit quantum processors: Simulating fermions, bosons, and gauge fields},
  author={Crane, Eleanor and Smith, Kevin C and Tomesh, Teague and Eickbusch, Alec and Martyn, John M and K{\"u}hn, Stefan and Funcke, Lena and DeMarco, Michael Austin and Chuang, Isaac L and Wiebe, Nathan and others},
  year={2024},
  eprint = {2409.03747},
  archivePrefix = {arXiv},
}

@article{chen2025genesis,
  title={Genesis: A Compiler Framework for Hamiltonian Simulation on Hybrid CV-DV Quantum Computers},
  author={Chen, Zihan and Li, Jiakang and Guo, Minghao and Chen, Henry and Li, Zirui and Bierman, Joel and Huang, Yipeng and Zhou, Huiyang and Liu, Yuan and Zhang, Eddy Z},
  year={2025},
  eprint = {2505.13683},
  archivePrefix = {arXiv},
}

@article{decker2025symbolic,
  title={Symbolic Hamiltonian Compiler for Hybrid Qubit-Boson Processors},
  author={Decker, Ethan and Gustafson, Erik and McKinney, Evan and Jones, Alex K and Goetz, Lucas and Li, Ang and Schuckert, Alexander and Stein, Samuel and Li, Gushu and Crane, Eleanor},
  year={2025},
  eprint = {2506.00215},
  archivePrefix = {arXiv},
}

@article{singh2025towards,
  title={Non-Abelian Quantum Signal Processing: A Composite Pulse for Fast Analytic Control of Hybrid Oscillator-Qubit Processors},
  author={Singh, Shraddha and Royer, Baptiste and Girvin, Steven M.},
  year={2025},
  eprint={2504.19992},
  archivePrefix={arXiv},
  primaryClass={quant-ph},
}

@article{kumar2025digital,
  title={Digital-analog quantum computing of fermion-boson models in superconducting circuits},
  author={Kumar, Shubham and Hegade, Narendra N and Visuri, Anne-Maria and Bhargava, Balaganchi A and Hernandez, Juan FR and Solano, Enrique and Albarr{\'a}n-Arriagada, Francisco and Barrios, G Alvarado},
  journal={npj Quantum Information},
  volume={11},
  number={1},
  pages={43},
  year={2025},
  publisher={Nature Publishing Group UK London},
  doi = {10.1038/s41534-025-01001-4},
}

@article{tarabunga2025efficientmutualmagicmagic,
  title={Efficient mutual magic and magic capacity with matrix product states},
  author={Tarabunga, Poetri Sonya and Haug, Tobias},
  journal={SciPost Phys.},
  volume={19},
  number={4},
  pages={085},
  year={2025},
  doi={10.21468/SciPostPhys.19.4.085},
  eprint={2504.07230},
  archivePrefix={arXiv},
  primaryClass={quant-ph},
}

@misc{saner2025realtimeobservationaharonovbohminterference,
      title={Real-Time Observation of Aharonov-Bohm Interference in a $\mathbb{Z}_2$ Lattice Gauge Theory on a Hybrid Qubit-Oscillator Quantum Computer},
      author={S. Saner and O. Băzăvan and D. J. Webb and G. Araneda and C. J. Ballance and R. Srinivas and D. M. Lucas and A. Bermúdez},
      year={2025},
      eprint={2507.19588},
      archivePrefix={arXiv},
      primaryClass={quant-ph},
      url={https://arxiv.org/abs/2507.19588},
}

@article{cahill1969ordered,
  title={Ordered expansions in boson amplitude operators},
  author={Cahill, Kevin E and Glauber, Roy J},
  journal={Physical Review},
  volume={177},
  number={5},
  pages={1857--1881},
  year={1969},
  publisher={APS},
  doi = {10.1103/PhysRev.177.1857},
  url = {https://link.aps.org/doi/10.1103/PhysRev.177.1857},
}

@article{martinezazcona2025magic,
  title={Magic Steady State Production: Non-Hermitian, Dissipative, and Stochastic Pathways},
  author={Martinez-Azcona, Pablo and Sarkis, Matthieu and Tkatchenko, Alexandre and Chenu, Aur{\'e}lia},
  year={2025},
  eprint = {2507.08676},
  archivePrefix = {arXiv},
}

@article{crew2025magic,
  title={Magic Entropy in Hybrid Spin-Boson Systems},
  author={Crew, Samuel and Li, Ying-Lin and Li, Heng-Hsi and Chang, Po-Yao},
  year={2025},
  eprint = {2508.06018},
  archivePrefix = {arXiv},
}

@article{tarabunga2023many,
  title={Many-body magic via pauli-markov chains—from criticality to gauge theories},
  author={Tarabunga, Poetri Sonya and Tirrito, Emanuele and Chanda, Titas and Dalmonte, Marcello},
  journal={PRX Quantum},
  volume={4},
  number={4},
  pages={040317},
  year={2023},
  publisher={APS},
  doi = {10.1103/prxquantum.4.040317},
}

@article{hoshino2025stabilizer,
  title={Stabilizer R{\'e}nyi Entropy and Conformal Field Theory},
  author={Hoshino, Masahiro and Oshikawa, Masaki and Ashida, Yuto},
  journal = {Phys. Rev. X},
  year = {2026},
  volume = {16},
  pages = {011037},
  doi = {10.1103/ylsz-dm3y},
  eprint = {2503.13599},
  archivePrefix = {arXiv},
}

@article{witten1982supersymmetry,
  title={Supersymmetry and Morse theory},
  author={Witten, Edward},
  journal={Journal of differential geometry},
  volume={17},
  number={4},
  pages={661--692},
  year={1982},
  publisher={Lehigh University},
  doi = {10.4310/jdg/1214437492},
}

@article{malvimat2026multipartite,
  title={Multipartite Non-local Magic and {SYK} Model},
  author={Malvimat, Vinay and Sarkis, Matthieu and Suk, Yena and Yoon, Junggi},
  year={2026},
  eprint={2601.03076},
  archivePrefix={arXiv},
  primaryClass={quant-ph},
}

@misc{leone2026unbearablehardnessdecidingmagic,
      title={The unbearable hardness of deciding about magic},
      author={Lorenzo Leone and Jens Eisert and Salvatore F. E. Oliviero},
      year={2026},
      eprint={2602.22330},
      archivePrefix={arXiv},
      primaryClass={quant-ph},
      url={https://arxiv.org/abs/2602.22330},
}

@article{tarabunga_nonstab_24,
  title = {Nonstabilizerness via Matrix Product States in the Pauli Basis},
  author = {Tarabunga, Poetri Sonya and Tirrito, Emanuele and Ba\~nuls, Mari Carmen and Dalmonte, Marcello},
  journal = {Phys. Rev. Lett.},
  volume = {133},
  issue = {1},
  pages = {010601},
  numpages = {7},
  year = {2024},
  month = {Jul},
  publisher = {American Physical Society},
  doi = {10.1103/PhysRevLett.133.010601},
  url = {https://link.aps.org/doi/10.1103/PhysRevLett.133.010601}
}

@Article{Vlastakis2015,
author={Vlastakis, Brian
and Petrenko, Andrei
and Ofek, Nissim
and Sun, Luyan
and Leghtas, Zaki
and Sliwa, Katrina
and Liu, Yehan
and Hatridge, Michael
and Blumoff, Jacob
and Frunzio, Luigi
and Mirrahimi, Mazyar
and Jiang, Liang
and Devoret, M. H.
and Schoelkopf, R. J.},
title={Characterizing entanglement of an artificial atom and a cavity cat state with Bell's inequality},
journal={Nature Communications},
year={2015},
month={Nov},
day={27},
volume={6},
number={1},
pages={8970},
issn={2041-1723},
doi={10.1038/ncomms9970},
url={https://doi.org/10.1038/ncomms9970}
}

@article{hybrid_resource,
  title={A Resource Theory of Non-Stabilizerness for Hybrid Boson-Fermion Systems},
  author={Sarkis, M. and Martinez-Azcona, P.},
  journal={To Appear},
  volume={},
  number={},
  pages={},
  year={2026},
  nolink = {},
}

@article{wang2016schrodinger,
  title={A Schr{\"o}dinger cat living in two boxes},
  author={Wang, Chen and Gao, Yvonne Y and Reinhold, Philip and Heeres, Reinier W and Ofek, Nissim and Chou, Kevin and Axline, Christopher and Reagor, Matthew and Blumoff, Jacob and Sliwa, KM and others},
  journal={Science},
  volume={352},
  number={6289},
  pages={1087--1091},
  year={2016},
  publisher={American Association for the Advancement of Science},
  doi = {10.1126/science.aaf2941},
}

@article{calajo2024digital,
  title={Digital quantum simulation of a {(1+1)D} {SU}(2) lattice gauge theory with ion qudits},
  author={Calaj{\`o}, Giuseppe and Magnifico, Giuseppe E and Edmunds, Claire J and Ringbauer, Martin and Montangero, Simone and Silvi, Pietro},
  journal={PRX Quantum},
  volume={5},
  number={4},
  pages={040309},
  year={2024},
  publisher={APS},
  doi = {10.1103/PRXQuantum.5.040309},
}

@Article{meth2023simulating,
author={Meth, Michael
and Zhang, Jinglei
and Haase, Jan F.
and Edmunds, Claire
and Postler, Lukas
and Jena, Andrew J.
and Steiner, Alex
and Dellantonio, Luca
and Blatt, Rainer
and Zoller, Peter
and Monz, Thomas
and Schindler, Philipp
and Muschik, Christine
and Ringbauer, Martin},
title={Simulating two-dimensional lattice gauge theories on a qudit quantum computer},
journal={Nature Physics},
year={2025},
month={Apr},
day={01},
volume={21},
number={4},
pages={570-576},
issn={1745-2481},
doi={10.1038/s41567-025-02797-w},
url={https://doi.org/10.1038/s41567-025-02797-w}
}

@article{farrell2023preparations,
  title = {Preparations for quantum simulations of quantum chromodynamics in $1+1$ dimensions. I. Axial gauge},
  author = {Farrell, Roland C. and Chernyshev, Ivan A. and Powell, Sarah J. M. and Zemlevskiy, Nikita A. and Illa, Marc and Savage, Martin J.},
  journal = {Phys. Rev. D},
  volume = {107},
  issue = {5},
  pages = {054512},
  numpages = {43},
  year = {2023},
  month = {Mar},
  publisher = {American Physical Society},
  doi = {10.1103/PhysRevD.107.054512},
  url = {https://link.aps.org/doi/10.1103/PhysRevD.107.054512}
}

@article{farrell2024quantum,
  title = {Quantum simulations of hadron dynamics in the Schwinger model using 112 qubits},
  author = {Farrell, Roland C. and Illa, Marc and Ciavarella, Anthony N. and Savage, Martin J.},
  journal = {Phys. Rev. D},
  volume = {109},
  issue = {11},
  pages = {114510},
  numpages = {52},
  year = {2024},
  month = {Jun},
  publisher = {American Physical Society},
  doi = {10.1103/PhysRevD.109.114510},
  url = {https://link.aps.org/doi/10.1103/PhysRevD.109.114510}
}

@article{bejan2024dynamical,
  title = {Dynamical Magic Transitions in Monitored Clifford+$T$ Circuits},
  author = {Bejan, Mircea and McLauchlan, Campbell and B\'eri, Benjamin},
  journal = {PRX Quantum},
  volume = {5},
  issue = {3},
  pages = {030332},
  numpages = {30},
  year = {2024},
  month = {Aug},
  publisher = {American Physical Society},
  doi = {10.1103/PRXQuantum.5.030332},
  url = {https://link.aps.org/doi/10.1103/PRXQuantum.5.030332}
}

@Article{niroula2024phase,
author={Niroula, Pradeep
and White, Christopher David
and Wang, Qingfeng
and Johri, Sonika
and Zhu, Daiwei
and Monroe, Christopher
and Noel, Crystal
and Gullans, Michael J.},
title={Phase transition in magic with random quantum circuits},
journal={Nature Physics},
year={2024},
month={Nov},
day={01},
volume={20},
number={11},
pages={1786-1792},
issn={1745-2481},
doi={10.1038/s41567-024-02637-3},
url={https://doi.org/10.1038/s41567-024-02637-3}
}

@article{zhang2024quantummagicdynamics,
  title={Quantum magic dynamics in random circuits},
  author={Zhang, Yuzhen and Gu, Yingfei},
  journal={npj Quantum Inf.},
  volume={12},
  number={1},
  pages={87},
  year={2026},
  doi={10.1038/s41534-026-01253-8},
  eprint={2410.21128},
  archivePrefix={arXiv},
}

@article{rattacaso2023stabilizer,
  title = {Stabilizer entropy dynamics after a quantum quench},
  author = {Rattacaso, Davide and Leone, Lorenzo and Oliviero, Salvatore F. E. and Hamma, Alioscia},
  journal = {Phys. Rev. A},
  volume = {108},
  issue = {4},
  pages = {042407},
  numpages = {9},
  year = {2023},
  month = {Oct},
  publisher = {American Physical Society},
  doi = {10.1103/PhysRevA.108.042407},
  url = {https://link.aps.org/doi/10.1103/PhysRevA.108.042407}
}

@article{frau2024nonstabilizerness,
  title = {Nonstabilizerness versus entanglement in matrix product states},
  author = {Frau, M. and Tarabunga, P. S. and Collura, M. and Dalmonte, M. and Tirrito, E.},
  journal = {Phys. Rev. B},
  volume = {110},
  issue = {4},
  pages = {045101},
  numpages = {13},
  year = {2024},
  month = {Jul},
  publisher = {American Physical Society},
  doi = {10.1103/PhysRevB.110.045101},
  url = {https://link.aps.org/doi/10.1103/PhysRevB.110.045101}
}

@article{robin2025stabilizeraccelerated,
  title = {Stabilizer-accelerated quantum many-body ground-state estimation},
  author = {Robin, Caroline E. P.},
  journal = {Phys. Rev. A},
  volume = {112},
  issue = {5},
  pages = {052408},
  numpages = {21},
  year = {2025},
  month = {Nov},
  publisher = {American Physical Society},
  doi = {10.1103/5qr5-7jkz},
  url = {https://link.aps.org/doi/10.1103/5qr5-7jkz}
}

@Article{viscardi2025interplay,
	title={{Interplay of entanglement structures and stabilizer entropy in spin models}},
	author={Michele Viscardi and Marcello Dalmonte and Alioscia Hamma and Emanuele Tirrito},
	journal={SciPost Phys. Core},
	volume={9},
	number={1},
	pages={012},
	year={2026},
	publisher={SciPost},
	doi={10.21468/SciPostPhysCore.9.1.012},
	url={https://scipost.org/10.21468/SciPostPhysCore.9.1.012},
}

@article{falcao2025nonstabilizerness,
  title = {Nonstabilizerness in U(1) lattice gauge theory},
  author = {Falc\~ao, Pedro R. Nic\'acio and Tarabunga, Poetri Sonya and Frau, Martina and Tirrito, Emanuele and Zakrzewski, Jakub and Dalmonte, Marcello},
  journal = {Phys. Rev. B},
  volume = {111},
  issue = {8},
  pages = {L081102},
  numpages = {7},
  year = {2025},
  month = {Feb},
  publisher = {American Physical Society},
  doi = {10.1103/PhysRevB.111.L081102},
  url = {https://link.aps.org/doi/10.1103/PhysRevB.111.L081102}
}

@misc{santra2025quantumresources,
  title={Quantum Resources in Non-Abelian Lattice Gauge Theories: Nonstabilizerness, Multipartite Entanglement, and Fermionic Non-Gaussianity},
  author={Santra, Gopal Chandra and Mildenberger, Julius and Ballini, Edoardo and Bottarelli, Alberto and Wauters, Matteo M. and Hauke, Philipp},
  year={2025},
  eprint={2510.07385},
  archivePrefix={arXiv},
  primaryClass={quant-ph},
  url={https://arxiv.org/abs/2510.07385}
}

@article{esposito2025magic,
  title={Magic of discrete lattice gauge theories},
  author={Esposito, Gianluca and Cepollaro, Simone and Cappiello, Luigi and Hamma, Alioscia},
  journal={Int. J. Geom. Methods Mod. Phys.},
  volume={22},
  number={06},
  pages={2550003},
  year={2024},
  doi={10.1142/S0219887825500033},
}

@article{robin2024magic,
  title={Quantum complexity fluctuations from nuclear and hypernuclear forces},
  author={Robin, Caroline E. P. and Savage, Martin J.},
  journal={Phys. Rev. C},
  volume={112},
  number={4},
  pages={044004},
  year={2025},
  doi={10.1103/r8rq-y9tb},
  eprint={2405.10268},
  archivePrefix={arXiv},
  primaryClass={nucl-th},
}

@article{brokemeier2025quantum,
  title = {Quantum magic and multipartite entanglement in the structure of nuclei},
  author = {Br\"okemeier, Florian and Hengstenberg, S. Momme and Keeble, James W. T. and Robin, Caroline E. P. and Rocco, Federico and Savage, Martin J.},
  journal = {Phys. Rev. C},
  volume = {111},
  issue = {3},
  pages = {034317},
  numpages = {30},
  year = {2025},
  month = {Mar},
  publisher = {American Physical Society},
  doi = {10.1103/PhysRevC.111.034317},
  url = {https://link.aps.org/doi/10.1103/PhysRevC.111.034317}
}

@book{ryan2002introduction,
  author    = {Ryan, Raymond A.},
  title     = {Introduction to Tensor Products of Banach Spaces},
  volume    = {73},
  publisher = {Springer},
  year      = {2002},
  doi = {10.1007/978-1-4471-3903-4},
}

@article{Rudolph2002,
  title={Further results on the cross norm criterion for separability},
  author={Rudolph, Oliver},
  journal={Quantum Inf. Process.},
  volume={4},
  number={3},
  pages={219--239},
  year={2005},
  doi={10.1007/s11128-005-5664-1},
  eprint={quant-ph/0202121},
  archivePrefix={arXiv},
  primaryClass={quant-ph},
}

\appendix
\onecolumngrid
\newpage
\section{Grassmann algebra}
\label{app:grassmann}

Though mainly standard material, we collect here for the reader's convenience our notations and definitions regarding Grassmann variables. We also define the notion of $L_p$ norm of functions (both even and odd) of c-number and Grassmann variables, to be used in the definition of the fermionic and hybrid magic.

\subsection{Grassmann variables and Berezin integral}

The Grassmann algebra with $2n$ (real) Grassmann variable can be viewed as the exterior algebra of an $2n$-dimensional vector space (that we will take to be real) $V$, namely as the tensor algebra:
\begin{equation}
    T(V) = \bigoplus_{k=0}^\infty V^{\otimes k} = \mathbb R\oplus V\oplus (V\otimes V)\oplus(V\otimes V\otimes V)\oplus\dots
\end{equation}
modded out by the ideal $\mathcal I = \left\langle \left\{v\otimes v\ |\ v\in V \right\}\right\rangle$:
\begin{equation}
    \text{Grass}(V) = T(V) / \mathcal I \equiv \bigoplus_{k=0}^{2n}\Lambda^k(V)
\end{equation}
The Grassmann algebra is finitely generated. Let us define the exterior product as $v\wedge v = v\otimes v$ mod$(\mathcal I)$. Note then the dimension of the $k^{\text{th}}$ summand in the Grassmann algebra is:
\begin{equation}
    \text{dim}\left(\Lambda^k(V)\right) = \binom{2n}{k}\,.
\end{equation}
Given a basis $\left\{\vartheta_1,\dots, \vartheta_{2n}\right\}$ of $V$, $\text{Grass}_k(V)$ can be viewed as being generated by the elements of the form:
\begin{equation}
    \vartheta_{i_1}\wedge\dots\wedge \vartheta_{i_k}\,, \text{ with } i_l\in\{1,\dots,2n\}\,.
\end{equation}
In the bulk of the paper, we denote the basis elements as $\vartheta_{j}$, and discard the wedge symbol whenever its presence is obvious. It is also convenient to introduce the complex Grassmann variables $\theta_j = \vartheta_{2j-1} + i\vartheta_{2j}$ and $\bar\theta_j = \vartheta_{2j-1} - i\vartheta_{2j}$

By construction, the Grassmann variables satisfy the anticommutation relations $\vartheta_i\vartheta_j=-\vartheta_j\vartheta_i$ and $\vartheta_j^2=0$.
Berezin integration is defined by $\int d\vartheta_j\,1=0$, $\int d\vartheta_j\,\vartheta_j=1$, and for a complex pair we set
\begin{equation}
    d^2\theta_j:=d\theta_j\,d\bar\theta_j,\qquad \int d^2\theta_j\,\bar\theta_j\,\theta_j = 1\,.
\end{equation}
Finally note that complex conjugation flips the order of factors, $\overline{\theta_{i_1}\cdots\theta_{i_k}}=\bar\theta_{i_k}\cdots\bar\theta_{i_1}$.

\subsection{$L^p$ norms on superspace}\label{app:superLp}

We briefly formalize the notion of the $L^p$ norm for functions on a superspace $\mathbb{R}^{M|N}$ with
$M$ commuting (``bosonic'') coordinates $x=(x_1,\dots,x_M)$ and $N$ real Grassmann (``fermionic'') coordinates
$\vartheta=(\vartheta_1,\dots,\vartheta_N)$.  Any \emph{even} Grassmann-valued function $f:\mathbb{R}^{M|N}\to \Lambda_N$ can always be expanded as\footnote{The normalization of the coefficients of the expansion is for convenience, cf. bulk of the paper.}
\begin{equation}
    f(x,\vartheta)=\frac{1}{2^{N}}\sum_{I\subset\{1,\dots,N\}} i^{\omega(I)}\, f_I(x)\, \vartheta_I,\qquad
    \vartheta_I\equiv \vartheta_{i_1}\cdots\vartheta_{i_{|I|}},\ \ i_1<\cdots<i_{|I|},
\end{equation}
where the numerical phase $i^{\omega(I)}$, a fourth root of unity, makes each monomial self-adjoint under complex
conjugation.  The coefficients $f_I:\mathbb{R}^M\to\mathbb{C}$ are ordinary (commuting) functions.

For $1\le p<\infty$, we define the $L^p$ norm of $f$ is
\begin{equation}\label{eq:superLp}
    \|f\|_{L^p(\mathbb{R}^{M|N})}
    \ :=\ \Bigg(\sum_{I\subset\{1,\dots,N\}} \ \big\| f_I \big\|_{L^p(\mathbb{R}^M)}^{\,p}\Bigg)^{\!1/p}\,.
\end{equation}
For $p=\infty$ we set $\|f\|_{L^\infty(\mathbb{R}^{M|N})}:=\max_{I}\|f_I\|_{L^\infty(\mathbb{R}^M)}$.

Note that: (i) If $M=0$ (purely fermionic case), Eq.~\eqref{eq:superLp} reduces to the usual $\ell^p$ norm of the finite coefficient vector $(f_I)_I$.
(ii) The norm is independent of the particular choice of real Grassmann basis for $p=2$: under any orthogonal change of variables
$\vartheta\mapsto O\,\vartheta$ with $O\in O(N)$, the coefficient vector is rotated, leaving
the right-hand side of \eqref{eq:superLp} invariant.
(iii) For $p=2$ the induced inner product is $\langle f,g\rangle=\sum_I\int_{\mathbb{R}^M}\! f_I(x)^{*}g_I(x)\,dx$, so that
$\|f\|_{L^2(\mathbb{R}^{M|N})}^2=\sum_I\|f_I\|_{L^2(\mathbb{R}^M)}^2$.
(iv) In the hybrid Wigner-function setting used in the paper, $f_I(x)$ are the (bosonic) phase-space coefficient functions multiplying
the real-Grassmann monomials; Eq.~\eqref{eq:superLp} is precisely the prescription employed to define the $p$-fermionic and $p$-hybrid magic.

Note that one could generalize the notion of norm on superspace as follows:
\begin{equation}\label{eq:superLp_generalized}
    \|f\|_{L^{p,q}(\mathbb{R}^{M|N})}
    \ :=\ \Bigg(\sum_{I\subset\{1,\dots,N\}} \ \big\| f_I \big\|_{L^q(\mathbb{R}^M)}^{\,p}\Bigg)^{\!1/p}\,.
\end{equation}
We point to the reader the subtle usage of a $L^q$ norm for the components $f_I$ instead of a $L^p$, allowing for slightly more flexibility in the definition of the hybrid boson-fermion magic, $q$ pertaining to the bosonic part, and $p$ to the fermionic part.

\section{Derivation of the hybrid magic of the dressed cat state}
\label{app:cat}

    We consider the following family of states:
    \begin{equation}
        |\psi(\beta)\rangle = \frac{|\beta\rangle\otimes|0\rangle+|-\beta\rangle\otimes|1\rangle}{\sqrt{2}}\,,
    \end{equation}
    describing an even bosonic cat state dressed by a fermionic degree of freedom.
    The bosonic $r$-ordered displacement operator reads:
    \begin{equation}
        D(\xi; r) = D(\xi; 0)\exp\left(\frac{r}{2}|\xi|^2\right)=\exp\left(\xi a^\dagger - \bar\xi a+\frac{r}{2}\,|\xi|^2\right)\,.
    \end{equation}
    The phase-point operator is then defined as the symplectic Fourier transform of the displacement operator:
    \begin{equation}
    \label{eq:bosonic_phase_point}
        \begin{aligned}
            \Upsilon(\alpha; r) &= \int \frac{d^2\xi}{\pi}\,\exp\left(\alpha\bar\xi-\bar\alpha\xi\right)D(\xi; r)\,, \\
            &= \int \frac{d^2\xi}{\pi}\,\exp\left(\xi(a^\dagger - \bar\alpha)\right)\exp\left(-\bar\xi(a - \alpha)\right)\exp\left(\frac{r-1}{2}\,|\xi|^2\right)\,,
        \end{aligned}
    \end{equation}
    leading to the following expression of the overlaps:
    \begin{equation}
        \begin{aligned}
            \langle\beta|\Upsilon(\alpha; r)|\gamma\rangle &= \langle\beta|\gamma\rangle\int \frac{d^2\xi}{\pi}\,\exp\left(\xi(\bar\beta - \bar\alpha)-\bar\xi(\gamma - \alpha)-\frac{1-r}{2}\,|\xi|^2\right)\,, \\
            &= \frac{\langle\beta|\gamma\rangle}{\pi}\int_\mathbb R\exp\left(\frac{r-1}{2}\,|\xi_1|^2+\left(\beta_1-\gamma_1+i\left(\beta_2+\gamma_2-2\alpha_2\right)\right)\xi_1\right)\,d\xi_1\times \\
            &\ \ \ \ \ \ \ \ \,\times\int_\mathbb R\exp\left(\frac{r-1}{2}\,|\xi_2|^2+\left(\beta_2-\gamma_2+i\left(\beta_1+\gamma_1-2\alpha_1\right)\right)\xi_2\right)\,d\xi_2
        \end{aligned}
    \end{equation}
    The Gaussian integral is thoroughly evaluated, we obtain:
    \begin{equation}
        \langle\beta|\Upsilon(\alpha; r)|\gamma\rangle = \frac{2}{1-r}\,\exp \left[-\frac{1}{2} \left(| \beta | ^2+| \gamma | ^2-2 \gamma  \bar\beta+\frac{\left(\bar\beta+\gamma -2 \Re(\alpha )\right)^2-\left(\beta -\bar\gamma-2 i \Im(\alpha )\right)^2}{1-r}\right)\right]
    \end{equation}
    We therefore obtain the following overlaps:
    \begin{equation}
    \label{eq:overlaps_cat}
        \begin{aligned}
            \mathcal O_\beta(\alpha; r) &= \langle\beta|\Upsilon(\alpha; r)|\beta\rangle \ \ \ \,= \frac{2}{1-r}\,\exp\left[-\frac{2 |\alpha -\beta |^2 }{1-r}\right]\,, \\
            \tilde{\mathcal O}_\beta(\alpha; r) &= \langle\beta|\Upsilon(\alpha; r)|-\beta\rangle = \frac{2}{1-r}\,\exp \left[-\frac{2 \left(| \alpha | ^2-r | \beta | ^2+2 i\Im(\alpha  \beta )\right)}{1-r}\right]\,.
        \end{aligned}
    \end{equation}

    Equipped with these overlaps, we can compute the hybrid Wigner function components:
    \begin{equation}
        \begin{aligned}
            \langle\psi(\beta)|\Upsilon(\alpha; r)|\psi(\beta)\rangle &= \frac{\mathcal O_\beta(\alpha; r)+\mathcal O_{-\beta}(\alpha; r)}{2}\,, \\
            \langle\psi(\beta)|\gamma_1\Upsilon(\alpha; r)|\psi(\beta)\rangle &= \text{Re}\left[\tilde{\mathcal O}_\beta(\alpha; r)\right]\,, \\
            \langle\psi(\beta)|\gamma_2\Upsilon(\alpha; r)|\psi(\beta)\rangle &= \text{Im}\left[\tilde{\mathcal O}_\beta(\alpha; r)\right]\,, \\
            \left\langle\psi(\beta)\left|\left(i\gamma_1\gamma_2+s\right)\Upsilon(\alpha; r)\right|\psi(\beta)\right\rangle &= \frac{1-s}{2}\,\mathcal O_{\beta}(\alpha; r)+\frac{1+s}{2}\,\mathcal O_{-\beta}(\alpha; r)\,.
        \end{aligned}
    \end{equation}
    finally leading to:
    \begin{equation}
        \begin{aligned}
            \langle\psi(\beta)|\Upsilon(\alpha; r)|\psi(\beta)\rangle &= \frac{1}{1-r}\left[\exp\left[-\frac{2 |\alpha -\beta |^2 }{1-r}\right]+\exp\left[-\frac{2 |\alpha +\beta |^2 }{1-r}\right]\right]\,, \\
            \langle\psi(\beta)|\gamma_1\Upsilon(\alpha; r)|\psi(\beta)\rangle &= \frac{2}{1-r}\exp \left[-\frac{2 \left(| \alpha | ^2-r | \beta | ^2\right)}{1-r}\right]\cos\left(\frac{4}{1-r}\,\text{Im}\left(\alpha\beta\right)\right)\,, \\
            \langle\psi(\beta)|\gamma_2\Upsilon(\alpha; r)|\psi(\beta)\rangle &= \frac{2}{1-r}\exp \left[-\frac{2 \left(| \alpha | ^2-r | \beta | ^2\right)}{1-r}\right]\sin\left(\frac{4}{1-r}\,\text{Im}\left(\alpha\beta\right)\right)\,, \\
            \left\langle\psi(\beta)\left|\left(i\gamma_1\gamma_2+s\right)\Upsilon(\alpha; r)\right|\psi(\beta)\right\rangle &= -\frac{1-s}{1-r}\exp\left[-\frac{2 |\alpha -\beta |^2 }{1-r}\right]+\frac{1+s}{1-r}\,\exp\left[-\frac{2 |\alpha +\beta |^2 }{1-r}\right]\,.
        \end{aligned}
    \end{equation}

\section{Bosonic cat state magic}

    The magic of a bosonic cat state
    \begin{equation}
        |\phi(\beta)\rangle = \frac{1}{\mathcal N(\beta)}\,\left(\,|\beta\rangle+|-\beta\rangle\right)
    \end{equation}
    can be easily computed in terms of its Wigner function. The normalization factor is given by:
    \begin{equation}
        \mathcal N(\beta) = \sqrt{2\left(1+e^{-2|\beta|^2}\right)}
    \end{equation}
    We have:
    \begin{equation}
        \begin{aligned}
            \mathcal W_{|\phi(\beta)\rangle}(\alpha; r)&=\langle\phi(\beta)|\Upsilon(\alpha; r)|\phi(\beta)\rangle = \frac{1}{\mathcal N(\beta)^2}\left(\mathcal O_\beta(\alpha; r) + \mathcal O_{-\beta}(\alpha; r) + 2\text{Re}\left[\tilde{\mathcal O}_\beta(\alpha; r)\right]\right) \\
            &=\frac{1}{(1-r)\left(1+e^{-2|\beta|^2}\right)}\Bigg(\exp\left[-\frac{2 |\alpha -\beta |^2 }{1-r}\right]+\exp\left[-\frac{2 |\alpha +\beta |^2 }{1-r}\right]+\\
            &+2\exp \left[-\frac{2 \left(| \alpha | ^2-r | \beta | ^2\right)}{1-r}\right]\cos\left(\frac{4}{1-r}\,\text{Im}\left(\alpha\beta\right)\right)\Bigg)\,.
        \end{aligned}
    \end{equation}

\section{Magic in the Holstein model}
\label{app:holstein}

    Let us focus on the approximate ground state (\ref{eq:holstein_ground_state}). Note that the fermionic phase point operator for mode $n$ reads in real basis:
    \begin{equation}
        \Delta_n(\theta; s) = 
        \frac{1}{2}\Bigg\{i\gamma_{2n-1}\gamma_{2n} +s  - i\gamma_{2n}\vartheta_{2n-1} + i\gamma_{2n-1}\vartheta_{2n}+i\vartheta_{2n-1}\vartheta_{2n}\Bigg\}\,.
    \end{equation}
    The fermionic phase point operator for the two-site problem therefore reads:
    \begin{equation}
        \begin{aligned}
            &\Delta_1(\theta; s)\Delta_2(\theta; s) =\frac{1}{4} \begin{pmatrix}
                \left(i\gamma_{1}\gamma_{2} + s\right)\left(i\gamma_{3}\gamma_{4} + s\right) \\
                -\gamma_2\left(i\gamma_{3}\gamma_{4} + s\right) \\
                \gamma_1\left(i\gamma_{3}\gamma_{4} + s\right) \\
                -\gamma_4\left(i\gamma_{1}\gamma_{2} + s\right) \\
                \gamma_3\left(i\gamma_{1}\gamma_{2} + s\right) \\
                \left(i\gamma_{3}\gamma_{4} + s\right) \\
                \left(i\gamma_{1}\gamma_{2} + s\right) \\
                -i\gamma_2\gamma_4 \\
                i\gamma_2\gamma_3 \\
                i\gamma_1\gamma_4 \\
                -i\gamma_1\gamma_3 \\
                \gamma_2 \\
                -\gamma_1 \\
                \gamma_4 \\
                -\gamma_3 \\
                -1
            \end{pmatrix}\cdot\begin{pmatrix}
                1 \\
                i\vartheta_1 \\
                i\vartheta_2 \\
                i\vartheta_3 \\
                i\vartheta_4 \\
                i\vartheta_1\vartheta_2 \\
                i\vartheta_3\vartheta_4 \\
                i\vartheta_1\vartheta_3 \\
                i\vartheta_1\vartheta_4 \\
                i\vartheta_2\vartheta_3 \\
                i\vartheta_2\vartheta_4 \\
                \vartheta_1\vartheta_3\vartheta_4 \\
                \vartheta_2\vartheta_3\vartheta_4 \\
                \vartheta_1\vartheta_2\vartheta_3 \\
                \vartheta_1\vartheta_2\vartheta_4 \\
                \vartheta_1\vartheta_2\vartheta_3\vartheta_4
            \end{pmatrix}
        \end{aligned}
    \end{equation}
    Some factors of $i$ are simply ensuring Hemiticity, as usual. 
    We recall that our ground state (\ref{eq:holstein_ground_state}) reads:
    \begin{equation}
        |\psi_0\rangle
        = \frac{1}{\sqrt{2}}
        \Big(\left|\beta\right\rangle\otimes |10\rangle
        + \left|-\beta\right\rangle\otimes |01\rangle\Big)\,.
    \end{equation}
    Focusing on fermion number-preserving operators, we compute:
    \begin{equation}
        \begin{aligned}
            \left(i\gamma_{1}\gamma_{2} + s\right)\left(i\gamma_{3}\gamma_{4} + s\right)|\psi_0\rangle &= -\frac{1-s^2}{\sqrt{2}}
            \Big(\left|\beta\right\rangle\otimes |10\rangle
            + \left|-\beta\right\rangle\otimes |01\rangle\Big) \\
            \left(i\gamma_{3}\gamma_{4} + s\right)|\psi_0\rangle &=
            -\frac{1}{\sqrt{2}}
            \Big((1-s)\left|\beta\right\rangle\otimes |10\rangle
            - (1+s)\left|-\beta\right\rangle\otimes |01\rangle\Big) \\
            \left(i\gamma_{1}\gamma_{2} + s\right)|\psi_0\rangle &=
            \frac{1}{\sqrt{2}}
            \Big((1+s)\left|\beta\right\rangle\otimes |10\rangle
            + s\left|-\beta\right\rangle\otimes |01\rangle\Big) \\
            \gamma_2\gamma_4|\psi_0\rangle
            &= \frac{1}{\sqrt{2}}
            \Big(\left|\beta\right\rangle\otimes |01\rangle
            + \left|-\beta\right\rangle\otimes |10\rangle\Big) \\
            \gamma_2\gamma_3|\psi_0\rangle
            &= \frac{1}{i\sqrt 2}
            \Big(\left|\beta\right\rangle\otimes |01\rangle
            - \left|-\beta\right\rangle\otimes |10\rangle\Big) \\
            \gamma_1\gamma_4|\psi_0\rangle
            &= -\frac{1}{i\sqrt 2}
            \Big(\left|\beta\right\rangle\otimes |01\rangle
            - \left|-\beta\right\rangle\otimes |10\rangle\Big) \\
            \gamma_1\gamma_3|\psi_0\rangle
            &= \frac{1}{\sqrt 2}
            \Big(\left|\beta\right\rangle\otimes |01\rangle
            + \left|-\beta\right\rangle\otimes |10\rangle\Big) \\
            |\psi_0\rangle&= \frac{1}{\sqrt{2}}\Big(\left|\beta\right\rangle\otimes |10\rangle+ \left|-\beta\right\rangle\otimes |01\rangle\Big)\,.
        \end{aligned}
    \end{equation}
    This leads to the following non-zero components of the hybrid Wigner function:
    \begin{equation}
        \begin{aligned}
            \left\langle\psi_0\left|\Upsilon(\alpha; r)\left(i\gamma_{1}\gamma_{2} + s\right)\left(i\gamma_{3}\gamma_{4} + s\right)\right|\psi_0\right\rangle &= -\frac{1-s^2}{2}
            \Big(\mathcal O_\beta(\alpha; r)
            + \mathcal O_{-\beta}(\alpha; r)\Big) \\
            \left\langle\psi_0\left|\Upsilon(\alpha; r)\left(i\gamma_{3}\gamma_{4} + s\right)\right|\psi_0\right\rangle &=
            -\frac{1}{2}
            \Big((1-s)\mathcal O_\beta(\alpha; r)
            - (1+s)\mathcal O_{-\beta}(\alpha; r)\Big) \\
            \left\langle\psi_0\left|\Upsilon(\alpha; r) \left(i\gamma_{1}\gamma_{2} + s\right)\right|\psi_0\right\rangle &=
            \frac{1}{2}
            \Big((1+s)\mathcal O_\beta(\alpha; r)
            + s\mathcal O_{-\beta}(\alpha; r)\Big) \\
            \left\langle\psi_0\left|\Upsilon(\alpha; r)\gamma_2\gamma_4\right|\psi_0\right\rangle
            &= \frac{1}{2}
            \Big(\tilde{\mathcal O}_\beta(\alpha; r)+\overline{\tilde{\mathcal O}_\beta(\alpha; r)}\Big) \\
            \left\langle\psi_0\left|\Upsilon(\alpha; r)\gamma_2\gamma_3\right|\psi_0\right\rangle
            &= -\frac{1}{2i}
            \Big(
            \tilde{\mathcal O}_\beta(\alpha; r)-\overline{\tilde{\mathcal O}_\beta(\alpha; r)}\Big) \\
            \left\langle\psi_0\left|\Upsilon(\alpha; r)\gamma_1\gamma_4\right|\psi_0\right\rangle
            &= \frac{1}{2i}
            \Big(\tilde{\mathcal O}_\beta(\alpha; r)-\overline{\tilde{\mathcal O}_\beta(\alpha; r)}\Big) \\
            \left\langle\psi_0\left|\Upsilon(\alpha; r)\gamma_1\gamma_3\right|\psi_0\right\rangle
            &= \frac{1}{2}
            \Big(\tilde{\mathcal O}_\beta(\alpha; r)+\overline{\tilde{\mathcal O}_\beta(\alpha; r)}\Big) \\
            \left\langle\psi_0\left|\Upsilon(\alpha; r)\right|\psi_0\right\rangle
            &= \frac{1}{2}
            \Big(\mathcal O_\beta(\alpha; r)
            + \mathcal O_{-\beta}(\alpha; r)\Big)
        \end{aligned}
    \end{equation}
    where the overlap $\mathcal O$ and $\tilde{\mathcal O}$ are given in Eq. (\ref{eq:overlaps_cat}). The hybrid magic can then be computed straightforwardly.

\section{Details of the computation for the fermionic Jaynes-Cummings model}
\label{app:JC}

    \subsection{Unitary evolution operator}

        The unitary evolution operator reads:
        \begin{equation}
            U_n(t) = e^{-iH_n t} = e^{-in\omega_c t+i
            \frac{\Delta t}{2}}\begin{pmatrix}
                \cos\left(\frac{\Omega_n t}{2}\right)-\frac{i\Delta}{\Omega_n}\sin\left(\frac{\Omega_n t}{2}\right) & -\frac{2ig\sqrt n}{\Omega_n}\sin\left(\frac{\Omega_n t}{2}\right) \\
                -\frac{2ig\sqrt n}{\Omega_n}\sin\left(\frac{\Omega_n t}{2}\right) & \cos\left(\frac{\Omega_n t}{2}\right)-\frac{i\Delta}{\Omega_n}\sin\left(\frac{\Omega_n t}{2}\right)
            \end{pmatrix}\,,
        \end{equation}
        with the detuning and generalized Rabi frequency defined as:
        \begin{equation}
            \Delta = \omega_c - \omega_a \quad \text{and} \quad \Omega_n = \sqrt{\Delta^2 + 4g^2 n}\,.
        \end{equation}
        The full evolution operator then reads:
        \begin{equation}
            U(t) = \bigoplus_{n=0}^\infty U_n(t)\,,\quad U_0(t) = (1)\,.
        \end{equation}

    \subsection{Hybrid Wigner function components}
        Focusing on the fermion number-preserving operators, we have:
        \begin{equation}
            \begin{aligned}
                \left(i\gamma_{1}\gamma_{2} + s\right)\left(i\gamma_{3}\gamma_{4} + s\right)|\psi(t)\rangle &= -\left(1-s^2\right)\sum_{n=0}^\infty \Big(\alpha_n|n\rangle_c \otimes |10\rangle_a + \beta_n|n\rangle_c \otimes |01\rangle_a\Big) \\
                \left(i\gamma_{3}\gamma_{4} + s\right)|\psi(t)\rangle &= -\sum_{n=0}^\infty \Big(\left(1-s\right)\alpha_n|n\rangle_c \otimes |10\rangle_a - \left(1+s\right)\beta_n|n\rangle_c \otimes |01\rangle_a\Big) \\
                \left(i\gamma_{1}\gamma_{2} + s\right)|\psi(t)\rangle &= \sum_{n=0}^\infty \Big(\left(1+s\right)\alpha_n|n\rangle_c \otimes |10\rangle_a - \left(1-s\right)\beta_n|n\rangle_c \otimes |01\rangle_a\Big) \\
                \gamma_2\gamma_4|\psi(t)\rangle &= \sum_{n=0}^\infty \Big(\alpha_n|n\rangle_c \otimes |01\rangle_a + \beta_n|n\rangle_c \otimes |10\rangle_a\Big) \\
                \gamma_2\gamma_3|\psi(t)\rangle &= -i\sum_{n=0}^\infty \Big(\alpha_n|n\rangle_c \otimes |01\rangle_a - \beta_n|n\rangle_c \otimes |10\rangle_a\Big) \\
                \gamma_1\gamma_4|\psi(t)\rangle &= i\sum_{n=0}^\infty \Big(\alpha_n|n\rangle_c \otimes |01\rangle_a - \beta_n|n\rangle_c \otimes |10\rangle_a\Big) \\
                \gamma_1\gamma_3|\psi(t)\rangle &= \sum_{n=0}^\infty \Big(\alpha_n|n\rangle_c \otimes |01\rangle_a + \beta_n|n\rangle_c \otimes |10\rangle_a\Big) \\
                |\psi(t)\rangle &= \sum_{n=0}^\infty \Big(\alpha_n|n\rangle_c \otimes |10\rangle_a + \beta_n|n\rangle_c \otimes |01\rangle_a\Big)
            \end{aligned}
        \end{equation}
        let us define the following series:
        \begin{equation}
            \begin{aligned}
                S_\text{sym}(r) &= \sum_{m,n=0}^\infty \mathcal O_{m,n}(\alpha; r)\Big(\bar\alpha_m\alpha_n + \bar\beta_m\beta_n\Big) \\
                S_\text{asym}(r) &= \sum_{m,n=0}^\infty \mathcal O_{m,n}(\alpha; r) \Big(\bar\beta_m\alpha_n - \bar\alpha_m\beta_n\Big) \\
                T_\text{sym}(r) &= \sum_{m,n=0}^\infty \mathcal O_{m,n}(\alpha; r) \Big(\bar\beta_m\alpha_n + \bar\alpha_m\beta_n\Big) \\
                T_\text{asym}(r) &= i\sum_{m,n=0}^\infty \mathcal O_{m,n}(\alpha; r) \Big(\bar\beta_m\alpha_n - \bar\alpha_m\beta_n\Big)
            \end{aligned}
        \end{equation}
        in terms of which:
        \begin{equation}
            \begin{aligned}
                \langle\psi(t)|\Upsilon(\alpha; r)\left(i\gamma_{1}\gamma_{2} + s\right)\left(i\gamma_{3}\gamma_{4} + s\right)|\psi(t)\rangle &= -\left(1-s^2\right)S_\text{sym}(r) \\
                \langle\psi(t)|\Upsilon(\alpha; r)\left(i\gamma_{3}\gamma_{4} + s\right)|\psi(t)\rangle &= -S_\text{asym}(r) + s\, S_\text{sym}(r) \\
                \langle\psi(t)|\Upsilon(\alpha; r)\left(i\gamma_{1}\gamma_{2} + s\right)|\psi(t)\rangle &= S_\text{asym}(r) + s\, S_\text{sym}(r) \\
                \langle\psi(t)|\Upsilon(\alpha; r)\gamma_2\gamma_4|\psi(t)\rangle &= T_\text{sym}(r) \\
                \langle\psi(t)|\Upsilon(\alpha; r)\gamma_2\gamma_3|\psi(t)\rangle &= -T_\text{asym}(r) \\
                \langle\psi(t)|\Upsilon(\alpha; r)\gamma_1\gamma_4|\psi(t)\rangle &= T_\text{asym}(r) \\
                \langle\psi(t)|\Upsilon(\alpha; r)\gamma_1\gamma_3|\psi(t)\rangle &= T_\text{sym}(r) \\
                \langle\psi(t)|\Upsilon(\alpha; r)|\psi(t)\rangle &= S_\text{sym}(r)
            \end{aligned}
        \end{equation}

        \subsection{Bosonic phase point operators Fock matrix elements}
        Let us report here the overlap coefficients in the Fock basis explicitly. We have:
        \begin{equation}
            \begin{aligned}
                \mathcal O_{m,n}(\alpha; r) &=\langle m|\Upsilon(\alpha; r)|n\rangle \\
                &= \int_{\mathbb C} \frac{d^2\xi}{\pi}\,\exp\left(\alpha\bar\xi-\bar\alpha\xi\right)\langle m|D(\xi; r)|n\rangle \\
                &= \left(\frac{r+1}{r-1}\right)^m \sqrt{\frac{m!}{n!}} \left(\frac{2}{1-r}\right)^{n-m+1} \exp \left(-\frac{2 | \alpha | ^2}{1-r}\right)
                \bar\alpha^{n-m} L_m^{(n-m)}\left(\frac{4 | \alpha | ^2}{1-r^2}\right)
            \end{aligned}
        \end{equation}
        where on the fourth line we used the classic result of Cahill and Glauber \cite{cahill1969ordered}.

        \subsection{Maximum magic time for Fock states}\label{app:maxMagicTimeFock}

\begin{figure}[h]
            \centering
            \includegraphics[width=0.5\linewidth]{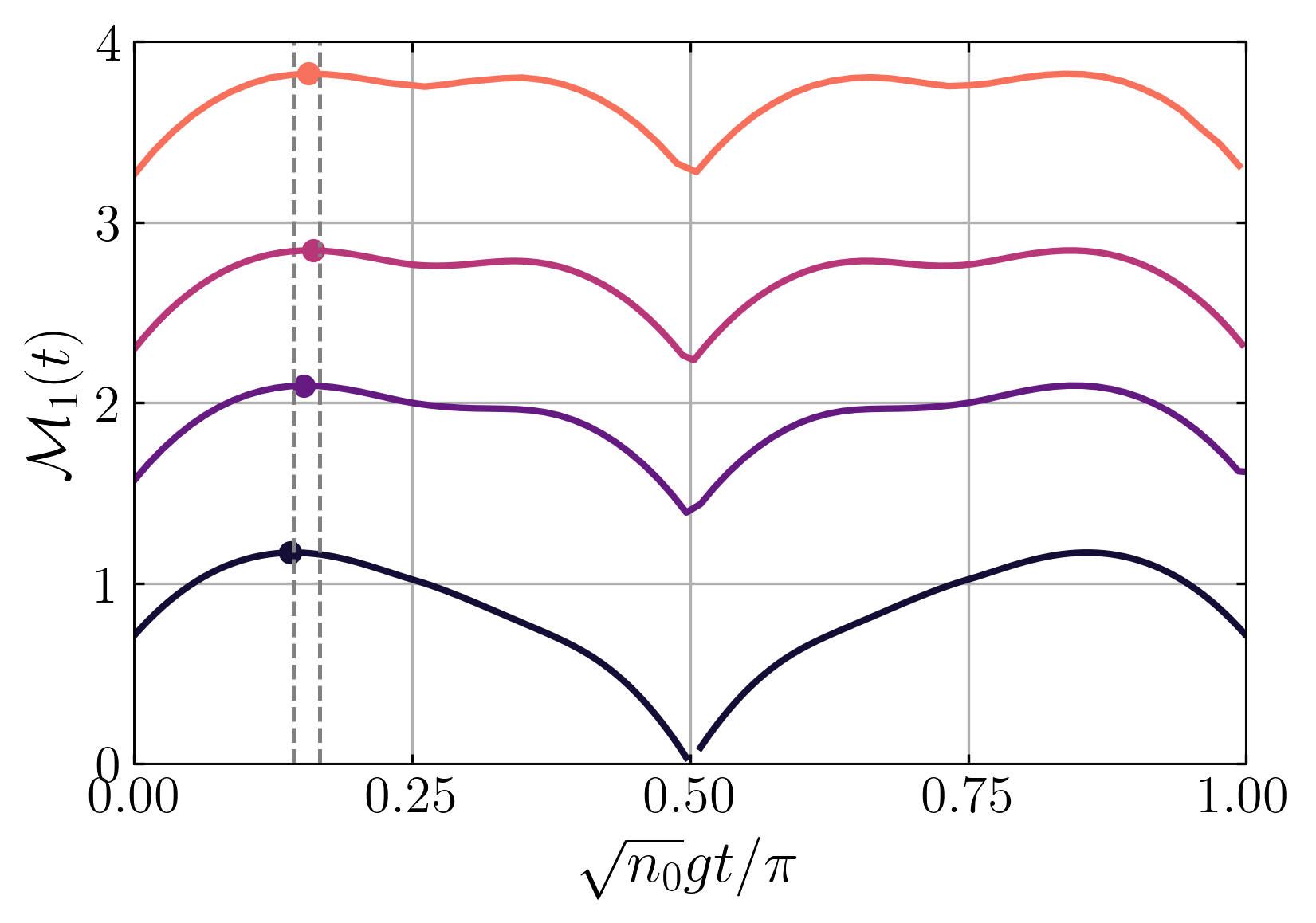}
            \caption{Hybrid magic as a function of rescaled time for different initial Fock states of the cavity $n_0=1,4, 10, 30$. The dashed vertical lines correspond to $\sqrt{n_0}g t /\pi = 1/7$ and $\sqrt{n_0}g t /\pi = 1/6$.}
            \label{fig:JC_hybMagFock_locMax}
        \end{figure}

        In this appendix we investigate the time at which the maximum value of the hybrid magic occurs for an initial Fock state of the cavity. Fig. \ref{fig:JC_hybMagFock_locMax} shows that, when rescaled by Rabi frequency, the maxima (circle) on the first period always occurs in the interval $\sqrt{n_0} g t \in [\frac{\pi}{7}, \frac{\pi}{6}]$. Another interesting observation from  this figure is that, as $n_0$ grows the hybrid magic starts to develop a second maximum in the first period, which is however still smaller than the first one.

\section{Magic cost of hybrid fermionic-oscillator gates}

    \subsection{Definition of simple hybrid gates}
    \label{app:hybrid_gates}

        We report here the list of hybrid gates that were proposed in \cite{liu2024hybrid}, translated to the fermionic language. We are only going to focus on three of them, but provide other examples of possibly interesting gates. In \cite{liu2024hybrid} are exhibited three universal \textit{instruction set architectures} (ISA):
        \begin{table}[H]
            \centering
            \begin{tabular}{ll}
            \multicolumn{2}{c}{\textbf{Universal sets of gate}} \\
            \hline
            Phase-Space ISA & $\{ \mathrm{CD}(\alpha),\, R_\varphi(\theta),\, \mathrm{BS}(\theta,\varphi) \}$ \\
            Fock-Space ISA & $\{ \mathrm{SQR}(\bm{\theta},\bm{\varphi}),\, D(\alpha),\, \mathrm{BS}(\theta,\varphi) \}$ \\
            Sideband ISA & $\{ R_\varphi(\theta),\, \mathrm{JC}(\theta),\, \mathrm{BS}(\theta,\varphi) \}$ \\
            \hline
            \end{tabular}
        \end{table}
        In the main text, we will focus on the \textit{Phase-Space} family of gates for concreteness. The explicit expression of the gates are:
        \begin{equation}
            \begin{aligned}
                D(\alpha) &= \exp\left[ \alpha a^\dagger - \alpha^* a \right] \\
                \mathrm{BS}(\theta, \varphi) &= \exp\left[ -i \frac{\theta}{2} \left( e^{i\varphi} a^\dagger b + e^{-i\varphi} a b^\dagger \right) \right] \\
                R_\varphi(\theta) &= \exp\left[-i \frac{\theta}{2} \left( c_e^\dagger c_g\, e^{-i\varphi} + c_g^\dagger c_e\, e^{i\varphi} \right)\right] \\
                \mathrm{SQR}(\bm{\theta}, \bm{\varphi}) &= \sum_n R_{\varphi_n}(\theta_n) \otimes |n\rangle\langle n| \\[1ex]
                \mathrm{JC}(\theta, \varphi) &= \exp\left[-i\theta \left(e^{i\varphi} c_g^\dagger c_e a^\dagger + e^{-i\varphi} c_e^\dagger c_g a \right)\right] \\[1ex]
                \mathrm{AJC}(\theta, \varphi) &= \exp\left[-i\theta \left(e^{i\varphi} c_e^\dagger c_g a^\dagger + e^{-i\varphi} c_g^\dagger c_e a \right)\right] \\[1ex]
                \mathrm{CD}(\alpha) &= \exp\left[(n_e - n_g) (\alpha a^\dagger - \alpha^* a)\right] \\[1ex]
            \end{aligned}
        \end{equation}

    \subsection{Computation of the non-stabilizer power of the conditional displacement gate}
    \label{app:non_stabilizer_power}

        We provide here the detailed derivation of the non-stabilizer power of the conditional displacement gate. We focus on the zero ordering parameters and $p=1$ case for concreteness for the hybrid magic.

        We are interested in computing the following expectated values $\langle\Upsilon(\beta)\otimes\gamma\rangle_{U(\alpha)|\psi\rangle}$ for all $\beta\in\mathbb C$, $\gamma\in\Gamma$ and $|\psi\rangle\in\textsc{Stab}$. One has:
        \begin{equation}
            \begin{aligned}
                \langle\Upsilon(\beta)\otimes\gamma\rangle_{U(\alpha)|\psi\rangle} &= \langle\psi|U(\alpha)^\dagger(\Upsilon(\beta)\otimes\gamma)U(\alpha)|\psi\rangle \\
                &= \sum_{m, n\in\{-1, 0, 1\}}\langle\psi_\text{g}|D(m\alpha)^\dagger\Upsilon(\beta)D(n\alpha)|\psi_\text{g}\rangle\langle\phi|\Pi_m\gamma\Pi_n|\phi\rangle \\
                &= \sum_{m, n\in\{-1, 0, 1\}}e^{-2i(m-n)\text{Im}(\alpha\bar\beta)}\,G\left(\beta-\frac{m+n}{2}\,\alpha; |\psi_\text{g}\rangle\right) F_{m,n}(\gamma; |\phi\rangle) \\
            \end{aligned}
        \end{equation}
        with the fermionic and bosonic kernels:
        \begin{equation}
            \begin{aligned}
                F_{m,n}(\gamma; |\phi\rangle) &= \langle\phi|\Pi_m\gamma\Pi_n|\phi\rangle \\
                G(\tau; |\psi_\text{g}\rangle) &= \langle\psi_\text{g}|\Upsilon(\tau)|\psi_\text{g}\rangle
            \end{aligned}
        \end{equation}
        and where we used the fact that:
        \begin{equation}
            D(m\alpha)^\dagger\Upsilon(\beta)D(n\alpha) = e^{-2i(m-n)\text{Im}(\alpha\bar\beta)}\,\Upsilon\left(\beta-\frac{m+n}{2}\,\alpha\right)\,.
        \end{equation}
        The bosonic phase-point operator for the Weyl ordering (\ref{eq:bosonic_phase_point}) can be written as:
        \begin{equation}
            \Upsilon(\tau) = 2\,D(\tau)(-1)^{a^\dagger a}D(\tau)^\dagger\,.
        \end{equation}
        The bosonic kernel can then be computed:
        \begin{equation}
            \begin{aligned}
                G(\tau; |\psi_\text{g}\rangle) &= 2\,\langle 0|S(\zeta)^\dagger D(\delta)^\dagger D(\tau)(-1)^{a^\dagger a}D(\tau)^\dagger D(\delta)S(\zeta)|0\rangle \\
                &= 2\,\langle 0|S(\zeta)^\dagger D(\tau-\delta)(-1)^{a^\dagger a}D(\delta - \tau)^\dagger S(\zeta)|0\rangle \\
                &= 2\,\langle 0|S(\zeta)^\dagger (-1)^{a^\dagger a}D(2(\delta - \tau))^\dagger S(\zeta)|0\rangle \\
                &= 2\,\langle 0|(-1)^{a^\dagger a}S(\zeta)^\dagger D(2(\delta - \tau))^\dagger S(\zeta)|0\rangle \\
                &= 2\,\langle 0|S(\zeta)^\dagger D(2(\delta - \tau))^\dagger S(\zeta)|0\rangle \\
                &= 2\,\langle 0|D(-2\mu(\tau-\delta) - 2\nu(\bar\tau-\bar\delta))|0\rangle \\
                &= 2\,\exp\left[-\frac{1}{2}\left|2\mu(\tau-\delta) + 2\nu(\bar\tau-\bar\delta)\right|^2\right]\,.
            \end{aligned}
        \end{equation}
        Let us now move to the fermionic sector. We have the following non-zero components of the fermionic kernel:
        \begin{center}
            \captionsetup{type=table}
            \label{tab:stabilizer-states}
            \footnotesize
            \setlength{\tabcolsep}{6pt}
            \renewcommand{\arraystretch}{1.2}
            \begin{tabular}{@{} l l l c l @{}}
            \toprule
            \textbf{Family} & \textbf{Parity} & \textbf{State} & \textbf{$F$} & $(\mathbb 1, P, B_1, B_2, B_3, B_4, B_5, B_6)$ \\
            \midrule
            \multirow{4}{*}{Product} & \multirow{2}{*}{Even}
              & $|00\rangle$ & $F_{00}$ & $(1,1,1,1,0,0,0,0)$ \\
             &  & $|11\rangle$ & $F_{00}$ & $(1,1,-1,-1,0,0,0,0)$ \\
             & \multirow{2}{*}{Odd}
              & $|01\rangle$ & $F_{++}$ & $(1,-1,1,-1,0,0,0,0)$ \\
             &  & $|10\rangle$ & $F_{--}$ & $(1,-1,-1,1,0,0,0,0)$ \\
            \midrule
            \multirow{5}{*}{Bell} & Even
              & $\dfrac{|00\rangle + t|11\rangle}{\sqrt{2}}$ & $F_{00}$ &
              $\bigl(1,1,0,0,\operatorname{Im}(t),-\operatorname{Im}(t),\operatorname{Re}(t),\operatorname{Re}(t)\bigr)$ \\
             & \multirow{4}{*}{Odd}
              & \multirow{4}{*}{$\dfrac{|01\rangle + s|10\rangle}{\sqrt{2}}$}
                & $F_{++}$ & $\tfrac{1}{2}(1,-1,1,-1,0,0,0,0)$ \\
             &  &  & $F_{--}$ & $\tfrac{1}{2}(1,-1,-1,1,0,0,0,0)$ \\
             &  &  & $F_{+-}$ & $\tfrac{s}{2}(0,0,0,0,-i,-i,-1,1)$ \\
             &  &  & $F_{-+}$ & $\tfrac{s}{2}(0,0,0,0,i,i,-1,1)$ \\
            \bottomrule
            \end{tabular}
            \end{center}

        Equipped with the bosonic and fermionic kernels, we can now explicitely compute $\sum_\gamma \lVert\langle\Upsilon(\beta)\otimes\gamma\rangle_{U(\alpha)|\psi\rangle}\rVert_1$ for all Majorana stabilizer states. Note that the bosonic kernel is normalized so that
        \begin{equation}
            \int_{\mathbb C}\left|G(\beta - \beta_0)\right|\frac{\text{d}^2\beta}{\pi} = 1\,,\quad\forall\beta_0\in\mathbb C\,.
        \end{equation}
        We therefore see easily that for product and for even parity Bell Majorana stabilizer states, independently of the choice of bosonic Gaussian state, we have:
        \begin{equation}
            \sum_\gamma \lVert\langle\Upsilon(\beta)\otimes\gamma\rangle_{U(\alpha)|\psi\rangle}\rVert_1 = 4\,.
        \end{equation}
        The odd parity Bell Majorana stabilizer states are more involved, but using the fact that the $L_1$ norm of a difference of two Gaussians corresponds to the total variation distance of the corresponding two probability densities, we obtain after dust settles down:
        \begin{equation}
            \sum_\gamma \lVert\langle\Upsilon(\beta)\otimes\gamma\rangle_{U(\alpha)|\psi\rangle}\rVert_1 = 2\left[1+\text{erf}\left(\sqrt 2|\mu\alpha+\nu\bar\alpha|\right) + \mathbb E_{\Theta\sim\mathcal N\left(4\text{Im}\left(\alpha\bar\delta\right), 4|\mu\alpha+\nu\bar\alpha|^2\right)}\left[\left|\sin\Theta\right|+\left|\cos\Theta\right|\right]\right]
        \end{equation}
        We are now ready to extract the hybrid magic and average over the Majorana stabilizer states. Let us introduce a finite measure $\mathfrak{m}$ on the space of pure Gaussian states (that we parameterized by the displacement $\delta$ and the squeezing $\zeta$). We finally obtain the non-stabilizer power of the conditional displacement gate:
        \begin{equation}
            \text{Power}_\mathfrak{m}(\text{CD}(\alpha)) = \frac{2}{3}\,\mathbb E_\mathfrak{m}\log\left\{\frac{1+\text{erf}\left(\sqrt 2|\mu\alpha+\nu\bar\alpha|\right) + \mathbb E_{\Theta}\left[\left|\sin\Theta\right|+\left|\cos\Theta\right|\right]}{2}\right\}
        \end{equation}
        The space of pure Gaussian states being of course non-compact, one needs to adjoin a physical cutoff in order to be able to define a finite measure. One can for instance use a cutoff on the squeezing parameter, corresponding to an energy constraint. The cutoff can be a hard cutoff or a quickly decaying smooth cutoff. One can allow for displacement or not. Note that the law of $\Theta$ itself depends on the bosonic Gaussian state instance. For simplicity, let us pick a Dirac measure on the Fock vacuum. The expectation value inside the logarithm reads then:
        \begin{equation}
            \mathbb E_{\Theta\sim\mathcal N\left(0, |2\alpha|^2\right)}\left[\left|\sin\Theta\right|+\left|\cos\Theta\right|\right] = \frac{4}{\pi}
            - \frac{8}{\pi}\sum_{n=1}^{\infty}
              \frac{e^{-8n^2|\alpha|^2}}{16 n^{2}-1}
        \end{equation}
        where we used the following series representation:
        \begin{equation}
            \begin{aligned}
            \lvert \sin x\rvert
            &= \frac{2}{\pi}
               - \frac{4}{\pi}\sum_{n=1}^{\infty}
                 \frac{\cos(2 n x)}{4 n^{2}-1},\\[6pt]
            \lvert \cos x\rvert
            &= \frac{2}{\pi}
               + \frac{4}{\pi}\sum_{n=1}^{\infty}
                 \frac{(-1)^{\,n+1}\cos(2 n x)}{4 n^{2}-1}.
            \end{aligned}
        \end{equation}
        We finally obtain:
        \begin{equation}
            \text{Power}(\text{CD}(\alpha)) = \frac{2}{3}\,\log\left\{\frac{1+\text{erf}\left(\sqrt 2|\alpha|\right)}{2}+\frac{2}{\pi}
            - \frac{4}{\pi}\sum_{n=1}^{\infty}
              \frac{e^{-8n^2|\alpha|^2}}{16 n^{2}-1}\right\}
        \end{equation}

\section{Marginal magic for the JC model}
\label{app:mutual_jc_magic}
    The state of the joint cavity-atom system at time $t$ is given by eq. (\ref{eq:JC_state}):
    \begin{equation}
        |\psi(t)\rangle = \sum_{n=0}^\infty |n\rangle_c \otimes \Big(\alpha_n(t) |10\rangle_a + \beta_n(t) |01\rangle_a\Big)\,,
    \end{equation}
    from which we obtain the following partial traces:
    \begin{equation}
        \begin{aligned}
            \rho_a(t) &= \text{Tr}_c\left(|\psi(t)\rangle\langle\psi(t)|\right) \\
            &= \sum_{n=0}^\infty \left(|\alpha_n(t)|^2 |10\rangle\langle 10| + |\beta_n(t)|^2 |01\rangle\langle 01| + \alpha_n(t)\bar\beta_n(t) |10\rangle\langle 01| + \bar\alpha_n(t)\beta_n(t) |01\rangle\langle 10|\right) \\
            \rho_c(t) &= \text{Tr}_a\left(|\psi(t)\rangle\langle\psi(t)|\right) \\
            &= \sum_{m,n=0}^\infty \left(\alpha_n(t)\bar\alpha_m(t) +\beta_n(t)\bar\beta_m(t)\right) |n\rangle\langle m|
        \end{aligned}
    \end{equation}
    By linearity of the Wigner function, we have for the cavity:
    \begin{equation}
        W_c(\alpha, t, r) = \sum_{m,n=0}^\infty \left(\alpha_n(t)\bar\alpha_m(t) +\beta_n(t)\bar\beta_m(t)\right) \mathcal O_{m,n}(\alpha; r)
    \end{equation}
    with the overlap coefficients given by eq. (\ref{eq:overlap_coefficients_JC}). Setting the ordering parameter to zero, we therefore obtain for the mana of the cavity\footnote{The global factor of $2$ is simply conventional and matches our hybrid magic definition.}:
    \begin{equation}
        \textsc{Mana}_1(\rho_c(t))=2\int\left|\sum_{m,n=0}^\infty \left(\alpha_n(t)\bar\alpha_m(t) +\beta_n(t)\bar\beta_m(t)\right) \mathcal O_{m,n}(\alpha; 0)\right|\frac{\text{d}^2\alpha}{\pi}
    \end{equation}
    Concerning the atom, we can compute the 2-fermionic mode SRE. First we compute:
    \begin{equation}
        \begin{array}{@{}l@{\qquad\qquad}l@{}}
            \langle10|(i\gamma_{1}\gamma_{2}+s)(i\gamma_{3}\gamma_{4}+s)|10\rangle = -(1-s^{2}) &
            \langle01|(i\gamma_{1}\gamma_{2}+s)(i\gamma_{3}\gamma_{4}+s)|10\rangle = 0 \\
            \langle10|(i\gamma_{3}\gamma_{4}+s)|10\rangle = -(1-s) &
            \langle01|(i\gamma_{3}\gamma_{4}+s)|10\rangle = 0 \\
            \langle10|(i\gamma_{1}\gamma_{2}+s)|10\rangle = 1+s &
            \langle01|(i\gamma_{1}\gamma_{2}+s)|10\rangle = 0 \\
            \langle10|\gamma_{2}\gamma_{4}|10\rangle = 0 &
            \langle01|\gamma_{2}\gamma_{4}|10\rangle = 1 \\
            \langle10|\gamma_{2}\gamma_{3}|10\rangle = 0 &
            \langle01|\gamma_{2}\gamma_{3}|10\rangle = -i \\
            \langle10|\gamma_{1}\gamma_{4}|10\rangle = 0 &
            \langle01|\gamma_{1}\gamma_{4}|10\rangle = i \\
            \langle10|\gamma_{1}\gamma_{3}|10\rangle = 0 &
            \langle01|\gamma_{1}\gamma_{3}|10\rangle = 1 \\
            \langle10|10\rangle = 1 &
            \langle01|10\rangle = 0
        \end{array}
    \end{equation}

    \begin{equation}
        \begin{array}{@{}l@{\qquad\qquad}l@{}}
            \langle01|(i\gamma_{1}\gamma_{2}+s)(i\gamma_{3}\gamma_{4}+s)|01\rangle = -(1-s^{2}) &
            \langle10|(i\gamma_{1}\gamma_{2}+s)(i\gamma_{3}\gamma_{4}+s)|01\rangle = 0 \\
            \langle01|(i\gamma_{3}\gamma_{4}+s)|01\rangle = 1+s &
            \langle10|(i\gamma_{3}\gamma_{4}+s)|01\rangle = 0 \\
            \langle01|(i\gamma_{1}\gamma_{2}+s)|01\rangle = -(1-s) &
            \langle10|(i\gamma_{1}\gamma_{2}+s)|01\rangle = 0 \\
            \langle01|\gamma_{2}\gamma_{4}|01\rangle = 0 &
            \langle10|\gamma_{2}\gamma_{4}|01\rangle = 1 \\
            \langle01|\gamma_{2}\gamma_{3}|01\rangle = 0 &
            \langle10|\gamma_{2}\gamma_{3}|01\rangle = i \\
            \langle01|\gamma_{1}\gamma_{4}|01\rangle = 0 &
            \langle10|\gamma_{1}\gamma_{4}|01\rangle = -i \\
            \langle01|\gamma_{1}\gamma_{3}|01\rangle = 0 &
            \langle10|\gamma_{1}\gamma_{3}|01\rangle = 1 \\
            \langle01|01\rangle = 1 &
            \langle10|01\rangle = 0
        \end{array}
    \end{equation}
    Setting the ordering parameter to zero and $p=1$ to match the mana definition of the bosonic sector, we obtain:
    \begin{equation}
        \begin{aligned}
            \text{SRE}_{\frac{1}{2}}(\rho_a(t)) &= 2\log\left(\frac{1}{4}\sum_{I}\left|\text{Tr}\left(\rho_a(t)\Gamma_I\right)\right|\right) \\
            &= 2\log\Bigg\{\left|\sum_{n=0}^\infty\left(|\alpha_n(t)|^2+|\beta_n(t)|^2\right)\right|+\left|\sum_{n=0}^\infty\left(|\alpha_n(t)|^2-|\beta_n(t)|^2\right)\right|\\
            &\ +\left|\sum_{n=0}^\infty\left(\alpha_n(t)\bar\beta_n(t)+\bar\alpha_n(t)\beta_n(t)\right)\right|+\left|\sum_{n=0}^\infty i\left(\alpha_n(t)\bar\beta_n(t)-\bar\alpha_n(t)\beta_n(t)\right)\right|\Bigg\} - 2\log 2
        \end{aligned}
    \end{equation}
   We present in the main text the mutual hybrid magic for the Jaynes-Cummings (JC) model, as a function of time $t$. The mutual hybrid magic quantifies the non-classical correlations between the bosonic and fermionic sectors in the JC model.

\section{Bounds on the hybrid magic}\label{sec:hybrid-bounds}

\subsection{Lower bounds}

Let $\rho$ be a density operator on $\mathcal{H}_{\mathrm b}\otimes\mathcal{H}_{\mathrm f}$.
For Weyl ordering (i.e.\ $r=s=0$), the hybrid Wigner function admits the real-Grassmann expansion
(cf.\ Eq.~(23) in the main text)
\begin{equation}
W_\rho(\alpha,\vartheta)=\frac{1}{2^N}\sum_I i^{\omega(I)}\,w_I(\alpha)\,\vartheta_I,
\qquad
w_I(\alpha)=\mathrm{Tr}\!\left[\rho\big(\Upsilon(\alpha)\otimes\Gamma_I\big)\right].
\end{equation}
Recall that the superspace $L_p$ norm $\|W_\rho\|_{L_p(\mathbb R^{2M|2N})}$ is defined from the
(coefficient) functions $w_I(\alpha)$ as in Supplemental Sec.~A.2, Eq.~(88).

\paragraph{$L_\infty$ bound.}
For all $\alpha$ and $I$ one has $|w_I(\alpha)|\le 2^M$. Consequently,
\begin{equation}
\|W_\rho\|_{L_\infty(\mathbb R^{2M|2N})}\le 2^{M}.
\end{equation}
Indeed, set $A=\Upsilon(\alpha)\otimes \Gamma_I$. Since $\rho\geq0$ and $\mathrm{Tr}\rho=1$,
$|\mathrm{Tr}(\rho A)|\le \|A\|_\infty$. By unitarity of displaced parity, $\|\Upsilon(\alpha)\|_\infty=2^M$,
and each Majorana string satisfies $\|\Gamma_I\|_\infty=1$. Hence $|w_I(\alpha)|\le 2^M$.
The claimed $\|W_\rho\|_{L_\infty}$ bound then follows from the definition of the superspace $L_\infty$ norm.

\paragraph{$L_2$ identity.}
The Weyl-ordered hybrid Wigner function satisfies
\begin{equation}
\|W_\rho\|_{L_2(\mathbb R^{2M|2N})}^2=2^{M+N}\,\mathrm{Tr}(\rho^2).
\end{equation}
Indeed, expand $\rho$ in the fermionic Hilbert--Schmidt orthonormal basis of Majorana strings:
\[
\rho=2^{-N}\sum_I \rho_I\otimes \Gamma_I,\qquad
\rho_I:=\mathrm{Tr}_{\mathrm f}\!\left[\rho(\mathbb I\otimes \Gamma_I)\right].
\]
Then, by orthogonality of the $\Gamma_I$'s, one has $w_I(\alpha)=\mathrm{Tr}_{\mathrm b}(\rho_I\Upsilon(\alpha))$.
By the bosonic Moyal identity,
\[
\int \Big|\mathrm{Tr}_{\mathrm b}(\rho_I\Upsilon(\alpha))\Big|^2\,\frac{d^{2M}\alpha}{\pi^M}
= 2^M\,\mathrm{Tr}_{\mathrm b}(\rho_I^2).
\]
Therefore,
\[
\|W_\rho\|_{2}^2
=\sum_I \int |w_I(\alpha)|^2\,\frac{d^{2M}\alpha}{\pi^M}
=2^M\sum_I \mathrm{Tr}_{\mathrm b}(\rho_I^2).
\]
On the other hand, using again orthogonality,
\[
\mathrm{Tr}(\rho^2)=2^{-N}\sum_I \mathrm{Tr}_{\mathrm b}(\rho_I^2).
\]
Combining yields $\|W_\rho\|_2^2=2^{M+N}\mathrm{Tr}(\rho^2)$.

\paragraph{$L_p$ bounds for $p\geq2$.}
For every $p\in[2,\infty]$,
\begin{equation}
\|W_\rho\|_{L_p(\mathbb R^{2M|2N})}
\le \|W_\rho\|_{L_2(\mathbb R^{2M|2N})}^{2/p}\,
\|W_\rho\|_{L_\infty(\mathbb R^{2M|2N})}^{1-2/p}
\le 2^{\,M(1-1/p)+N/p}\,\mathrm{Tr}(\rho^2)^{1/p}.
\end{equation}
Indeed, for $p\geq2$ and any measurable $f$ one has
$\|f\|_p^p=\int |f|^{p-2}|f|^2\le \|f\|_\infty^{p-2}\|f\|_2^2$.
Applying this to each coefficient $w_I$ and summing over $I$ yields
$\|W_\rho\|_p^p\le \|W_\rho\|_\infty^{p-2}\|W_\rho\|_2^2$,
hence the interpolation bound above.
In particular,
\begin{equation}
\|W_\rho\|_{L_p(\mathbb R^{2M|2N})}^{p}\le 2^{M(p-1)+N}\,\mathrm{Tr}(\rho^2).
\end{equation}

\paragraph{Consequence for the hybrid magic (Weyl ordering).}
Using the definition of the $p$-hybrid magic (Eq.~(24) in the main text),
\begin{equation}
M_p(\rho;\bm 0)=\frac{1}{1-\tfrac p2}\log\!\left(\frac{1}{2^N}\|W_\rho\|_{L_p(\mathbb R^{2M|2N})}^p\right),
\end{equation}
we obtain for every $p>2$ the lower bound
\begin{equation}
M_p(\rho;\bm 0)\ \ge\ \frac{1}{1-\tfrac p2}\,
\log\!\left(2^{M(p-1)}\,\mathrm{Tr}(\rho^2)\right).
\end{equation}
In particular, when $M=0$ (purely fermionic case), this reduces to
$M_p(\rho;\bm 0)\ge 0$, consistent with fermionic stabilizer states having zero magic.

\subsection{Upper bounds}
\label{app:upper_bounds_hybrid_magic}

We now derive the upper bounds quoted in Sec.~\ref{sec:resource_theory}. The cutoff needed in the argument is only a cutoff in the bosonic phase space. Indeed, for fixed $N$ the fermionic part of the hybrid expansion is a finite sum over $2^{2N}$ Majorana strings. We therefore avoid introducing a measure on the full hybrid phase space and work directly with the bosonic measure
\begin{equation}
    \mathrm d\mu_{\mathrm B}(\bm\alpha)
    =
    \frac{\mathrm d^{2M}\alpha}{\pi^M}.
\end{equation}
For Weyl ordering, recall that
\begin{equation}
    A_p(\rho)
    =
    \frac{1}{2^N}
    \sum_I
    \int_{\mathbb C^M}
    |w_I(\bm\alpha)|^p
    \mathrm d\mu_{\mathrm B}(\bm\alpha),
    \qquad
    \mathcal M_p(\rho;\bm 0)
    =
    \frac{1}{1-\frac p2}\log A_p(\rho).
\end{equation}
In particular,
\begin{equation}
    \mathcal M_1(\rho;\bm 0)=2\log A_1(\rho),
    \qquad
    \mathcal M_4(\rho;\bm 0)=-\log A_4(\rho).
\end{equation}
The sign difference between these two expressions is important. An upper bound on $\mathcal M_1$ follows from an upper bound on $A_1$, whereas an upper bound on $\mathcal M_4$ follows from a lower bound on $A_4$.

We use the Weyl-ordered $L_2$ identity in the form
\begin{equation}
    A_2(\rho)
    =
    \frac{1}{2^N}
    \sum_I
    \int_{\mathbb C^M}
    |w_I(\bm\alpha)|^2
    \mathrm d\mu_{\mathrm B}(\bm\alpha)
    =
    2^M\,\mathrm{Tr}(\rho^2).
\end{equation}
Assume that all coefficient functions $w_I$ are supported, or effectively truncated, in a common bosonic region $\Omega\subset\mathbb C^M$ of finite volume
\begin{equation}
    \mathcal V_{\mathrm B}(\Omega)
    =
    \int_\Omega
    \mathrm d\mu_{\mathrm B}(\bm\alpha)
    <
    \infty.
\end{equation}
For $p=1$, Cauchy's inequality in the bosonic variables gives, for each fixed Majorana string $I$,
\begin{equation}
    \int_\Omega |w_I(\bm\alpha)|\,\mathrm d\mu_{\mathrm B}(\bm\alpha)
    \leq
    \mathcal V_{\mathrm B}(\Omega)^{1/2}
    \left(
    \int_\Omega |w_I(\bm\alpha)|^2\,\mathrm d\mu_{\mathrm B}(\bm\alpha)
    \right)^{1/2}.
\end{equation}
Applying Cauchy's inequality once more to the finite sum over the $2^{2N}$ Majorana strings yields
\begin{equation}
\begin{aligned}
    A_1(\rho)
    &\leq
    \frac{\mathcal V_{\mathrm B}(\Omega)^{1/2}}{2^N}
    \sum_I
    \left(
    \int_\Omega |w_I(\bm\alpha)|^2\,\mathrm d\mu_{\mathrm B}(\bm\alpha)
    \right)^{1/2} \\
    &\leq
    \frac{\mathcal V_{\mathrm B}(\Omega)^{1/2}}{2^N}
    \left(2^{2N}\right)^{1/2}
    \left(
    \sum_I
    \int_\Omega |w_I(\bm\alpha)|^2\,\mathrm d\mu_{\mathrm B}(\bm\alpha)
    \right)^{1/2} \\
    &=
    \left(
    2^{M+N}\,\mathcal V_{\mathrm B}(\Omega)\,\mathrm{Tr}(\rho^2)
    \right)^{1/2}.
\end{aligned}
\end{equation}
Since the prefactor in $\mathcal M_1=2\log A_1$ is positive, the inequality direction is preserved:
\begin{equation}
    \mathcal M_1(\rho;\bm 0)
    \leq
    \log\left(
    2^{M+N}\,\mathcal V_{\mathrm B}(\Omega)\,\mathrm{Tr}(\rho^2)
    \right).
\end{equation}

For $p=4$, one must keep track of the negative prefactor. Applying Cauchy's inequality to the finite collection of functions $|w_I|^2$ gives
\begin{equation}
\begin{aligned}
    \left(
    \sum_I
    \int_\Omega |w_I(\bm\alpha)|^2\,\mathrm d\mu_{\mathrm B}(\bm\alpha)
    \right)^2
    &\leq
    2^{2N}\,\mathcal V_{\mathrm B}(\Omega)
    \sum_I
    \int_\Omega |w_I(\bm\alpha)|^4\,\mathrm d\mu_{\mathrm B}(\bm\alpha).
\end{aligned}
\end{equation}
Equivalently,
\begin{equation}
    A_2(\rho)^2
    \leq
    2^N\,\mathcal V_{\mathrm B}(\Omega)\,A_4(\rho).
\end{equation}
Thus
\begin{equation}
    A_4(\rho)
    \geq
    \frac{A_2(\rho)^2}{2^N\,\mathcal V_{\mathrm B}(\Omega)}
    =
    \frac{2^{2M}\,\mathrm{Tr}(\rho^2)^2}
    {2^N\,\mathcal V_{\mathrm B}(\Omega)}.
\end{equation}
Now $x\mapsto -\log x$ is decreasing. Consequently, the lower bound on $A_4$ becomes an upper bound on $\mathcal M_4$:
\begin{equation}
    \mathcal M_4(\rho;\bm 0)
    =
    -\log A_4(\rho)
    \leq
    \log\left(
    \frac{2^N\,\mathcal V_{\mathrm B}(\Omega)}
    {2^{2M}\,\mathrm{Tr}(\rho^2)^2}
    \right).
\end{equation}
For pure states, one obtains
\begin{equation}
\begin{aligned}
    \mathcal M_1(\ket{\psi};\bm 0)
    &\leq
    \log\left(2^{M+N}\,\mathcal V_{\mathrm B}(\Omega)\right), \\
    \mathcal M_4(\ket{\psi};\bm 0)
    &\leq
    \log\left(2^{N-2M}\,\mathcal V_{\mathrm B}(\Omega)\right).
\end{aligned}
\end{equation}
These are bosonic finite-volume, finite-cutoff, or energy-window estimates. Since the fermionic sector is finite-dimensional, no cutoff is needed there. For the full bosonic phase space one has $\mathcal V_{\mathrm B}(\Omega)=\infty$, and the above inequalities become vacuous; without a constraint on the region explored in bosonic phase space, the continuous-variable sector does not give a finite state-independent upper bound on the hybrid $p=1$ or $p=4$ magic.

\section{Outline of the pure-state faithfulness mechanism}
\label{app:faithfulness_outline}

We briefly outline the logic behind the pure-state faithfulness statement used in the main text. This is not a proof; the complete argument will appear in the companion paper~\cite{hybrid_resource}. The purpose of this appendix is only to explain why the condition of vanishing Weyl-ordered $p=1$ hybrid magic is strong enough to identify the free pure sector.

For Weyl ordering, write the dressed Wigner functions as
\begin{equation}
    w_I(\bm\alpha)
    =
    \mathrm{Tr}\big(
    \rho\,\Upsilon(\bm\alpha;0)\otimes\Gamma_I
    \big),
\end{equation}
and define the corresponding $p=1$ hybrid factor
\begin{equation}
    A_{\mathrm{hyb}}(\rho)
    =
    \frac{1}{2^N}
    \sum_I
    \int_{\mathbb C^M}
    |w_I(\bm\alpha)|
    \frac{\mathrm d^{2M}\alpha}{\pi^M}.
\end{equation}
Vanishing hybrid magic is the statement $A_{\mathrm{hyb}}(\rho)=1$. For a product free pure state, this is immediate: the bosonic Gaussian factor has a non-negative Wigner function with unit $L^1$ norm, while the Majorana stabilizer factor has exactly $2^N$ non-zero Majorana-string coefficients of unit modulus. Hence the normalization is chosen so that every pure state in
\begin{equation}
    \textsc{hStab}
    =
    \left\{
    |\psi_{\mathrm g}\rangle\otimes|\phi\rangle
    \;\middle|\;
    |\psi_{\mathrm g}\rangle\in\textsc{Gauss},\;
    |\phi\rangle\in\textsc{mStab}
    \right\}
\end{equation}
has zero hybrid magic.

The non-trivial direction is the converse. The key point is that the family $\{w_I\}_I$ is not just a list of unrelated functions: it is the Weyl-Majorana decomposition of a single hybrid density operator. Under a mild finite-energy regularity assumption, the super-$L^1$ norm controls a natural cross norm for the operator $\rho$ across the boson--fermion bipartition~\cite{ryan2002introduction,Rudolph2002}. For pure states, this cross norm detects Schmidt rank: its minimal value is attained only by product vectors. Thus
\begin{equation}
    A_{\mathrm{hyb}}(|\Psi\rangle\langle\Psi|)
    =
    1
    \quad\Longrightarrow\quad
    |\Psi\rangle
    =
    |\psi_{\mathrm B}\rangle\otimes|\phi_{\mathrm F}\rangle.
\end{equation}
Intuitively, zero hybrid magic leaves no room for hidden boson--fermion correlations, because any such correlation increases the amount of signed hybrid phase-space data needed to represent the state.

Once product structure is obtained, the problem separates into its two familiar local parts. On the bosonic side, the equality of the $L^1$ norm with the normalization of the Wigner function forces the Wigner function to be non-negative, and the pure-state Hudson theorem then identifies $|\psi_{\mathrm B}\rangle$ as Gaussian. On the fermionic side, the Majorana coefficients
\begin{equation}
    t_I
    =
    \langle\phi_{\mathrm F}|\Gamma_I|\phi_{\mathrm F}\rangle
\end{equation}
obey an extremality condition: the minimal possible value of $\sum_I |t_I|$ for a pure state is attained precisely by Majorana stabilizer states. Therefore $|\phi_{\mathrm F}\rangle$ must be a Majorana stabilizer state.

Putting these steps together gives the pure-state faithfulness statement
\begin{equation}
    \mathcal M_1(|\Psi\rangle\langle\Psi|;\bm 0)=0
    \quad\Longleftrightarrow\quad
    |\Psi\rangle\in\textsc{hStab},
\end{equation}
within the regular pure-state regime described above. The proof should be viewed as a sequence of filters: the cross-norm step removes hybrid correlations, Hudson's theorem identifies the bosonic factor, and Majorana-string extremality identifies the fermionic factor.

\end{document}